\documentclass[iop,twocolappendix,numberedappendix,appendixfloats]{emulateapj}

\usepackage{natbib}
\usepackage[bookmarks,breaklinks]{hyperref}
   \hypersetup{
     colorlinks,
     citecolor=blue,
     filecolor=magenta,
     linkcolor=blue,
     urlcolor=cyan
}
\usepackage{multirow}
\usepackage{longtable}
\usepackage{threeparttablex}
\usepackage{csquotes}

\newcommand{\OIII}{\mbox{[O\,\textsc{iii}]}}
\newcommand{\NII}{\mbox{[N\,\textsc{ii}]}}
\newcommand{\SII}{\mbox{[S\,\textsc{ii}]}}
\newcommand\tna{\,\tablenotemark{a}}
\newcommand\tnb{\,\tablenotemark{b}}
\newcommand\tnc{\,\tablenotemark{c}}
\newcommand\tnd{\,\tablenotemark{d}}
\newcommand\tne{\,\tablenotemark{e}}
\newcommand\tnf{\,\tablenotemark{f}}

\shorttitle{Structure of ionized gas outflows in Type 2 AGNs - II}
\shortauthors{Karouzos et al.}

\begin{document}


\title{Unravelling the complex structure of AGN-driven outflows:\\ II. Photoionization and energetics}


\author{Marios Karouzos$^{1}$, Jong-Hak Woo$^{1}$, Hyun-Jin Bae$^{2}$}
\affil{$^{1}$Astronomy Program, Department of Physics and Astronomy, Seoul National University, Seoul 151-742, Republic of Korea}
\affil{$^{2}$Department of Astronomy and Center for Galaxy EVolution Research, Yonsei University, Seoul 120-749, Republic of Korea}
\email{email: woo@astro.snu.ac.kr}




\begin{abstract}
Outflows have been shown to be prevalent in galaxies hosting luminous active galactic nuclei (AGNs) and present a physically plausible way to couple the AGN energy output with the interstellar medium of their hosts. Despite their prevalence, accurate characterization of these outflows has been challenging. In the second of a series of papers, we use Gemini Multi-Object Spectrograph IFU data of 6 local (z$<0.1$) and {moderate-luminosity} Type 2 AGNs to study the ionization properties and energetics of AGN-driven outflows. We find strong evidence that connect the extreme kinematics of the ionized gas with the AGN photoionization. The kinematic component related to the AGN-driven outflow is clearly separated from other kinematic components, such as virial motions or rotation, on the velocity and velocity dispersion diagram. Our spatially resolved kinematic analysis reveals that from 30\% up to 90\% of the total mass and kinetic energy of the outflow is contained within the central kpc of the galaxy. The spatially integrated mass and kinetic energy of the gas entrained in the outflow correlates well with the AGN bolometric luminosity and results in energy conversion efficiencies between 0.01\% and 1\%. Intriguingly, we detect ubiquitous signs of ongoing circumnuclear star formation. Their small size, the centrally contained mass and energy, and the universally detected circumnuclear star formation cast doubts on the potency of these AGN-driven outflows as agents of galaxy-scale negative feedback.
\end{abstract}



\section{Introduction}
\label{sec:intro}

Outflows in different phases of the interstellar medium (ISM) have been commonly found in the host galaxies of active galactic nuclei (AGNs), using spatially integrated (\citealt{Heckman1981,Nelson1995,Greene2005b,Mullaney2013,Bae2014,Woo2016,Harrison2016}) and spatially resolved spectroscopy (\citealt{Nesvadba2006,Storchi2010,Riffel2013,Rupke2013,Harrison2014,McElroy2015,Karouzos2016}). These outflows provide a plausible way of coupling the energy output of accreting supermassive black holes (SMBHs) with the ISM of their host galaxies. As a result, they have been invoked within the context of the co-evolution of SMBHs and galaxies (e.g., \citealt{Kauffmann2000,Kormendy2013}) as a way to explain the observed link between the mass of SMBHs and properties of their hosts (e.g., \citealt{Magorrian1998,Ferrarese2000,Gebhardt2000,Tremaine2002,Marconi2003,Woo2010,Woo2013,Woo2015}) and other apparent correlations between the luminosity and accretion rate of AGNs with the levels of star formation in their hosts (e.g., \citealt{Netzer2009,Woo2012,Karouzos2014b,Matsuoka2015}). {Outflows have been associated with AGN in a wide range of luminosities (e.g.,\citealt{MuellerSanchez2013,Carniani2015,Cheung2016}) and as such are often classified according to the physical mechanism thought to drive them. Roughly divided into mechanically and radiatively driven outflows (e.g., \citealt{Fabian2012}), the former are usually associated with low luminosity AGNs and are often powered by radio jets or radiatively inefficient accretion disks. In contrast, the latter are usually characteristic of high Eddington ratio accretion disks and high luminosity AGNs. Radiatively-driven outflows are further subdivided into energy or momentum-driven, reflecting the physical processes behind their launching and acceleration.}

It is established that AGN-driven outflows can accelerate the ISM to very large velocities that often exceed the velocity required to escape the gravitational pull of their hosts. Moreover, there are strong indications that most outflows suffer from some degree of dust extinction that results to predominantly blueshifted emission (e.g., \citealt{Karouzos2016}, hereafter Paper I). In the case of nearby Type 2 AGN, these outflows show clearly defined biconical structures of expanding ionized gas (e.g., \citealt{Evans1993,Crenshaw2000,Kaiser2000}). Observations of AGN at higher redshifts and higher luminosities reveal ionized gas moving at high velocities (e.g., \citealt{Liu2013,Harrison2014}) that is often more uniformly distributed. This may be an effect of limited spatial resolution and convolution of a central very bright source with the instruments point-spread-function (e.g., \citealt{Husemann2015}).

Beyond their phenomenological description, direct comparison of these outflows with (semi-)analytical AGN feedback models {(e.g., \citealt{Silk2010,Silk2013,Zubovas2014b, Costa2014,Croton2016})}, sub-grid AGN feedback prescriptions in large cosmological simulations (e.g., \citealt{Sijacki2015}), {and hydrodynamical simulations (e.g., \citealt{Bieri2016})} requires first to establish that they are in fact powered by the AGN. This is not a trivial question, especially in systems where nuclear accretion and star formation activity may contribute comparably to the total emission of the galaxy. The flux ratios of different emission lines (primarily H$\beta$, \OIII $\lambda 5007\AA$, \NII $\lambda 6584\AA$, and H$\alpha$) have been used to derive the so-called Baldwin-Phillips-Terlevich -- BPT -- classification (e.g., \citealt{Baldwin1981,Veilleux1987}) and characterize the photoionization field in the galaxy. Spatially resolved BPT maps have been used to distinguish between regions photoionized by the AGN and massive stars and therefore infer the processes behind the often co-spatial extreme kinematics characterizing the ionized gas (e.g., \citealt{Sharp2010,Liu2015,McElroy2015}). 

Second, observations need to constrain the mass and energy carried by these outflows. The former is often derived based on the luminosities of excited gas species (usually the bright H$\alpha$ line) and assuming standard photoionization theory (e.g., \citealt{Osterbrock1989}). However, the ionized gas mass is only a small part of the total gas mass and sub-mm observations have revealed orders of magnitude larger mass in the cold and dense molecular gas entrained in AGN outflows (e.g., \citealt{Cicone2012,Cicone2014}), which is also expected from theoretical models (e.g., \citealt{Zubovas2014a}). Given reasonable assumptions for the size (see Paper I), the geometry, the estimated gas mass, and the measured velocities of the outflow, several studies have derived the energy carried by these outflows. The way of estimating the outflow energy varies throughout the literature with many studies opting for a straightforward assumption of the kinetic energy of a spherically expanding outflow of constant velocity (e.g., \citealt{Holt2006,Harrison2014}), while others prefer more physically motivated models (e.g., \citealt{Heckman1990,Nesvadba2006}). Given the uncertainty of both the mass and the geometry of the outflow, energy is estimated with an order of magnitude uncertainty. The energy coupling efficiency (defined as the ratio of the outflow energy to the AGN bolometric luminosity) has been found to be {of the order of 0.001\% to 10\% (e.g., \citealt{Storchi2010,Mueller2011,Liu2013,Harrison2014,Schnorr2016})}, with mass outflow rates often order of magnitude higher than typical AGN accretion rates.

In Paper I, we presented the analysis of Gemini Multi-Object Spectrograph (GMOS) optical integral field unit (IFU) data for 6 {moderate-luminosity} Type 2 AGNs in the local Universe (z$<0.1$), that have been selected based on strong ionized gas outflow signatures in their Sloan Digital Sky Survey (SDSS; \citealt{Abazajian2009}) spectra. By performing a careful determination of their stellar kinematics and separating the gravitational from the non-gravitational component of the ionized gas, we showed that the narrow component of the emission lines is mostly gravitationally dominated and follows the stellar virial motion. The broad component on the other hand was found to be significantly offset from the systemic velocity, with \OIII\ velocity dispersions reaching up to 5 times the stellar velocity dispersion, $\sigma_{*}$. Moreover, we showed that there exists a clear negative gradient with radius for both velocity and velocity dispersion of the ionized gas. The kinematics of the broad components experience a plateau in the innermost region of the galaxy and then gradually converge to the kinematics of the stellar component. Based on our analysis, we defined the kinematic size of these outflows as the region within which the outflow dominates the dynamics of the gas and found it to range between 1.3 and 2.1 kpc. Photometrically-defined sizes based on the S/N of \OIII were shown to overestimate the actual outflow size by at least a factor of 2.

Here, in the second of a series of papers, we exploit the rich Gemini/GMOS dataset and the improved estimation of sizes and kinematics of these outflows to study what is driving the individual kinematic components, and determine the energetics of the outflows. In Section \ref{sec:sample} we briefly summarize the sample selection, observations, methodology, and the analysis we follow. In Section \ref{sec:results} we present the photoionisation properties  of the ionized gas and the connection with the gas kinematics. In Section \ref{sec:physical} we derive the physical properties of the outflows and combine previous literature samples to explore correlations between the outflow properties and AGN luminosity. In Sections \ref{sec:discussion} and \ref{sec:conclusions} we discuss our results and offer a summary and our conclusions, respectively. Finally, in the Appendix we present the radial profiles of the electron density and provide comments on individual sources. Throughout the paper we assume the following cosmological parameters for a cold $\Lambda$-CDM Universe: h$_{0}$=0.71, $\Omega_{m}$=0.27, and $\Omega_{\Lambda}$=0.73.

\section{Sample, observations, and methodology}
\label{sec:sample}

\subsection{Sample selection and observations}
Here we provide a brief summary of the sample selection, observations, and refer the reader to Paper I for a more detailed description. {Drawing from a large sample of $\sim$39,000 SDSS DR7 (\citealt{Abazajian2009}) Type 2 AGNs at $z<0.3$ (\citealt{Woo2016}), we select sources with dust-corrected\footnote{{Luminosities are corrected for dust extinction based on the Balmer decrement (assuming an intrinsic value of 2.89), measured in the SDSS spectra and reported in \citet{Bae2014} and \citet{Woo2016}, and the \citet{Calzetti1999} extinction law.}}} $L_{\mathrm{\OIII}}>10^{42}$ erg s$^{-1}$, \OIII\ velocity $|v_{\mathrm{\OIII}}|>250$ km s$^{-1}$, \OIII\ velocity dispersion $\sigma_{\mathrm{\OIII}}>400$ km s$^{-1}$, and redshift $<0.1$. These criteria select local Type 2 AGNs with extreme \OIII\ kinematics, indicative of strong outflows driven by the central AGN, which {comprise only 0.1\% of the total sample (at z$<0.1$) and} are arguably the best local candidates to study the properties of these outflows. {This statistically complete sample comprises 29 sources, 6 out of which where observed with Gemini-North and are the focus of this paper\footnote{{The remaining 23 sources were observed with a mix of other IFU instruments including IMACS on Magellan and SNIFS on the UH2.2m telescope.}} (see Table \ref{tab:sample}). While less luminous than the AGN typically observed at z$>$0.3, our sample is representative of the bulk of intermediately luminous AGN at higher redshifts.}

We used the GMOS IFU (\citealt{Allington2002}) on Gemini North during the 2015A semester (ID: GN-2015A-Q-204, PI: Woo) with the B600 disperser (R$\sim1400$ for 2 pixel spectral binning, translating to an instrumental resolution FWHM $\sim215$ km s$^{-1}$) centered at 6200\AA\ in 1-slit mode. The field of view (FoV) was $5\farcs0\times3\farcs5$, which corresponds to 3-9 kpc at the redshift of our targets, {with a spatial pixel (spaxel) scale of $\sim$ 0\farcs07}. All targets were observed under optimal conditions, with seeing between 0\farcs5 and 0\farcs8, corresponding to sub-kpc resolution for all targets ($\sim1$ kpc for our farthest target, J1720). {During the data reduction we resample the spaxel size to 0\farcs1$\times$0\farcs1, effectively oversampling our data by a factor of $\sim5-8$.}

\begin{deluxetable*}{c c c c c c c c c c c c c}
\tabletypesize{\footnotesize}
\tablecolumns{13}
\tablewidth{0pt}
\tablecaption{Our sample of 6 Type 2 AGNs with extreme \mbox{[O\,\textsc{iii}]} kinematics. \label{tab:sample}}
\tablehead{\colhead{ID}	&	\colhead{z}	& \colhead{Scale} &	\colhead{v$_{\mathrm{[OIII]}}$}	&	\colhead{$\sigma_{\mathrm{[OIII]}}$}	&	\colhead{$\log{L_{\mathrm{[OIII]}}}$} 	&	\colhead{$\log{L_{\mathrm{[OIII]}}^{\mathrm{cor}}}$}	&	\colhead{m$_{r}$}	& \colhead{R$_{out}$}	&	\colhead{v$_{\mathrm{[OIII]}}^{\mathrm{B;r}_{\mathrm{eff}}}$}	&	\colhead{$\sigma_{\mathrm{[OIII]}}^{\mathrm{B;r}_{\mathrm{eff}}}$}	&	\colhead{v$_{H\alpha}^{\mathrm{B;r}_{\mathrm{eff}}}$}	&	\colhead{$\sigma_{H\alpha}^{\mathrm{B;r}_{\mathrm{eff}}}$}	\\
\colhead{ } 	& \colhead{ }	& \colhead{[kpc/$\arcsec$]} &	\multicolumn{2}{c}{[km s$^{-1}$]}		&	\multicolumn{2}{c}{[erg s$^{-1}$]}	&	\colhead{[AB]}	&	\colhead{[kpc] } & 	\multicolumn{2}{c}{[km s$^{-1}$]} & 	\multicolumn{2}{c}{[km s$^{-1}$]}\\
\colhead{(1)} & \colhead{(2)} & \colhead{(3)} & \colhead{(4)} & \colhead{(5)} & \colhead{(6)} & \colhead{(7)} & \colhead{(8)} & \colhead{(9)} & \colhead{(10)} & \colhead{(11)} & \colhead{(12)} & \colhead{(13)}}
\startdata
J091808+343946 &	0.0973	& 1.78 &	-419	&	444	& 40.8	&	42.9	&	16.45	&	1.85 & -670$\pm$130 & 500$\pm$120 & -90$\pm$40     & 230$\pm$30	\\
J113549+565708 &	0.0515	& 0.99 &	-212	&	400	& 41.7	&	43.1	&	14.72	&	1.80 & -350$\pm$50	  & 460$\pm$40   & -80$\pm$120   & 200$\pm$120	\\
J140453+532332 &	0.0813	& 1.51 &	-303	&	392	& 41.4	&	42.6	&	16.11	&	2.07 & -320$\pm$20	  & 400$\pm$20   & -250$\pm$250 &	470$\pm$240\\
J160652+275539 &	0.0461	& 0.89 &	-268	&	324	& 40.8	&	42.2	&	15.22	&	1.29 & -380$\pm$70	  & 300$\pm$30   & -90$\pm$250   &	70$\pm$70\\
J162233+395650 &	0.0631	& 1.20 &	-34	&	556	& 41.6	&	42.4	&	15.61	&	1.58 & 4$\pm$130	  & 680$\pm$160 & -40$\pm$140   &	350$\pm$150\\
J172038+294112 &	0.0995	& 1.81 &	-65	&	401	& 41.3	&	42.3	&	16.73	&	1.92 & -140$\pm$20	  & 580$\pm$30   & 4$\pm$40        &	290$\pm$70\\
\enddata
\tablecomments{
Col. 1: target ID, Col. 2:  the redshift based on the stellar absorption lines using the SDSS spectra \citep{Bae2014}, {Col. 3: the spatial scale at the redshift of the source,} Col. 4: the \mbox{[O\,\textsc{iii}]} velocity shift with respect to the systemic velocity,
Col. 5: the \mbox{[O\,\textsc{iii}]} velocity dispersion measured from SDSS spectra, Col. 6: the measured \OIII\ luminosity from SDSS spectra, Col. 7: the extinction-corrected \mbox{[O\,\textsc{iii}]}  luminosity (see \citealt{Bae2014}), Col. 8: the \textit{r}-band SDSS magnitude, Col. 9: the kinematics-based size of the outflow from Paper I, Col. 10-13: the flux-weighted velocity and velocity dispersion of the broad \OIII component and the broad H$\alpha$ component within one effective radius, taken from Table 2 of Paper I (refer to that paper for more details on how these were calculated).}
\end{deluxetable*}%

\subsection{Extraction of emission-line properties}
We provide a summary of the method, which is described in detail in Paper I. We fit the AGN continuum together with the stellar spectrum using an IDL implementation of the penalized pixel-fitting method (pPXF; \citealt{Cappellari2004}) and derive the stellar velocity with respect to the systemic velocity\footnote{The systemic velocity is calculated from the pPXF fitting of the central 3\arcsec\ spatially integrated IFU spectrum of each source.} at each spaxel. Next, we subtract the best pPXF fit to acquire a pure emission line spectrum. We then perform a spectral fitting of the H$\beta$, \OIII\ doublet, \NII\ doublet, H$\alpha$, and \SII\ doublet. Each individual line is fit with up to two Gaussian components, employing an iterative fitting method to decide whether a secondary component is required based on its S/N. One of the sources, J1606, harbors a Type 1 core with a very broad Balmer like component ($\sigma_{\mathrm{H\alpha}}>1500$ km s$^{-1}$). This is accounted for by using a third very broad Gaussian component in the Balmer line fits of this source (see Paper 1 for details and examples).

For each emission line we calculate the following: the peak S/N, the first and second moment, $\lambda_{0}$ and $\sigma$, and flux of the total best-fit model and the central wavelength, dispersion, and flux of the narrow and broad components of the line separately. We also calculate the velocity of each line based on the systemic velocity. Velocity dispersions are corrected for instrumental broadening ($\sigma_{inst}$$\sim90$ km s$^{-1}$). We use Monte Carlo simulations to calculate the uncertainties of these values, randomizing each spaxel spectrum based on its error spectrum. Uncertainties are the standard deviations of the resulting distributions. A total of 100 random spectra per spaxel (total $\sim170000$ spectra per source) are generated.

\section{Results}
\label{sec:results}

\begin{figure*}[tbp]
\begin{center}
\raisebox{-0.5\height}{\includegraphics[width=0.8\textwidth,angle=0,trim={0 60 24 5},clip]{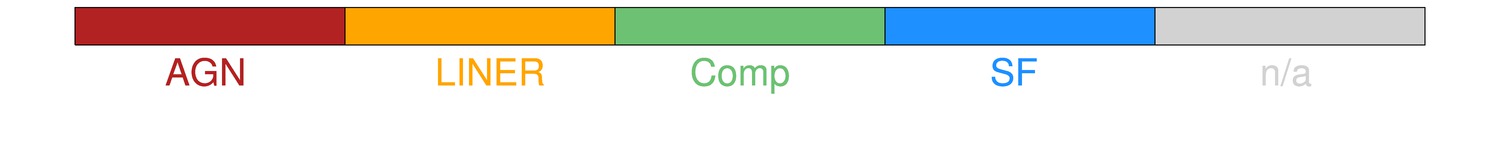}}\\
\vspace{5pt}
\raisebox{-0.5\height}{\includegraphics[width=0.16\textwidth,angle=0,trim={60 60 60 60},clip]{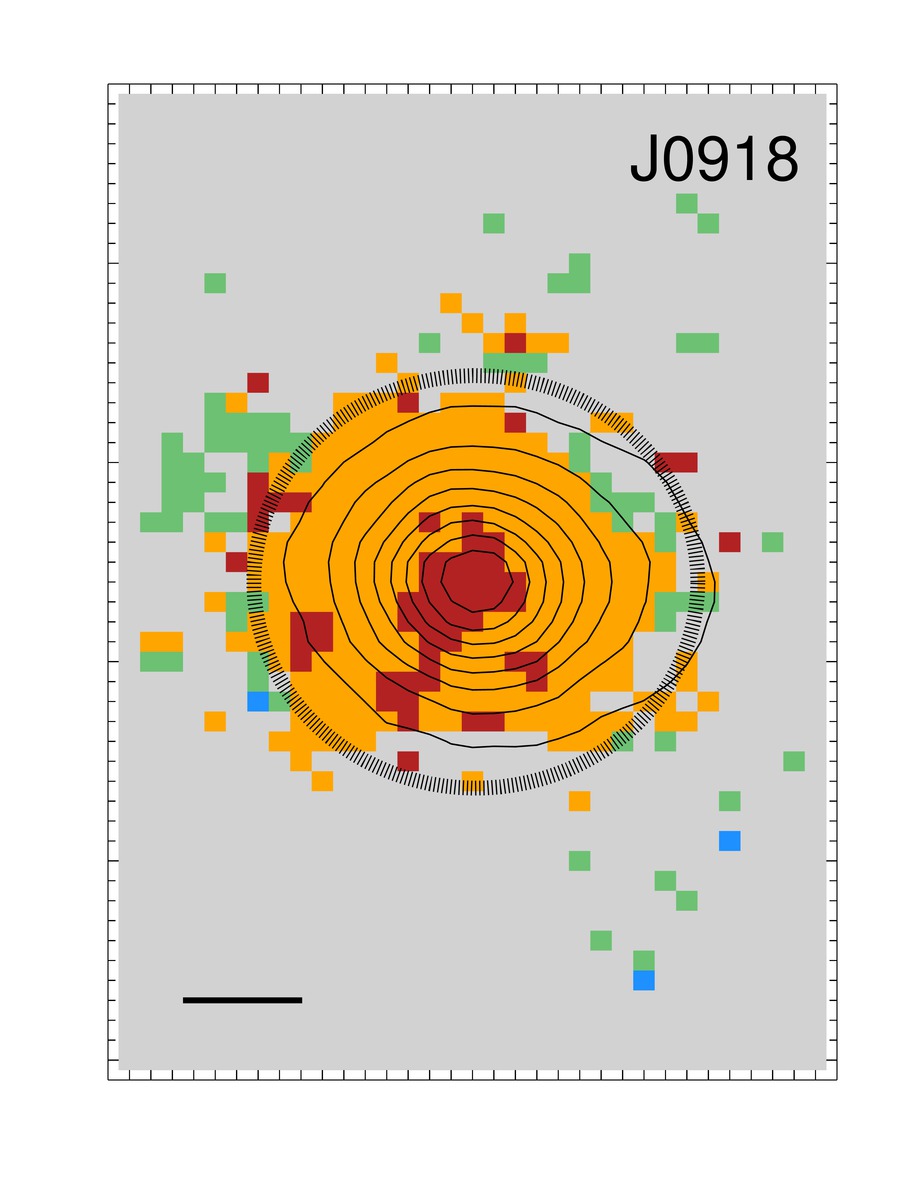}}
\raisebox{-0.5\height}{\includegraphics[width=0.16\textwidth,angle=0,trim={60 60 60 60},clip]{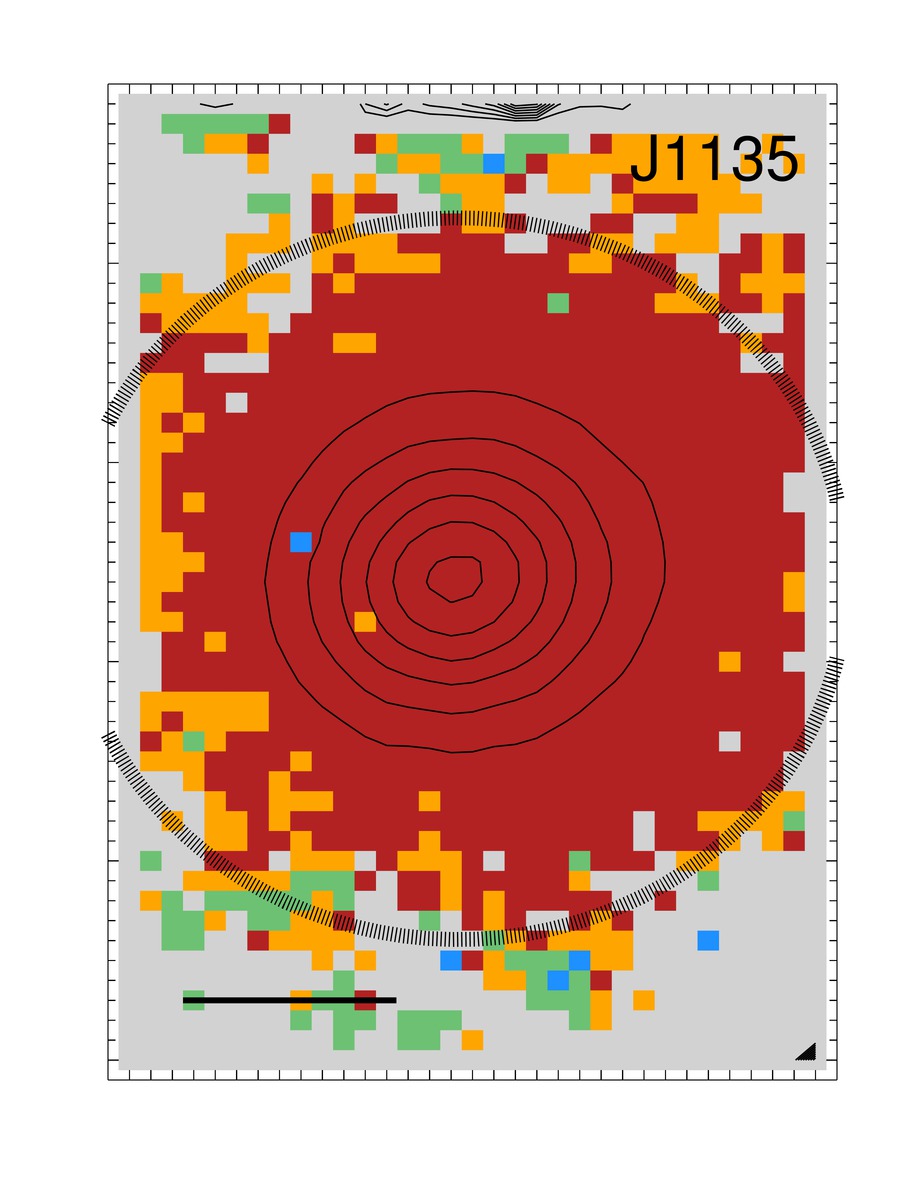}}
\raisebox{-0.5\height}{\includegraphics[width=0.16\textwidth,angle=0,trim={60 60 60 60},clip]{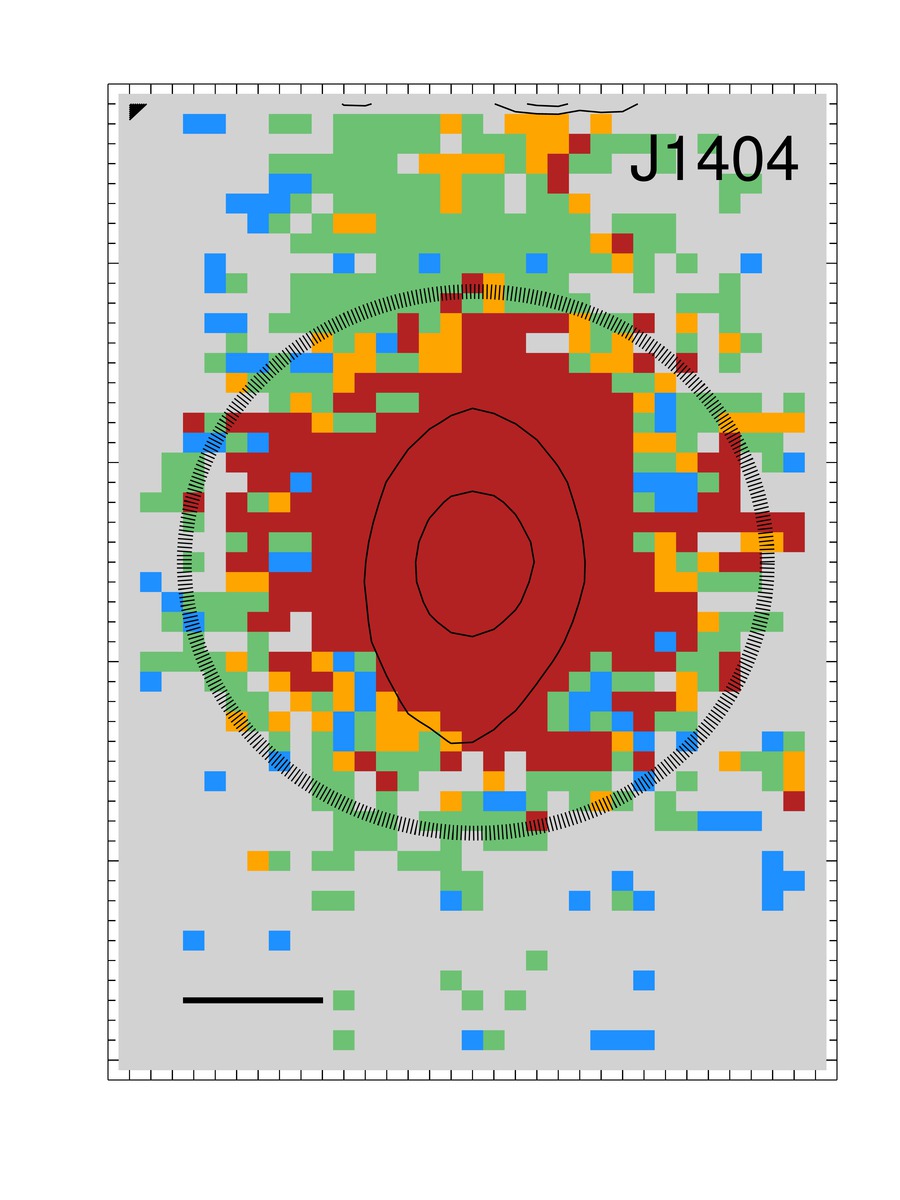}}
\raisebox{-0.5\height}{\includegraphics[width=0.16\textwidth,angle=0,trim={60 60 60 60},clip]{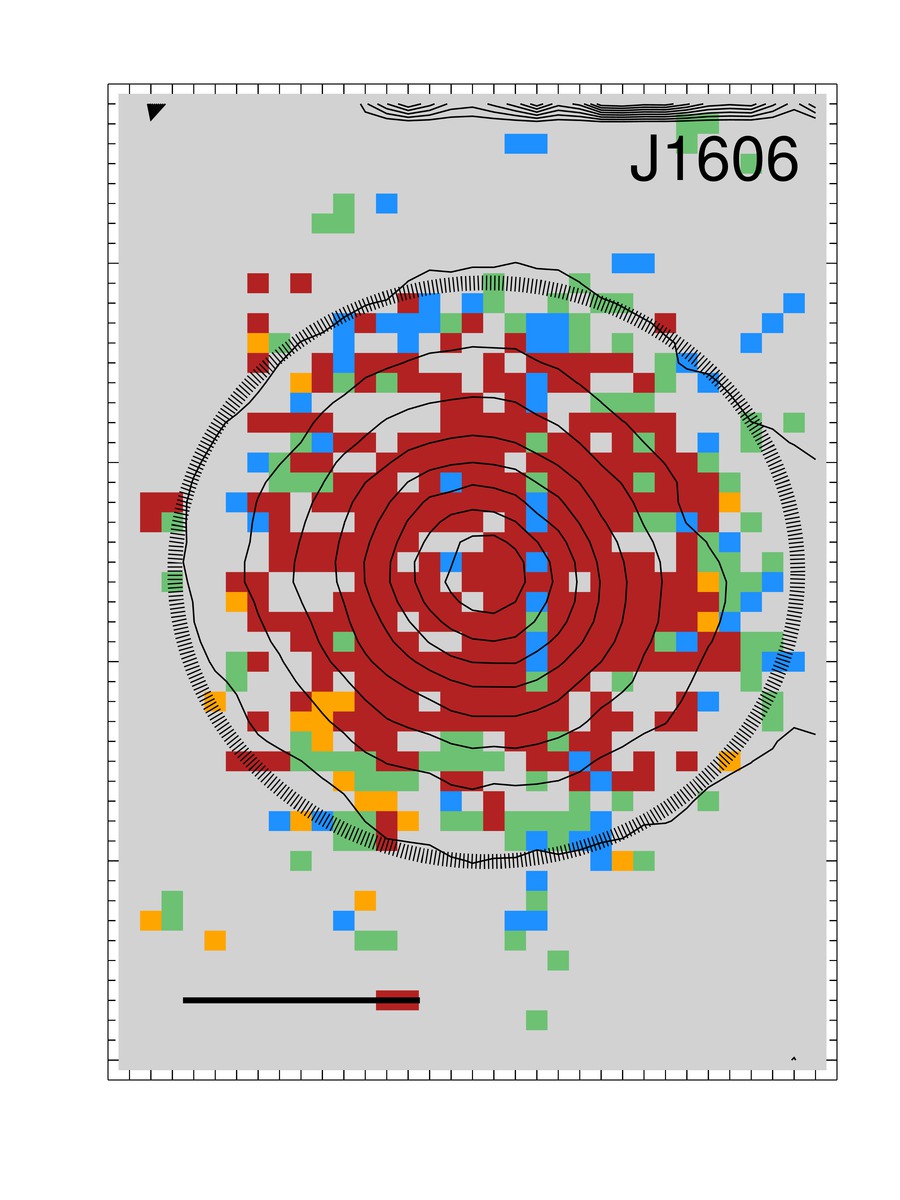}}
\raisebox{-0.5\height}{\includegraphics[width=0.16\textwidth,angle=0,trim={60 60 60 60},clip]{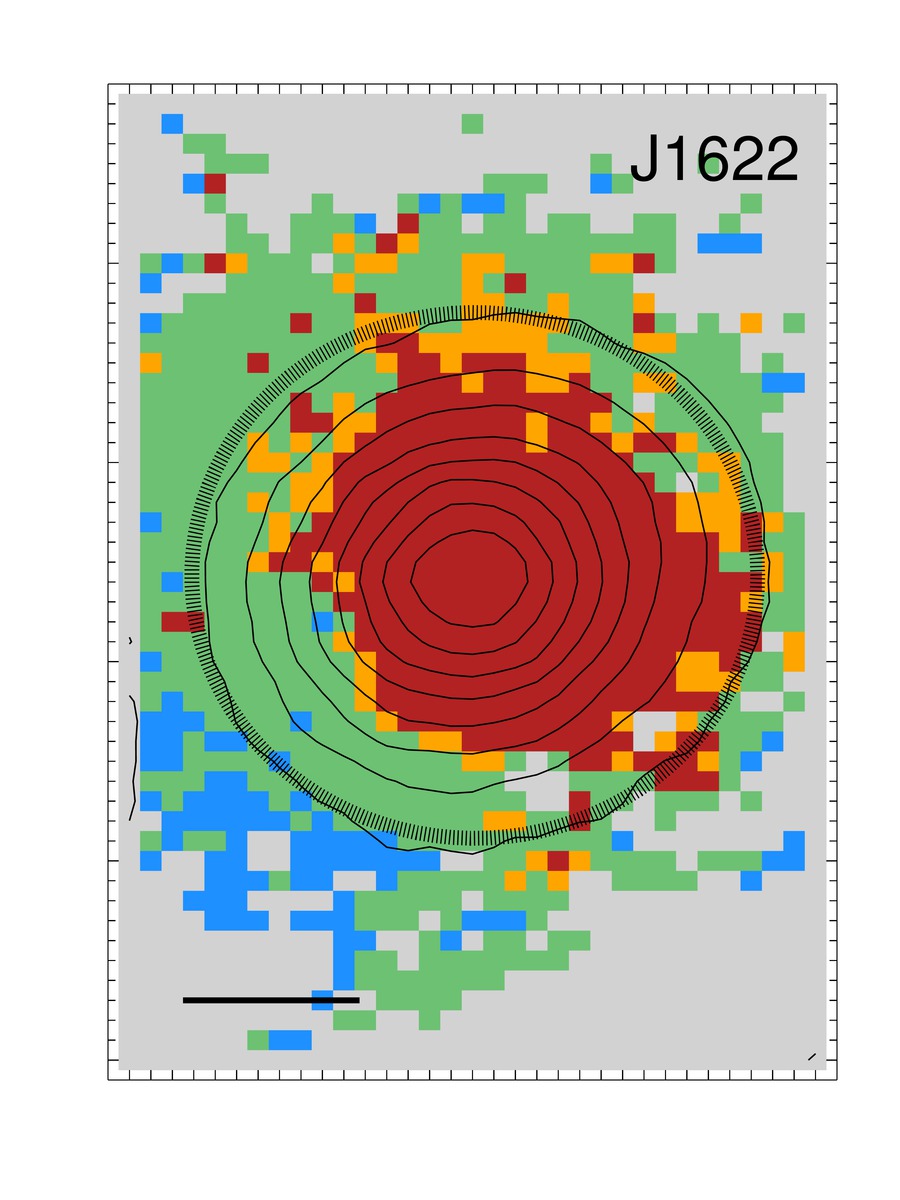}}
\raisebox{-0.5\height}{\includegraphics[width=0.16\textwidth,angle=0,trim={60 60 60 60},clip]{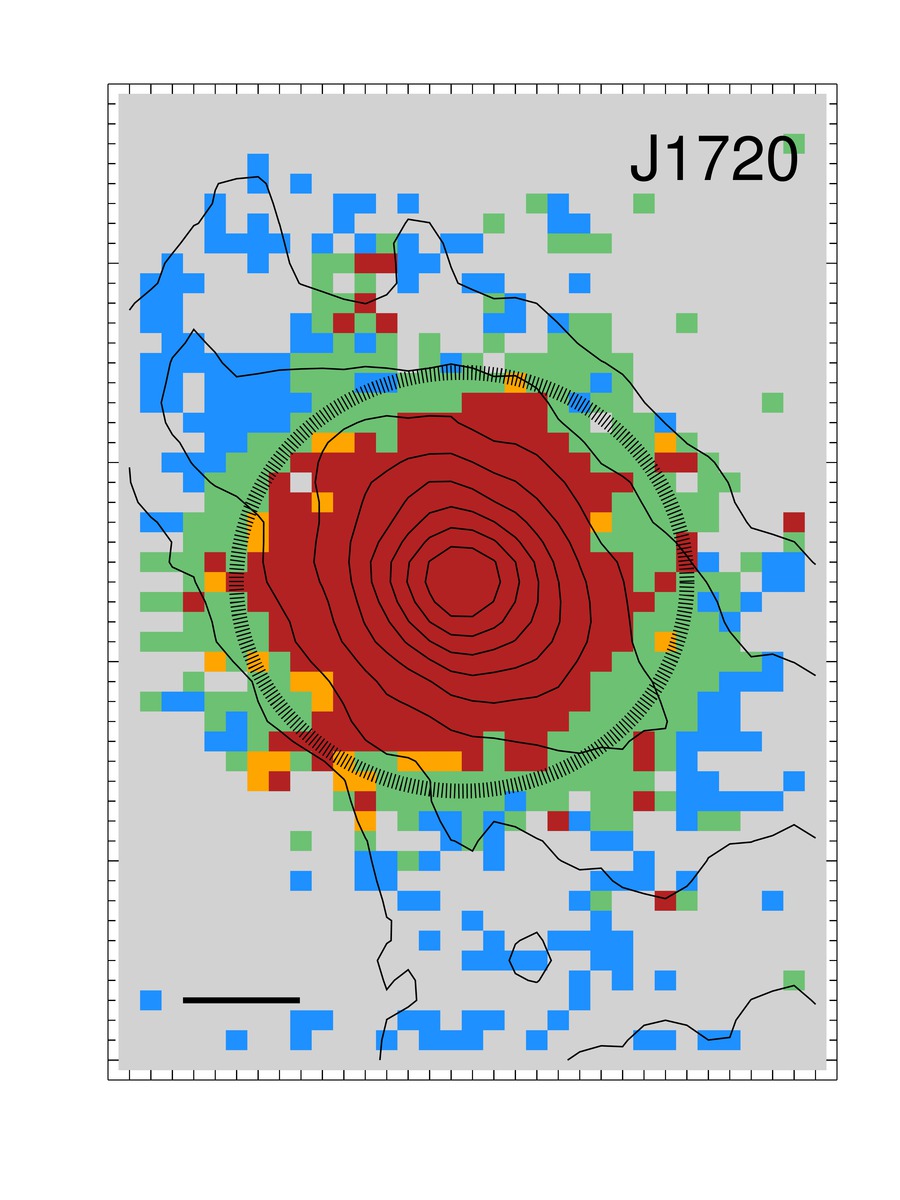}}\\
\raisebox{-0.5\height}{\includegraphics[width=0.16\textwidth,angle=0,trim={60 60 60 60},clip]{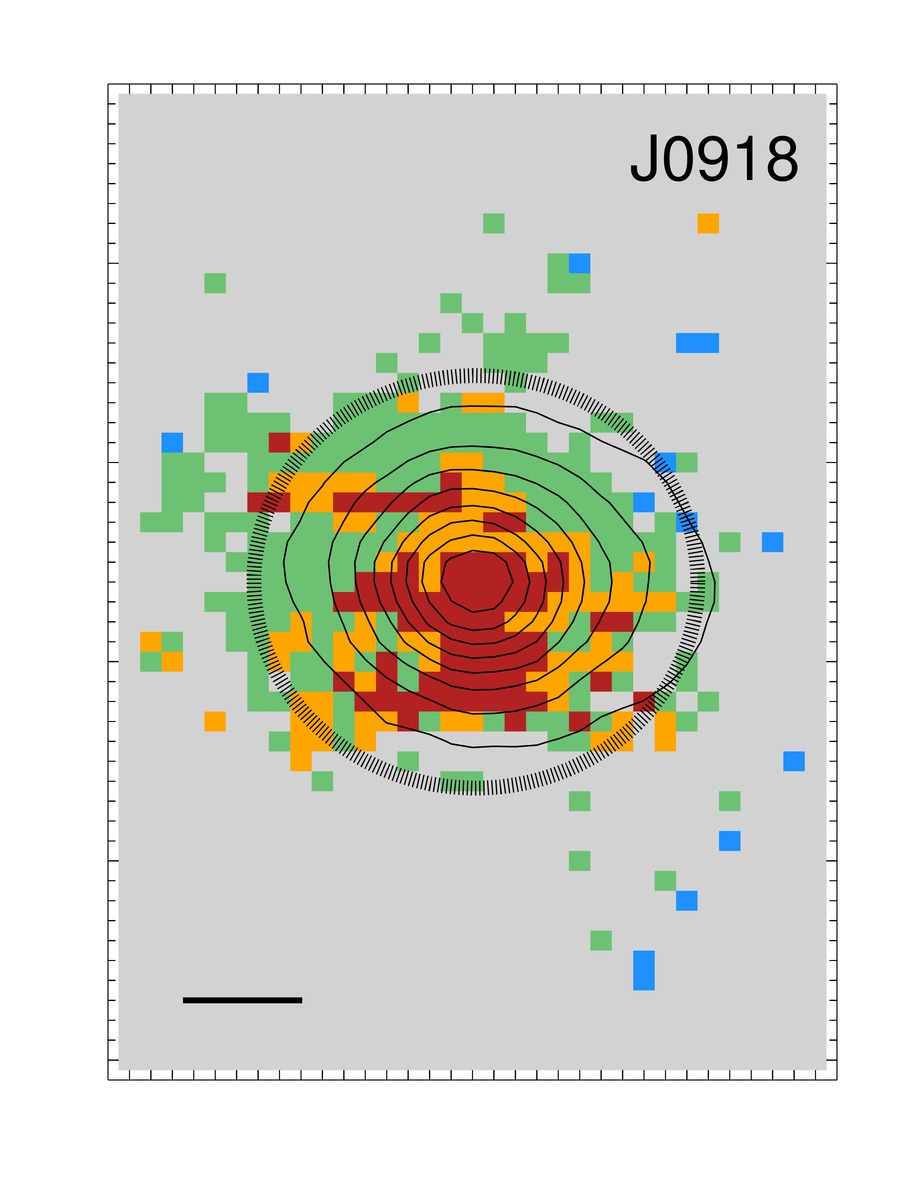}}
\raisebox{-0.5\height}{\includegraphics[width=0.16\textwidth,angle=0,trim={60 60 60 60},clip]{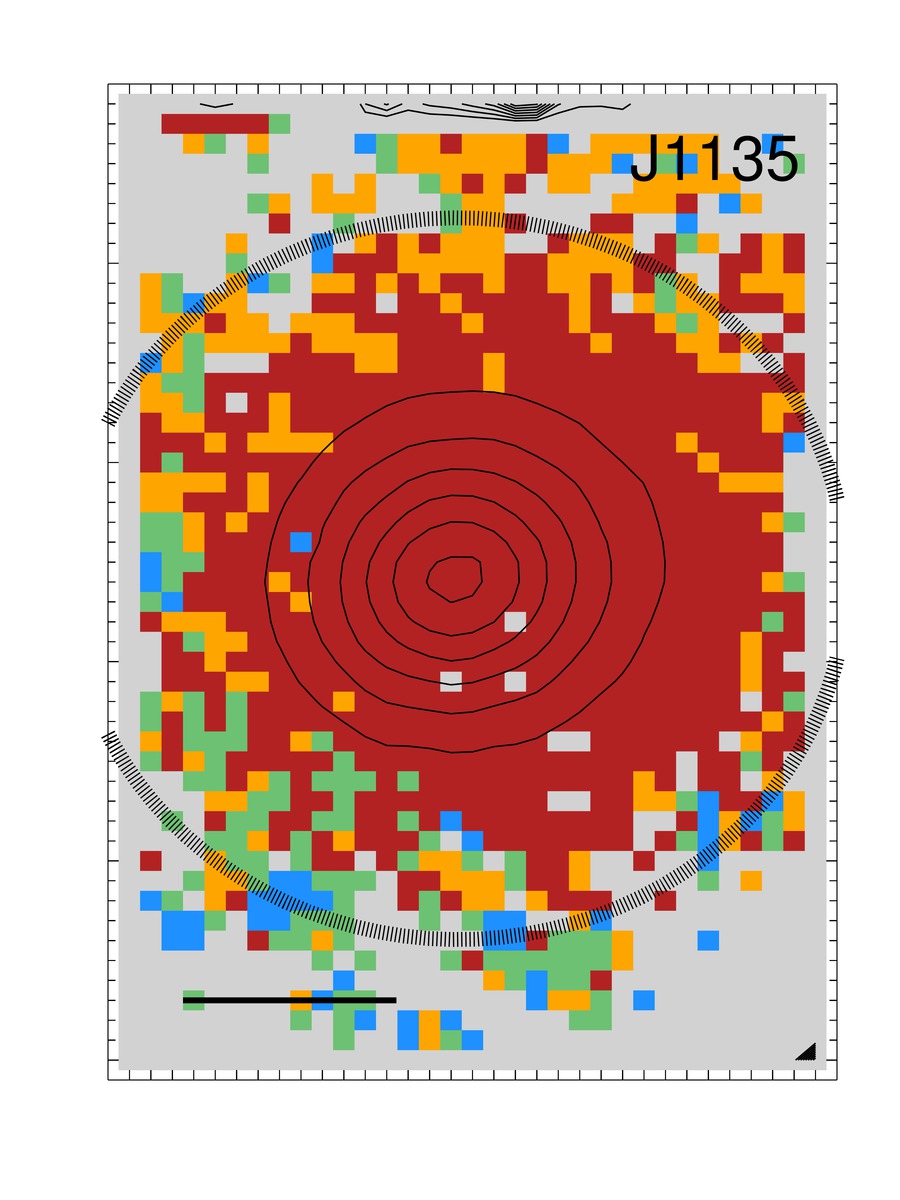}}
\raisebox{-0.5\height}{\includegraphics[width=0.16\textwidth,angle=0,trim={60 60 60 60},clip]{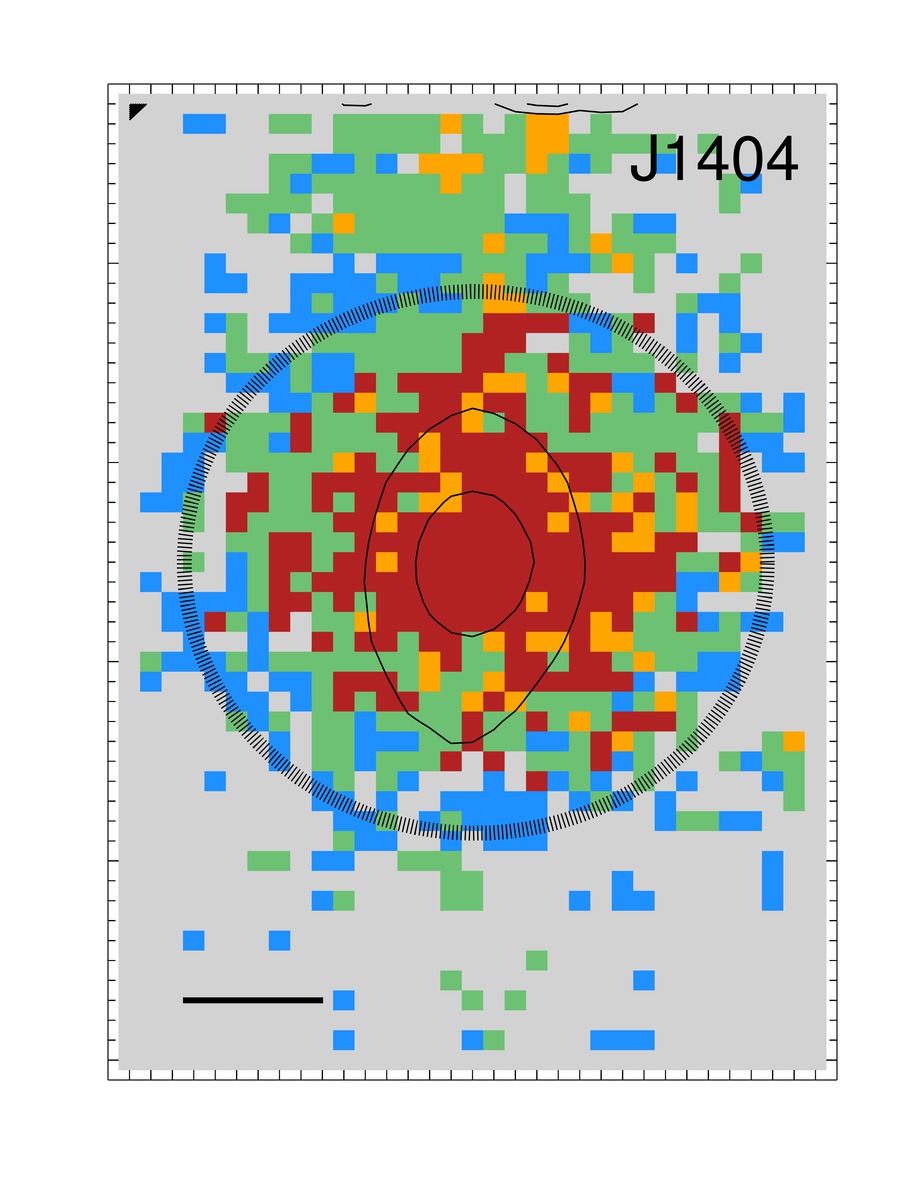}}
\raisebox{-0.5\height}{\includegraphics[width=0.16\textwidth,angle=0,trim={60 60 60 60},clip]{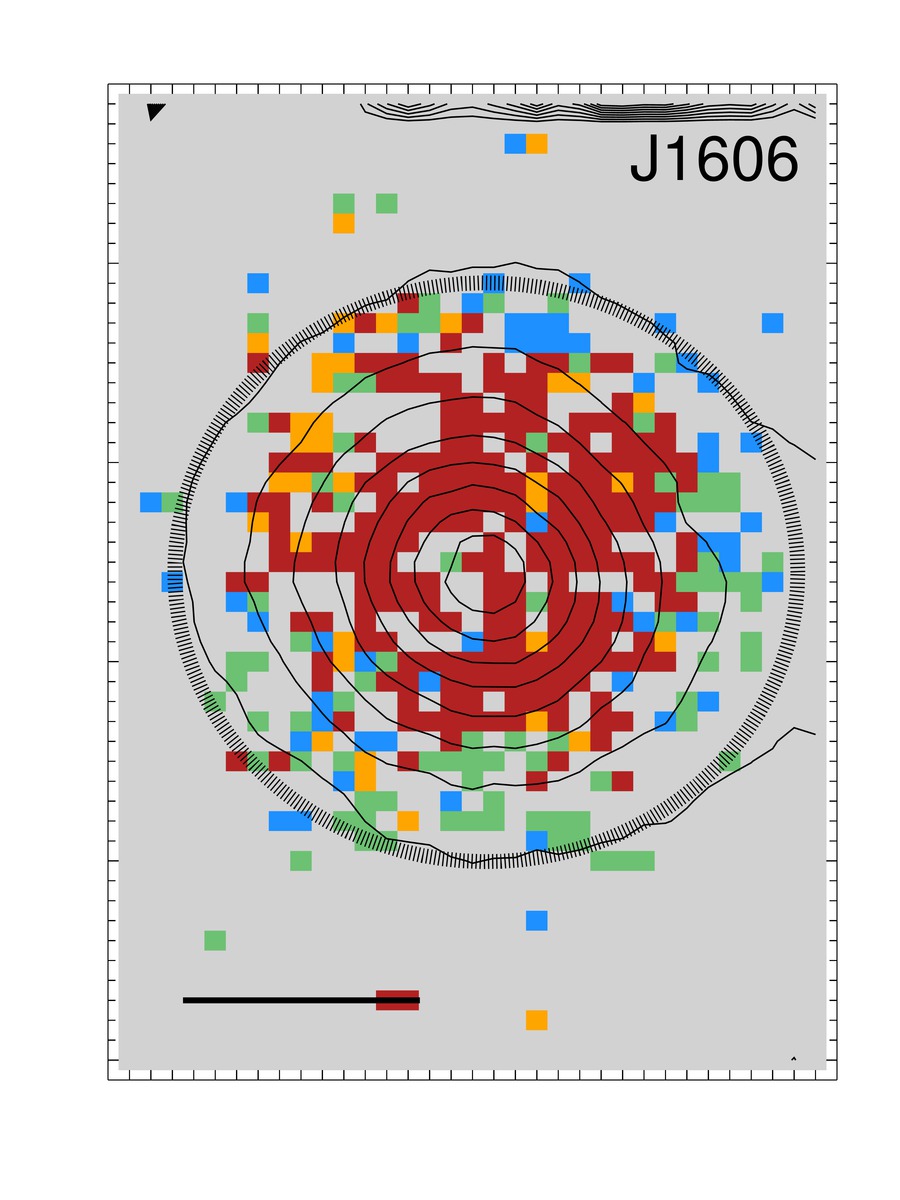}}
\raisebox{-0.5\height}{\includegraphics[width=0.16\textwidth,angle=0,trim={60 60 60 60},clip]{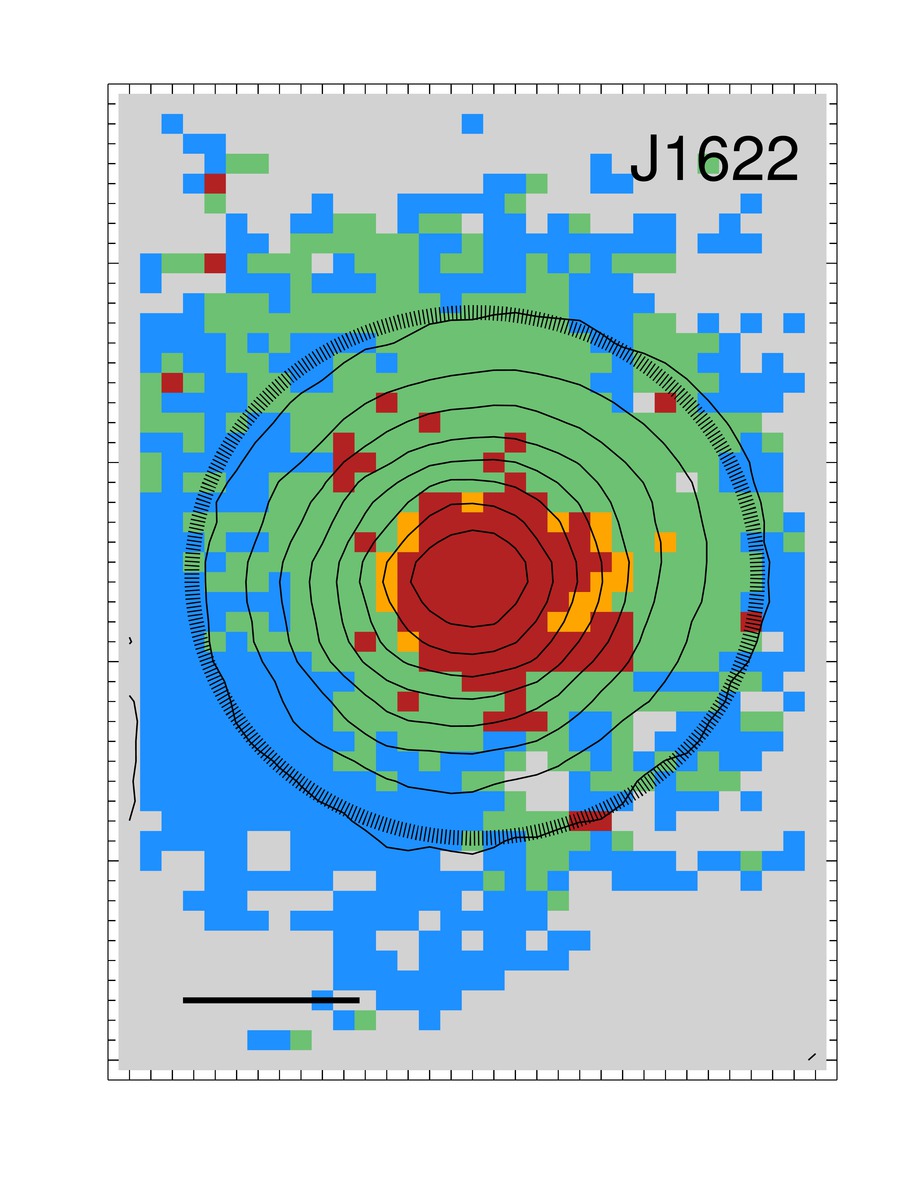}}
\raisebox{-0.5\height}{\includegraphics[width=0.16\textwidth,angle=0,trim={60 60 60 60},clip]{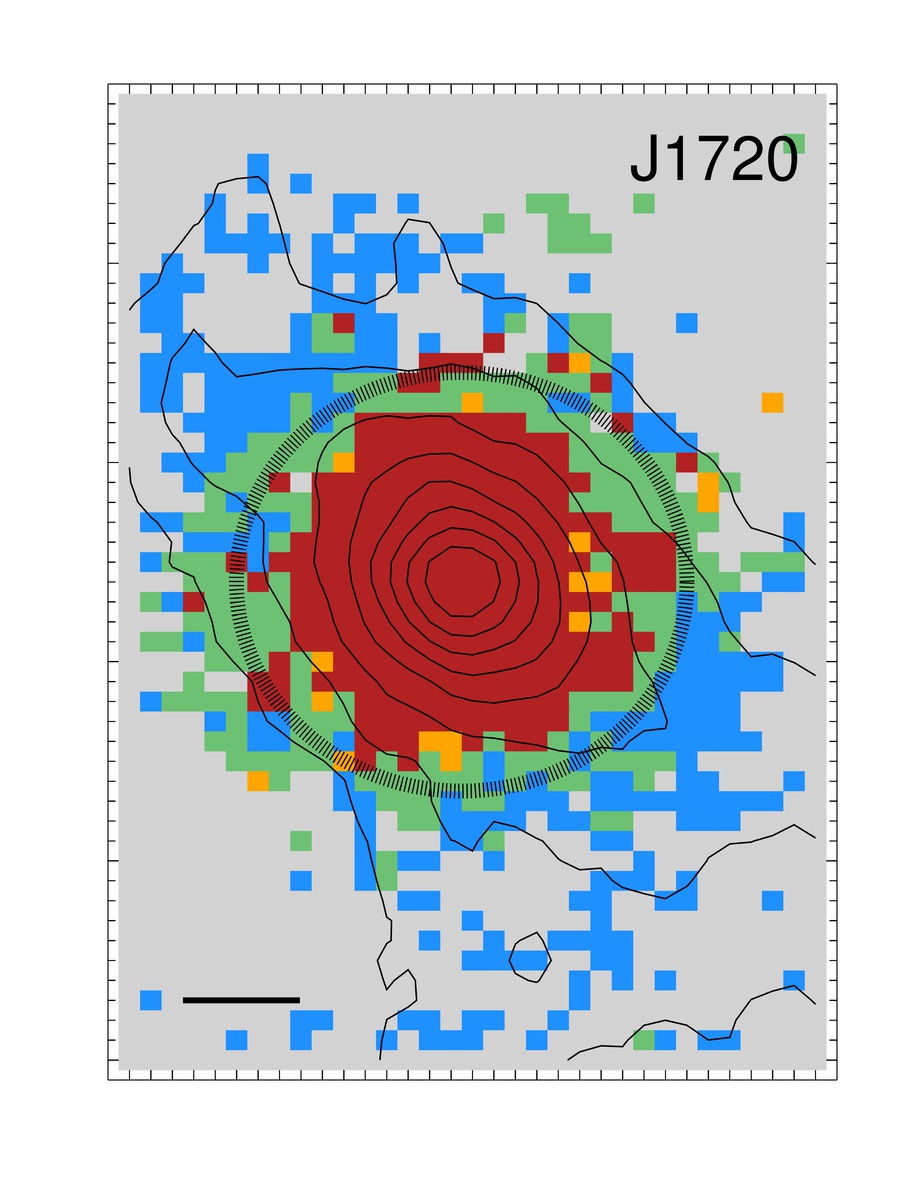}}\\
\raisebox{-0.5\height}{\includegraphics[width=0.16\textwidth,angle=0,trim={60 60 60 60},clip]{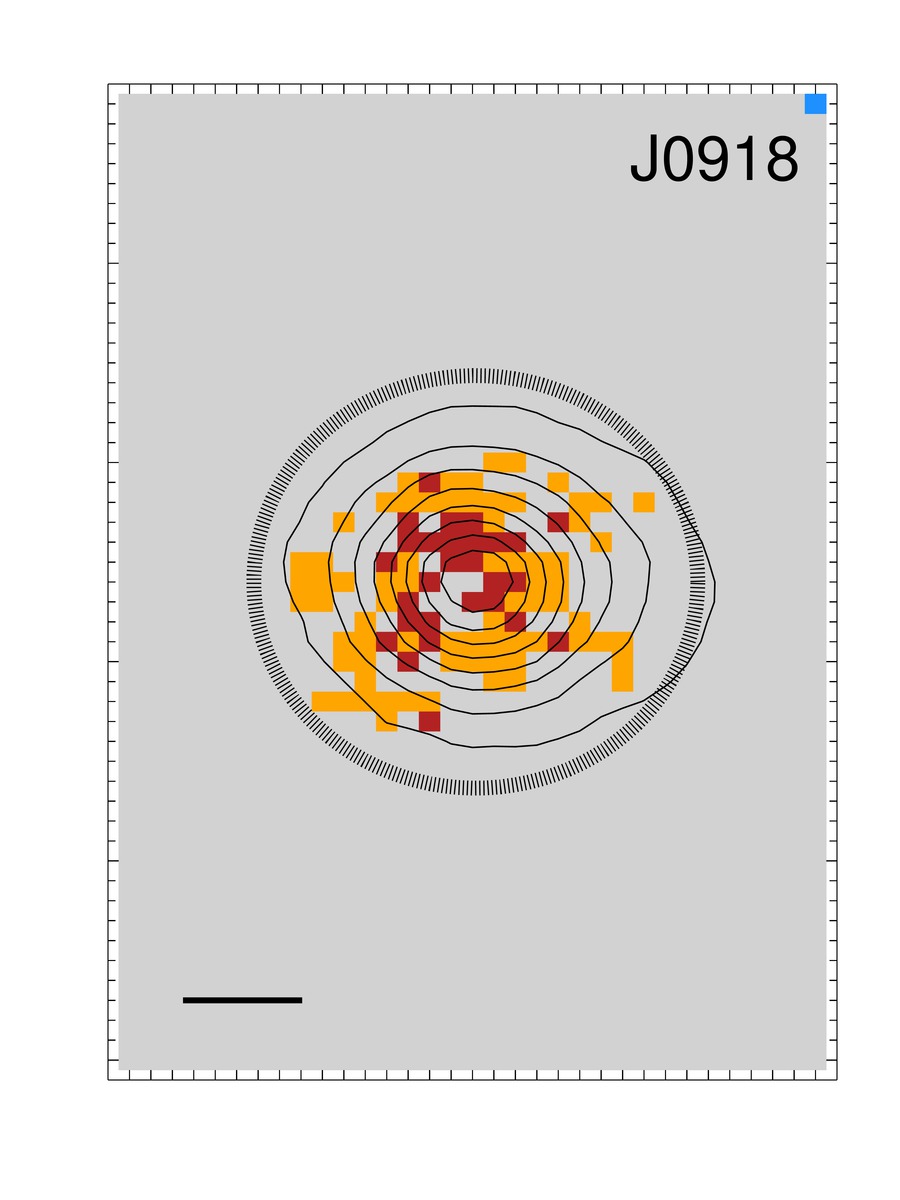}}
\raisebox{-0.5\height}{\includegraphics[width=0.16\textwidth,angle=0,trim={60 60 60 60},clip]{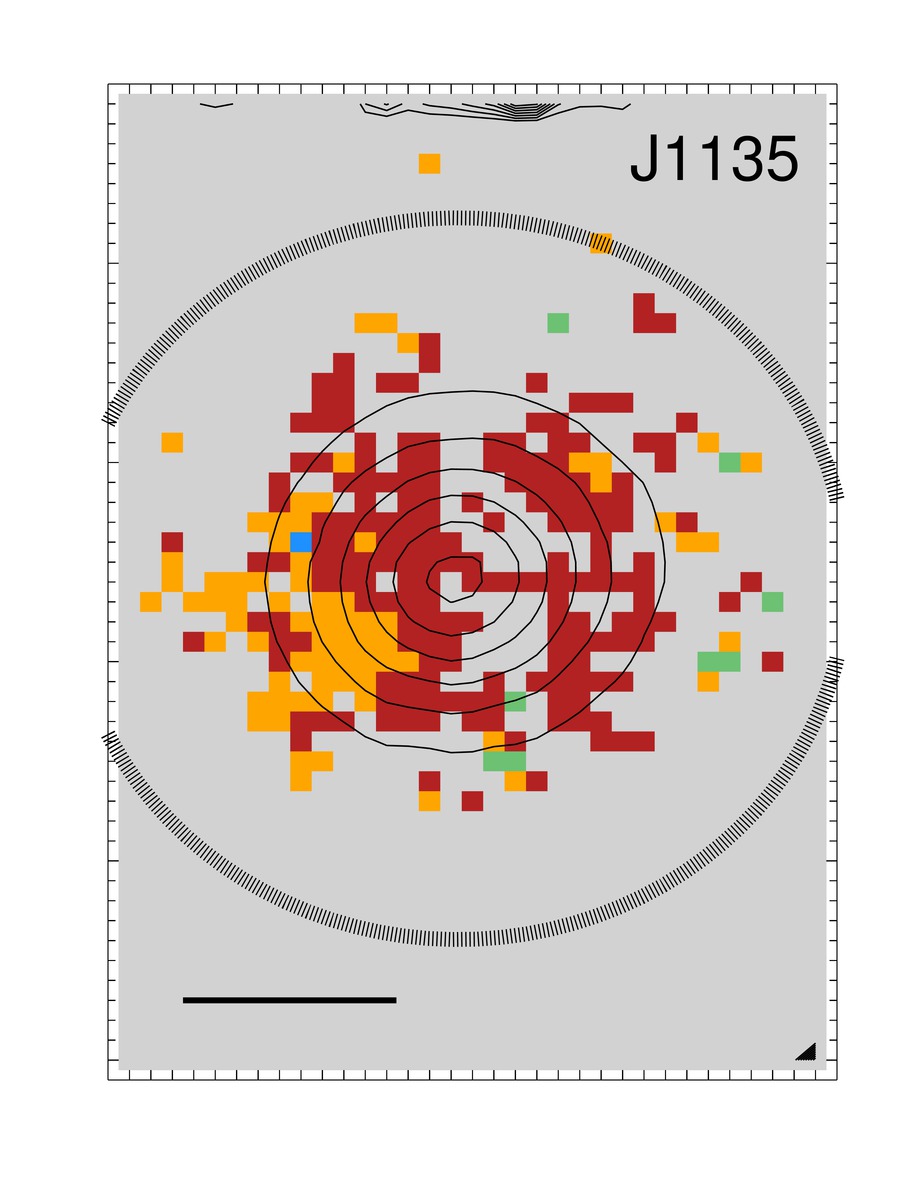}}
\raisebox{-0.5\height}{\includegraphics[width=0.16\textwidth,angle=0,trim={60 60 60 60},clip]{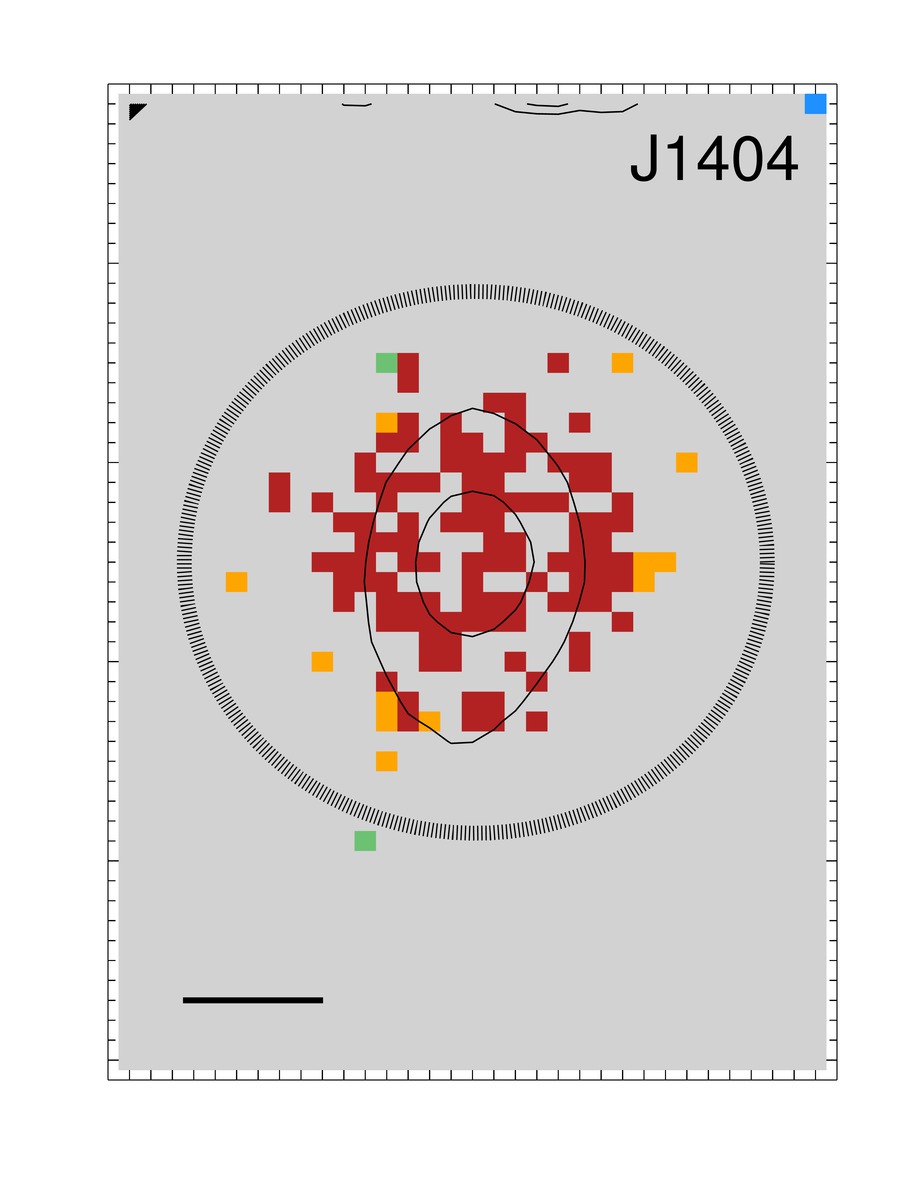}}
\raisebox{-0.5\height}{\includegraphics[width=0.16\textwidth,angle=0,trim={60 60 60 60},clip]{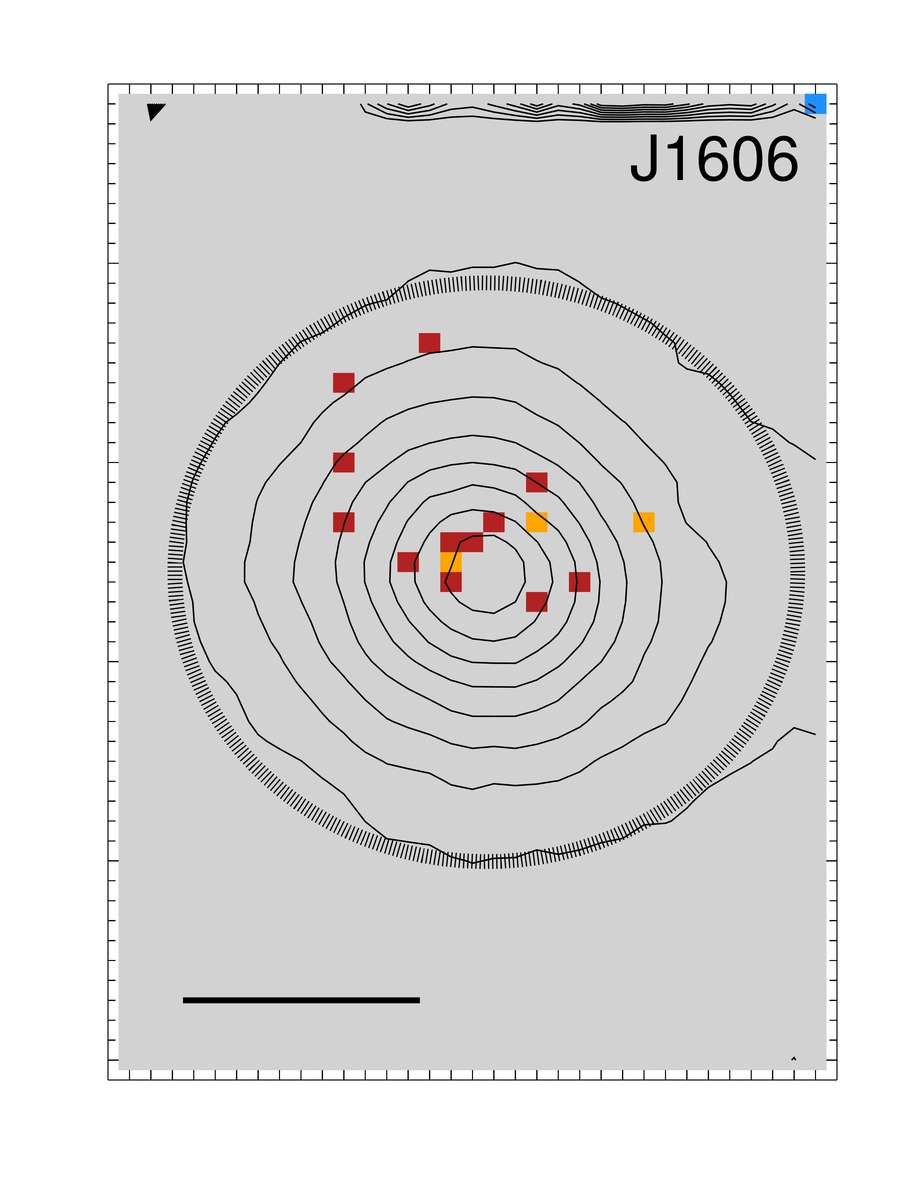}}
\raisebox{-0.5\height}{\includegraphics[width=0.16\textwidth,angle=0,trim={60 60 60 60},clip]{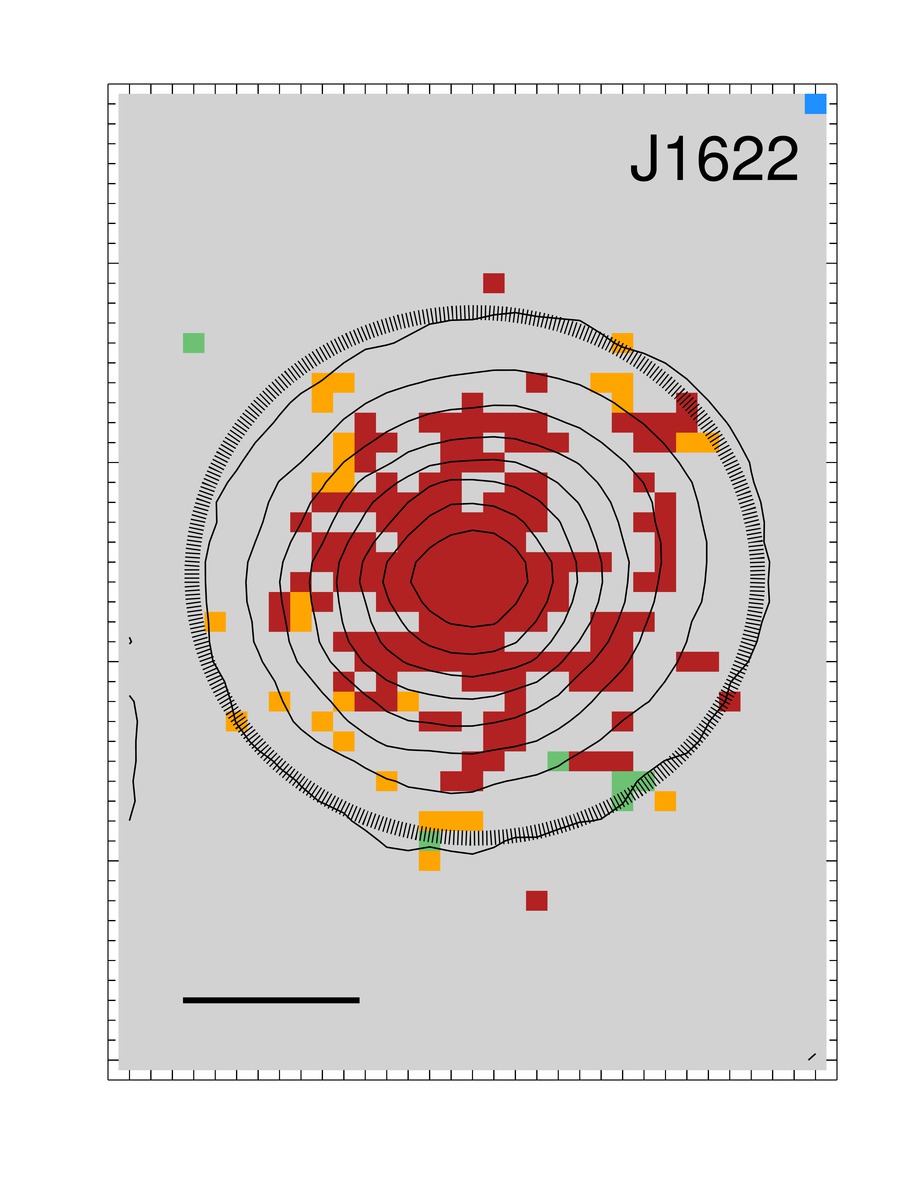}}
\raisebox{-0.5\height}{\includegraphics[width=0.16\textwidth,angle=0,trim={60 60 60 60},clip]{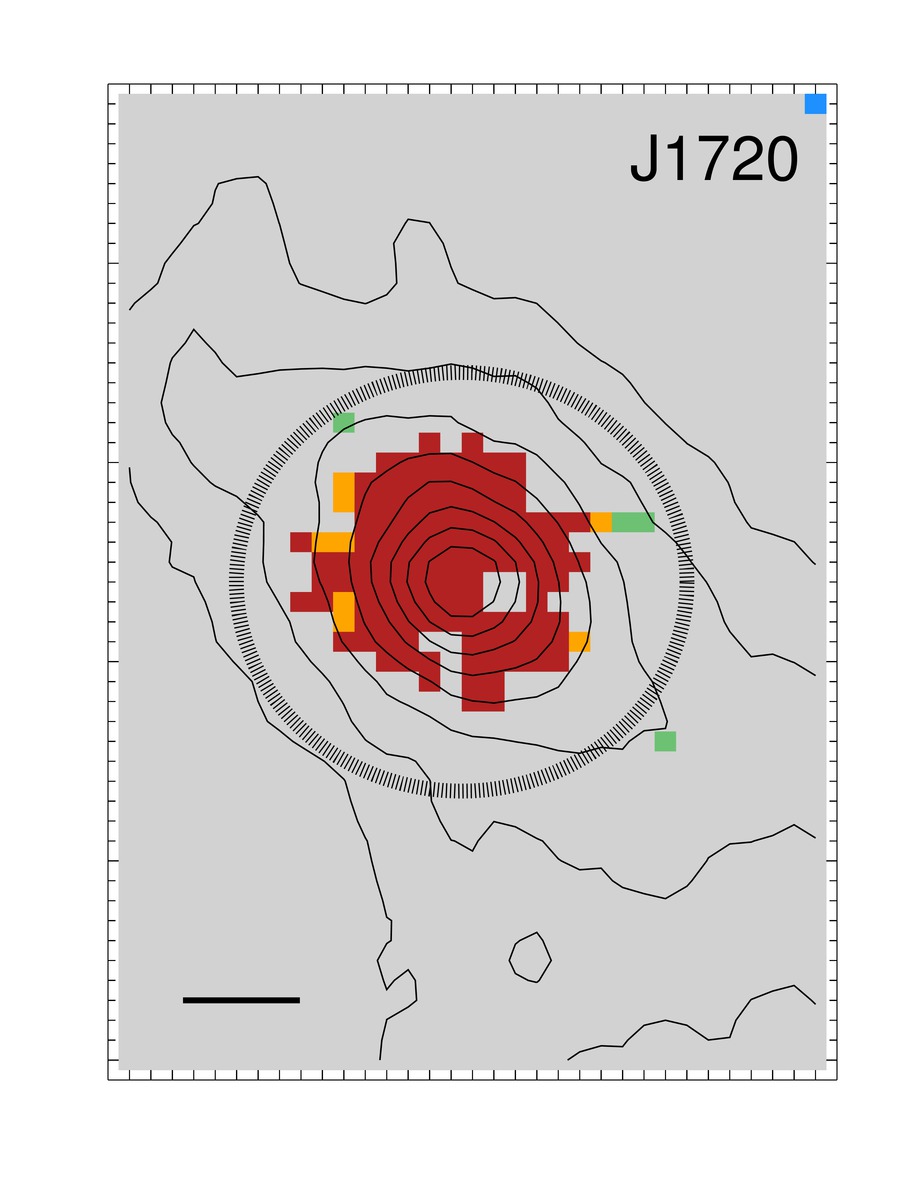}}\\
\caption{BPT classification maps based on the total emission line profiles (top), only the narrow emission line component (middle), and only the broad emission line component (bottom). Contours show the continuum flux at 10\% intervals from the peak. The adopted S/N limit is 1 for H$\beta$ and 3 for \OIII\, H$\alpha$, and \NII. Gray spaxels indicate weak or non-detected emission lines. Circles show the kinematically defined size of the outflow as calculated in Paper I. {Black bars in the lower panels mark a physical distance of 1 kpc at the redshift of each source. North is up and East is left for J0918, J1606, J1622, and J1720. North is left and East is up for J1135 and J1404.}}
\label{fig:all_bpt_map}
\end{center}
\end{figure*}

\subsection{What is ionizing the gas?}
\label{sec:bpt_map}

\subsubsection{BPT classification maps}

We first investigate flux ratios based on the total profile of the emission lines to classify the photoionizing mechanism based on the demarcation lines from \citet{Kauffmann2003} and \citet{Kewley2001,Kewley2006}. These define (empirically and from photoionization models, respectively) 4 regions on the BPT diagram: the star-forming region, where photoionization by young stars dominates, the composite region, where both young stars and AGN emission can operate, and the {non-star-forming region}. The latter is further divided into the Seyfert region (with $\log$(\OIII/H$\beta$) flux ratio $>$0.5) and the LINER region (Low-Ionization Nuclear Emission-line Region). {LINER-like emission can originate from a low luminosity/low Eddington ratio AGN, shock-ionized gas, or from hot, evolved (post-asymptotic giant branch, AGB) stars (e.g., \citealt{Belfiore2016})}. In the top panel of Fig. \ref{fig:all_bpt_map} we show the spatially resolved BPT  maps, based on the total profiles of \OIII\ and H$\beta$ and \NII\ and H$\alpha$. For the classification of each spaxel we require \OIII\ , \NII , and H$\alpha$ to have an S/N$\geqslant3$, while for H$\beta$ we relax this to S/N$>1$. The weak H$\beta$ emission is the main limitation in acquiring reliable BPT classification in the outer parts of the FoV (for a typical \OIII/H$\beta$ ratio of 10 in AGN-dominated regions, H$\beta$ S/N of 1 implies an effective S/N cut of 10 for \OIII). 

All sources show an AGN (Seyfert-like) core with various sizes (red color). At least 3 sources (J1404, J1622, J1720) show clear evidence of circumnuclear star formation, in the form of rings classified either as HII- or composite-like regions (respectively blue and green). One source, J1135, appears to be exclusively photoionized by the AGN, but for which we detect evidence for a transition to LINER-like emission (orange) at the edge of the FoV. J1606 is plagued by faint H$\beta$ emission and a challenging decomposition of \NII\ and H$\alpha$\footnote{We remind that J1606 harbors a Type 1-like very broad Balmer emission ($\sigma >1500$ km s$^{-1}$).} that results in more noisy BPT maps. Nevertheless, we see qualitatively the same behavior as for the other sources. The AGN region appears mostly concentric with the continuum peak, except for the case of J1622, where there appears to be a "protrusion" to the SouthWest. It coincides with the elongated feature of increased dispersion in the broad H$\alpha$ velocity dispersion map that was discussed in Paper I (see the Appendix for more details). The edge of the outflow (circles, marking the kinematic sizes reported in Paper I) is right beyond the edge of the area classified as Seyfert-like region based on the flux ratios of the total emission lines. J0918 is the only source for which most spaxels show LINER-like BPT classification and the edge of the outflow is at the edge of the LINER region. We note that a fraction of the spaxels classified as composite for J1622 and J1720 are within the outflow radius, which probably reflects the still significant AGN contribution in these spaxels.

In Paper I, we presented compelling evidence that the complex and asymmetric shape of the \OIII\ and H$\alpha$ emission lines can be understood as a superposition of two kinematic components, one driven by the gravitational potential of the galaxy and the other due to the AGN-driven outflow. This is more generally corroborated by statistical studies of the \OIII\ emission line profile shape (e.g., \citealt{Woo2016}), showing that it can be decomposed into a narrow and a broad component. The kinematics of the latter shows a broad correlation with the \OIII\ luminosity and the AGN Eddington ratio, reinforcing that it is physically linked to the AGN. We are therefore motivated to study the ionization properties and energetics of these two components separately, as we did for the kinematics in Paper I. We first focus on the narrow component of the emission lines, which in Paper I was shown to mostly follow the stellar kinematics with little spatial variation (e.g., see Fig. 8 of Paper I). 

If we consider the BPT classification based on the narrow component of the emission lines (Fig. \ref{fig:all_bpt_map}, middle row), the core is still dominated by the AGN photoionization, but in many cases the extent of the Seyfert-like region is decreased (J1404, J1622, J1720). Instead, strong composite and star-forming regions emerge. This corroborates our interpretation that there is a superposition of two different components, which are separated by both their kinematics (Paper I) and their ionization properties. The BPT maps of the narrow component additionally reinforce that all sources show circumnuclear star formation, which {would be otherwise mischaracterized (in terms of extent, level of star formation, and energetics) or even missed (e.g., in the case of J0918 or J1135)} if only the total line profile was considered. Interestingly, when comparing the location of star forming spaxels with the size of the outflow (thick circle) they fall within the outflow radius. Nonetheless, we cannot conclude whether this is due to a physical connection or simply a projection of physically distinct components onto each other. We similarly cannot exclude the possibility that there is ongoing star formation within the AGN-like region but is masked by the strong AGN emission. Unfortunately, the BPT classification is not possible at the edge of the IFU FoV due to the faint nature of H$\beta$ and the quickly decreasing surface brightness of \OIII\ at larger radii. Nonetheless, the fact that H$\alpha$ emission is clearly detected out to the edge of the FoV for most of our sources indicates that there is ongoing star formation within the gray area of the BPT maps shown in Fig. \ref{fig:all_bpt_map}.

The BPT classification based on the broad components of the emission lines (Fig. \ref{fig:all_bpt_map}, bottom row) is more challenging to constrain beyond the nuclear region. We first notice that the nuclear region (out to $\sim1$ kpc) is dominated by the AGN. Second, we detect both Seyfert-like and LINER-like emission, with the LINER emission usually found at the outer part of the nucleus. LINER-like emission can also be explained in terms of shock ionization (e.g., \citealt{Dopita1995}). We may therefore be seeing the region where the relatively lower density outflow driven by the AGN meets the denser circumnuclear ISM, leading to the gradual disruption of the outflow and the emergence of shocks due to the discontinuity of both the kinematics and the gas density. A rise of the electron density, N$_{e}$, is indeed observed for some of our sources (see Appendix) but without information on the molecular phase of the gas we can not draw definite conclusions. This scenario is however further corroborated by the fact that the edge of the BPT classification based on the broad component roughly coincides with the beginning of the spaxels classified as composite and star-forming in the BPT maps of the narrow component (Fig. \ref{fig:all_bpt_map}, middle row). {Alternatively, this extended ($\gtrsim1-2$ kpc, Fig. \ref{fig:all_bpt_map}) low-ionization emission-line region (eLIER, following the scheme of \citealt{Belfiore2016}) can be powered by post-AGB stars. As this region is bordering the composite and star-forming regions, a significant population of post-AGB stars but absence of ongoing star formation may imply that within the radiative sphere of influence of the AGN the formation of young stars has been recently suppressed.}

\subsubsection{Flux ratio radial profiles and maps}

\begin{figure}[tbp]
\begin{center}
\includegraphics[width=0.45\textwidth,angle=0]{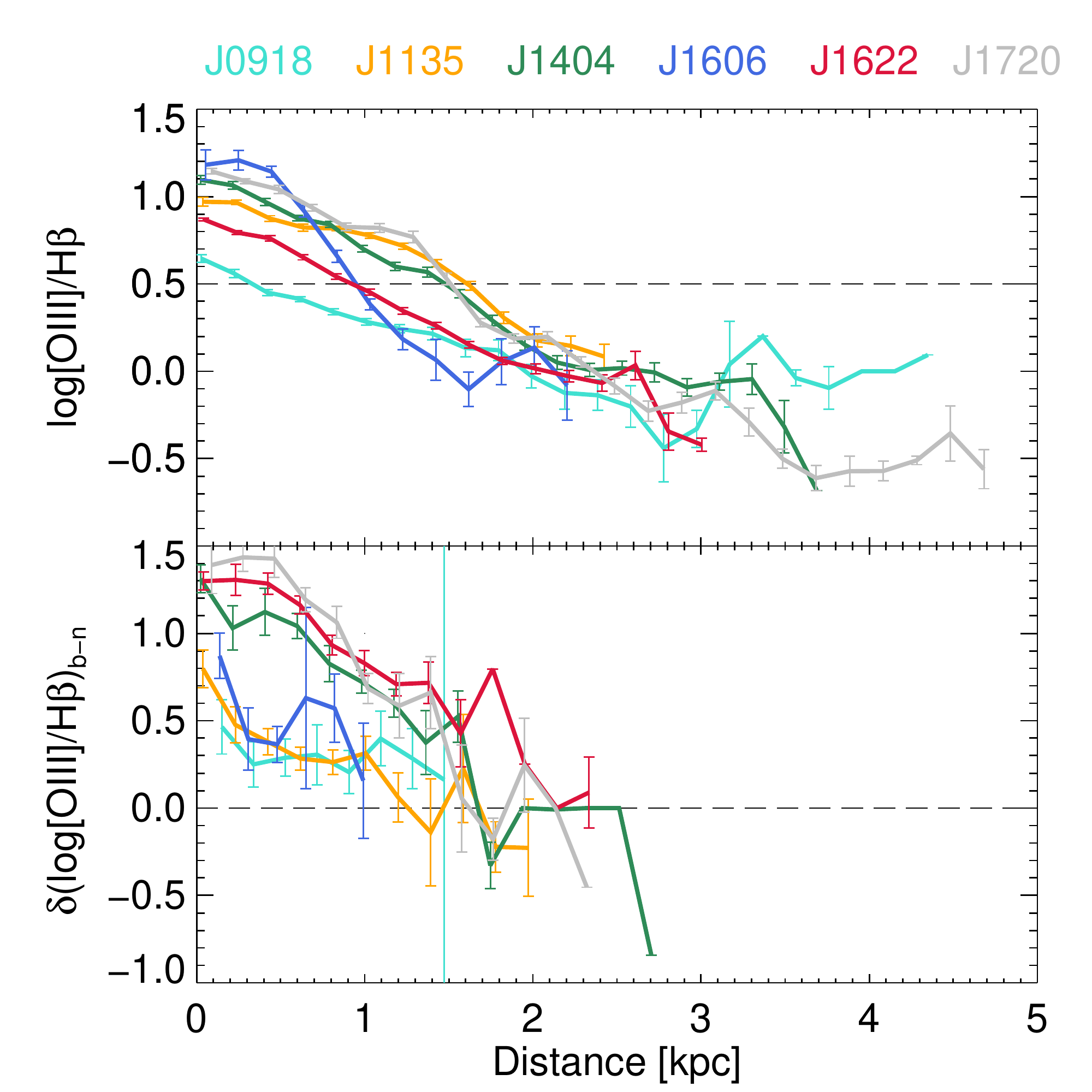}
\caption{Mean \OIII/H$\beta$ flux ratio based on the total profile (top) and mean \OIII/H$\beta$ flux ratio difference between the broad and narrow emission components (bottom) as a function of distance. The dashed horizontal line marks the boundary between the LINER-like and Seyfert-like regions on the BPT diagram in the top panel and a zero difference in the lower panel. Plotted uncertainties are the standard error of the mean.}
\label{fig:bpt_rad}
\end{center}
\end{figure}
\begin{figure}[tbp]
\begin{center}
\includegraphics[width=0.45\textwidth,angle=0]{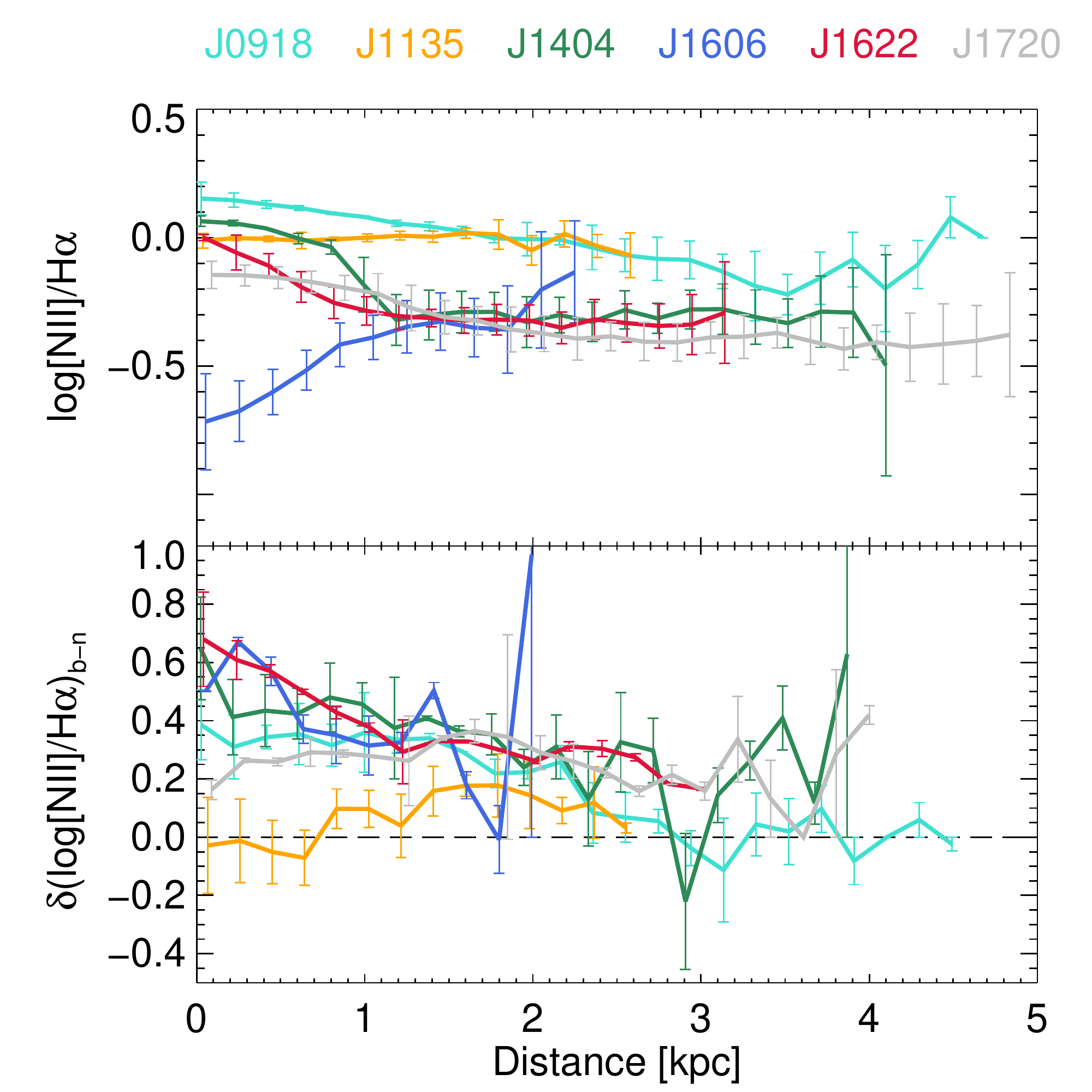}
\caption{Mean \NII/H$\alpha$ flux ratio based on the total profile (top) and mean \NII/H$\alpha$ flux ratio difference between the broad and narrow emission components (bottom) as a function of distance. The dashed horizontal line in the lower panel marks a zero difference. Plotted uncertainties are the standard error of the mean.}
\label{fig:bpt_rad2}
\end{center}
\end{figure}

\begin{figure*}[tbp]
\begin{center}
\raisebox{-0.5\height}{\includegraphics[width=0.16\textwidth,angle=0,trim={60 65 20 40},clip]{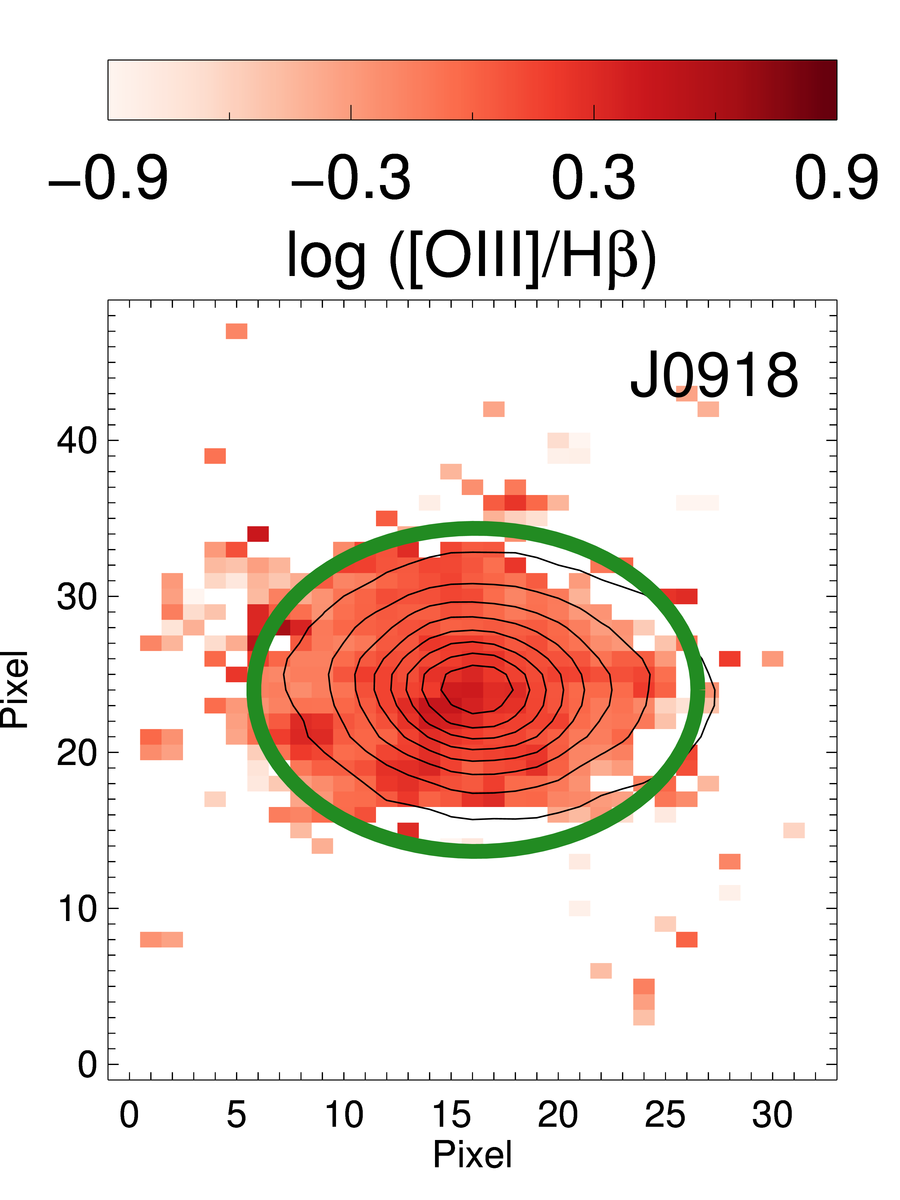}}
\raisebox{-0.5\height}{\includegraphics[width=0.16\textwidth,angle=0,trim={60 65 20 40},clip]{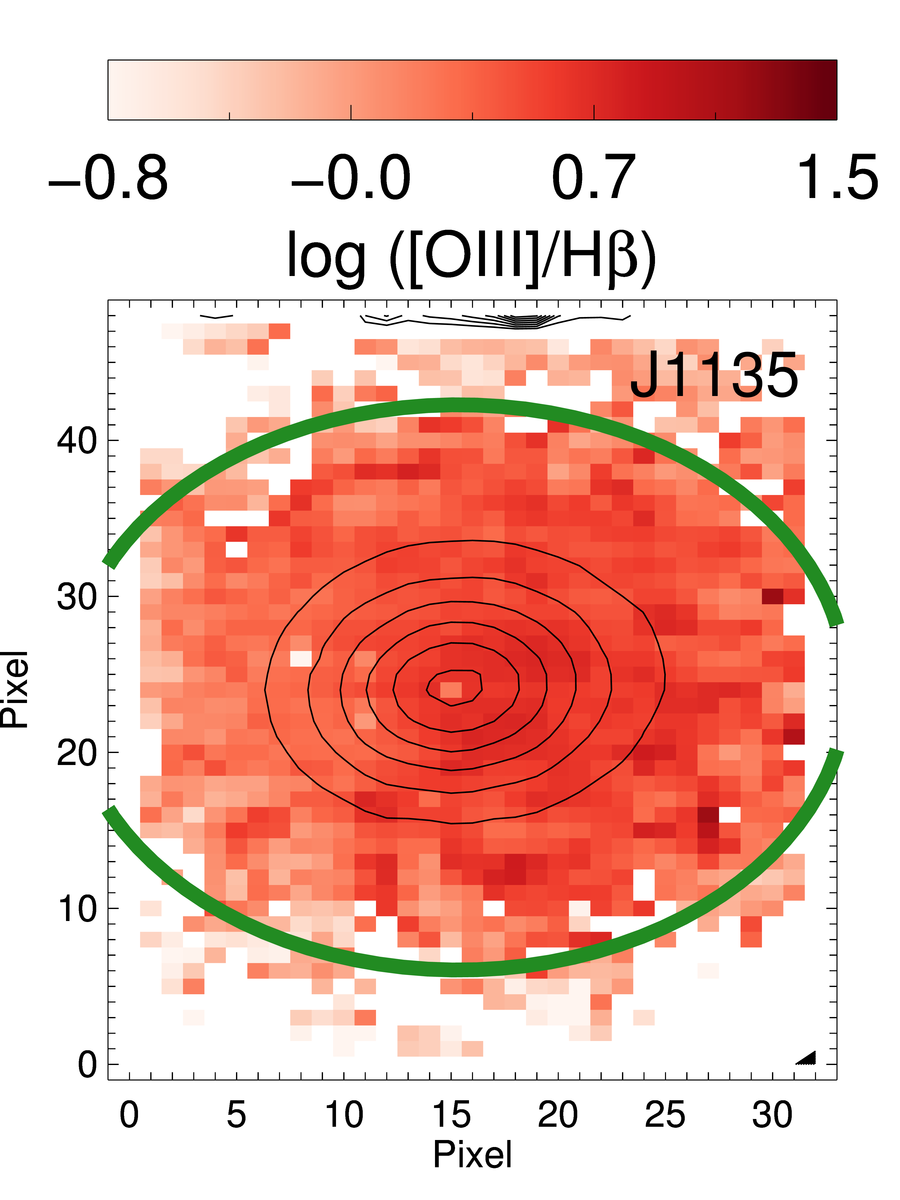}}
\raisebox{-0.5\height}{\includegraphics[width=0.16\textwidth,angle=0,trim={60 65 20 40},clip]{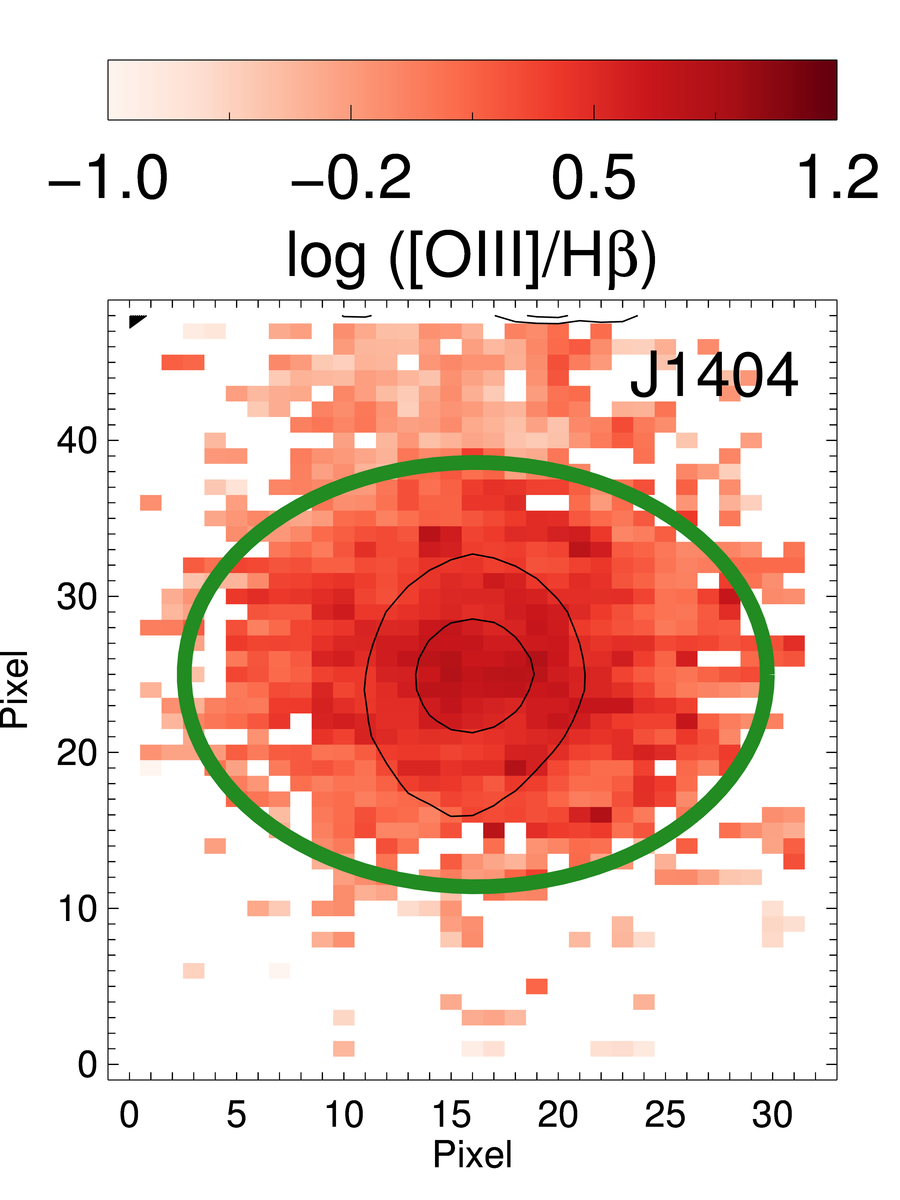}}
\raisebox{-0.5\height}{\includegraphics[width=0.16\textwidth,angle=0,trim={60 65 20 40},clip]{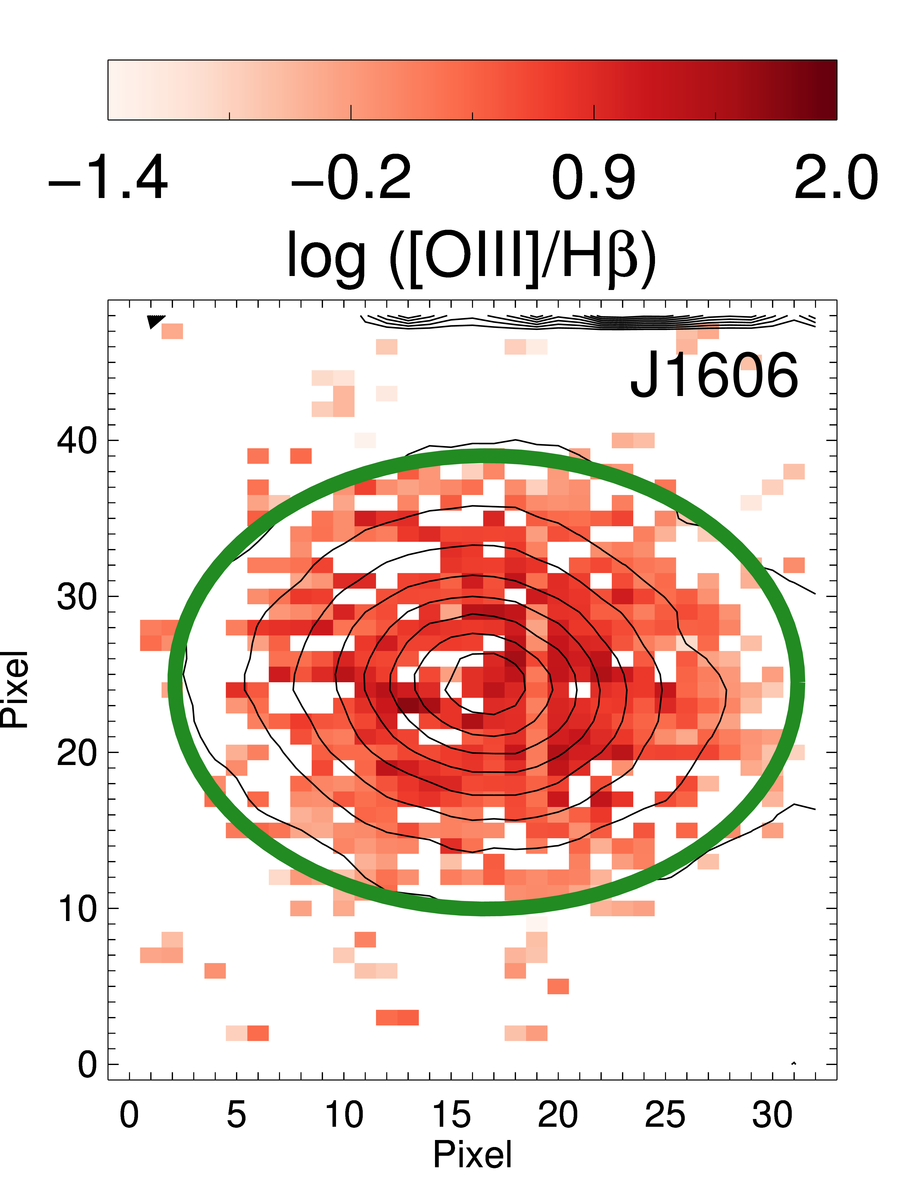}}
\raisebox{-0.5\height}{\includegraphics[width=0.16\textwidth,angle=0,trim={60 65 20 40},clip]{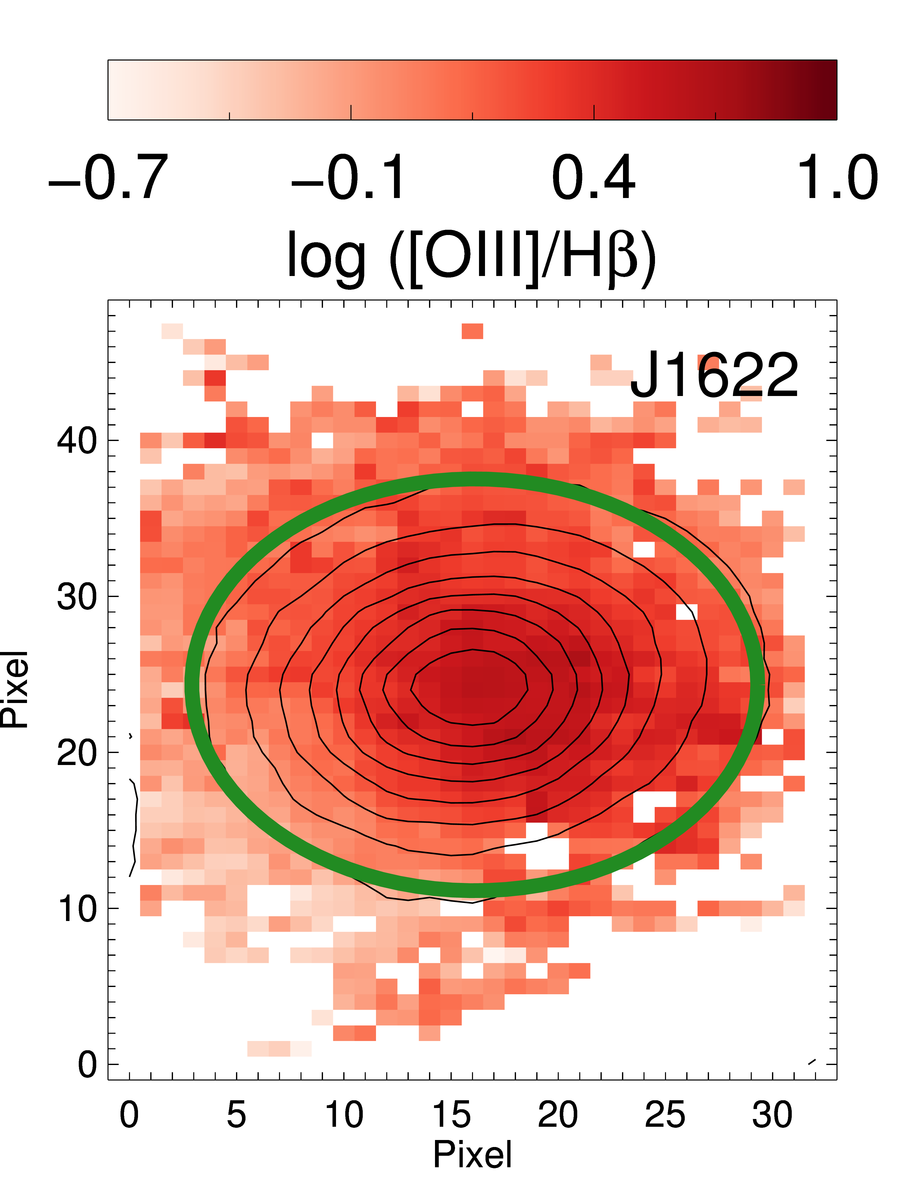}}
\raisebox{-0.5\height}{\includegraphics[width=0.16\textwidth,angle=0,trim={60 65 20 40},clip]{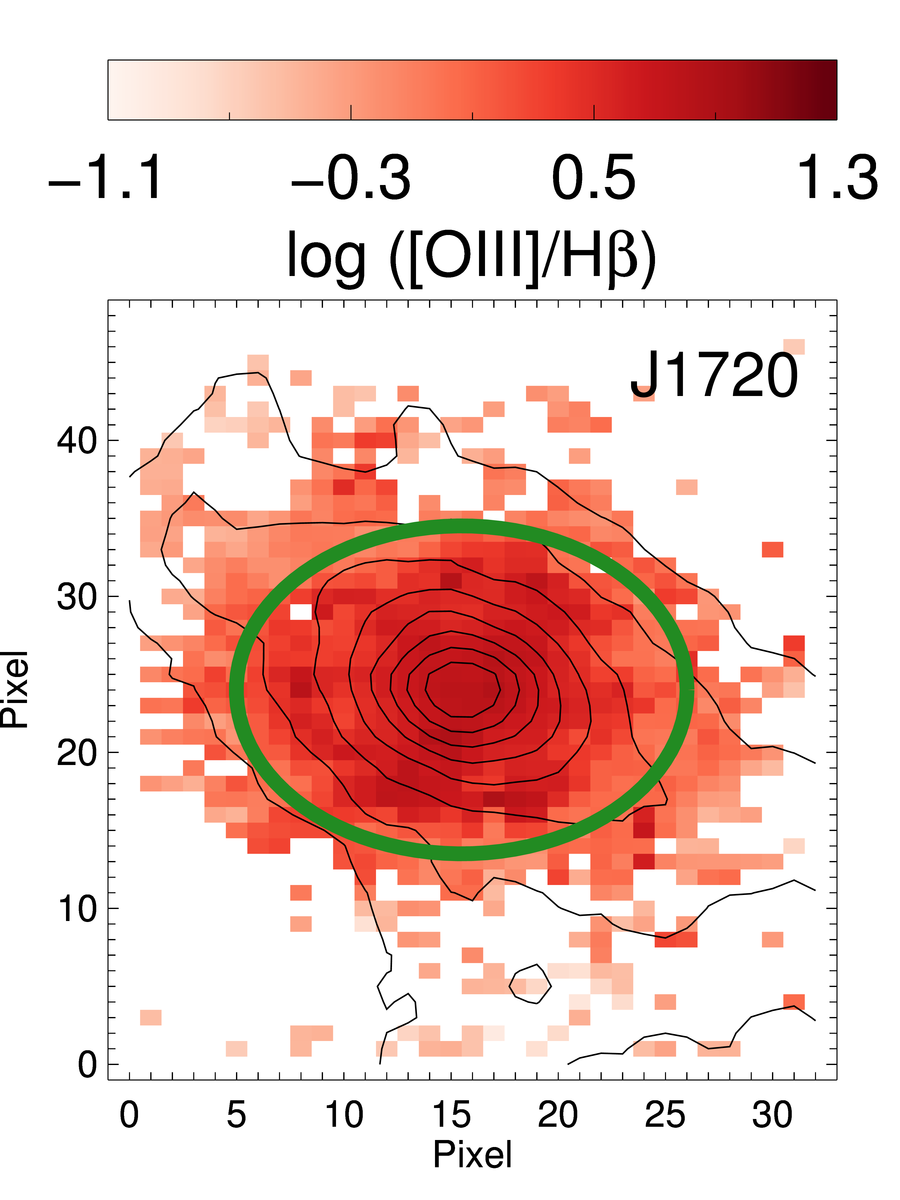}}\\
\raisebox{-0.5\height}{\includegraphics[width=0.16\textwidth,angle=0,trim={60 65 20 40},clip]{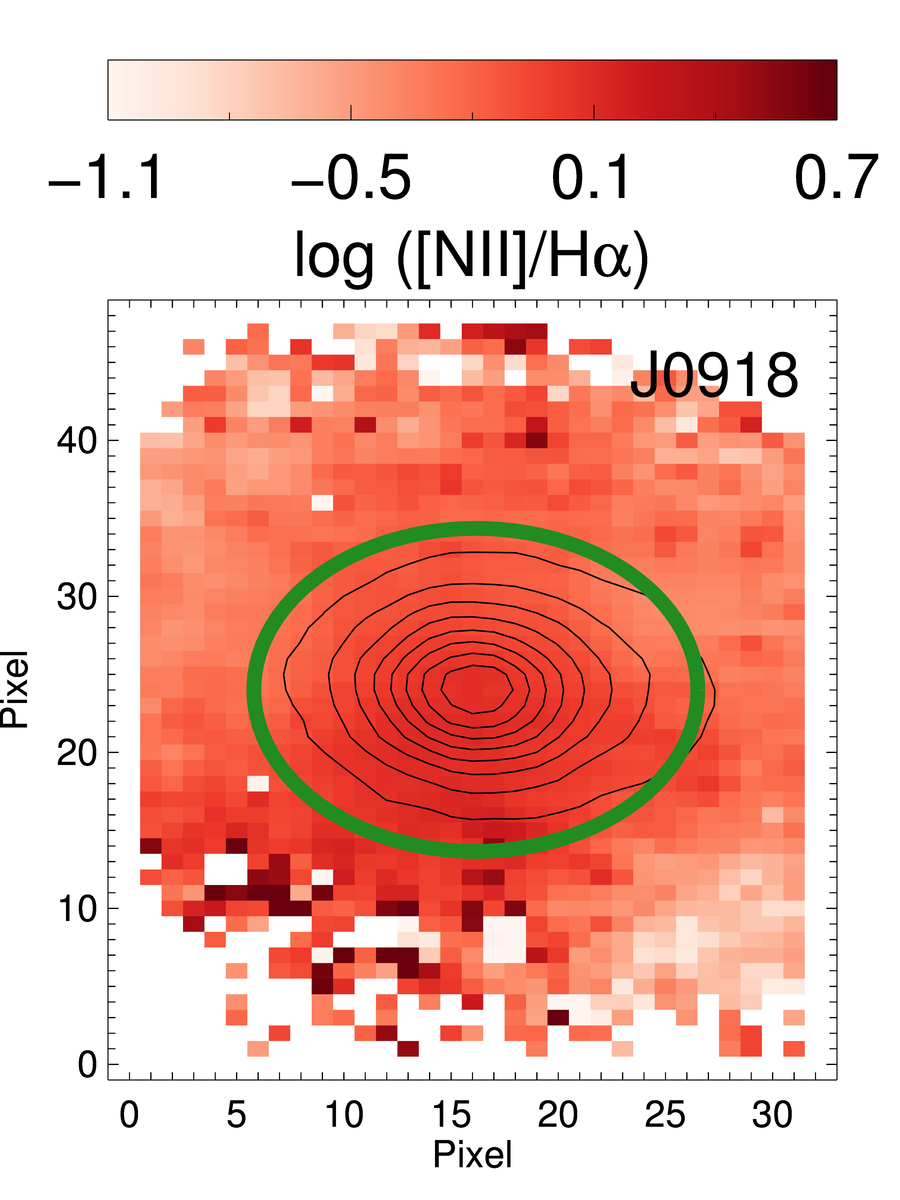}}
\raisebox{-0.5\height}{\includegraphics[width=0.16\textwidth,angle=0,trim={60 65 20 40},clip]{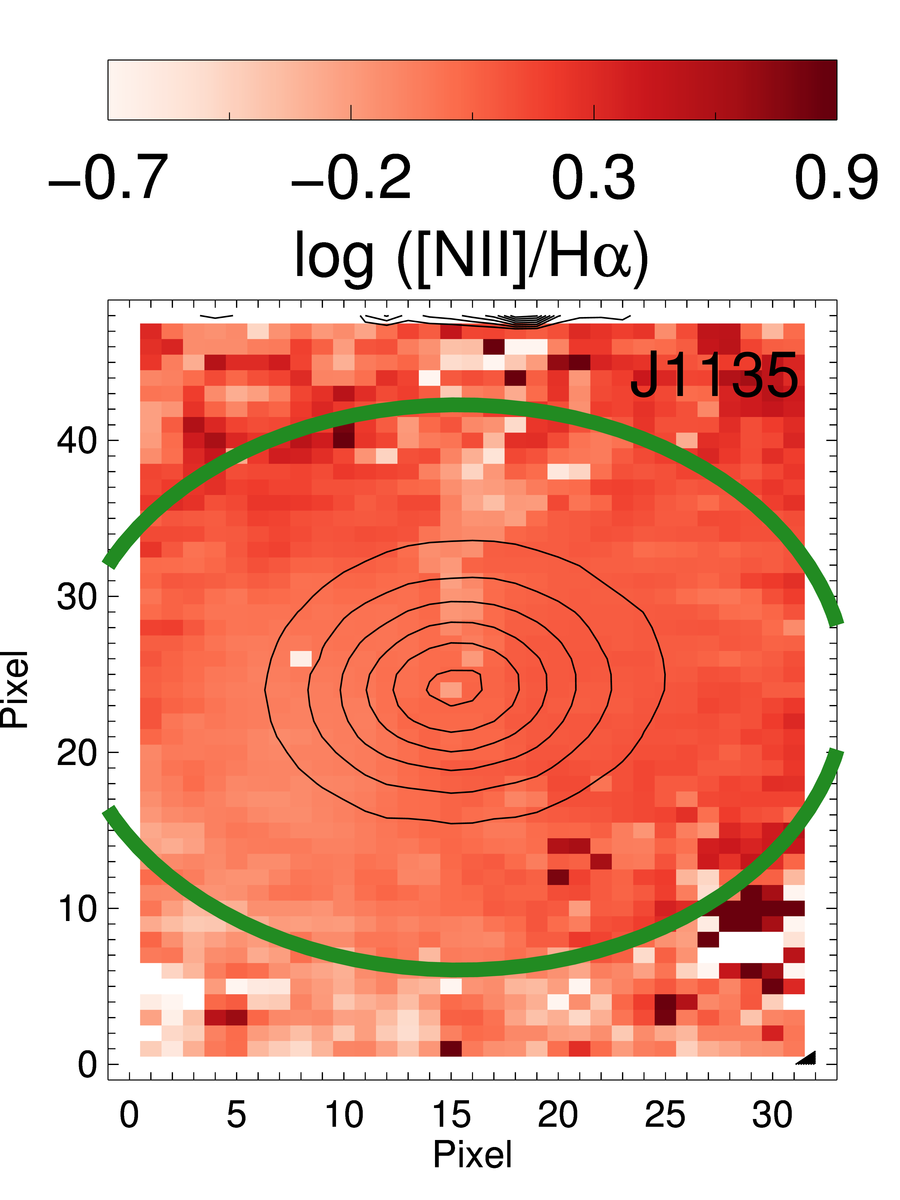}}
\raisebox{-0.5\height}{\includegraphics[width=0.16\textwidth,angle=0,trim={60 65 20 40},clip]{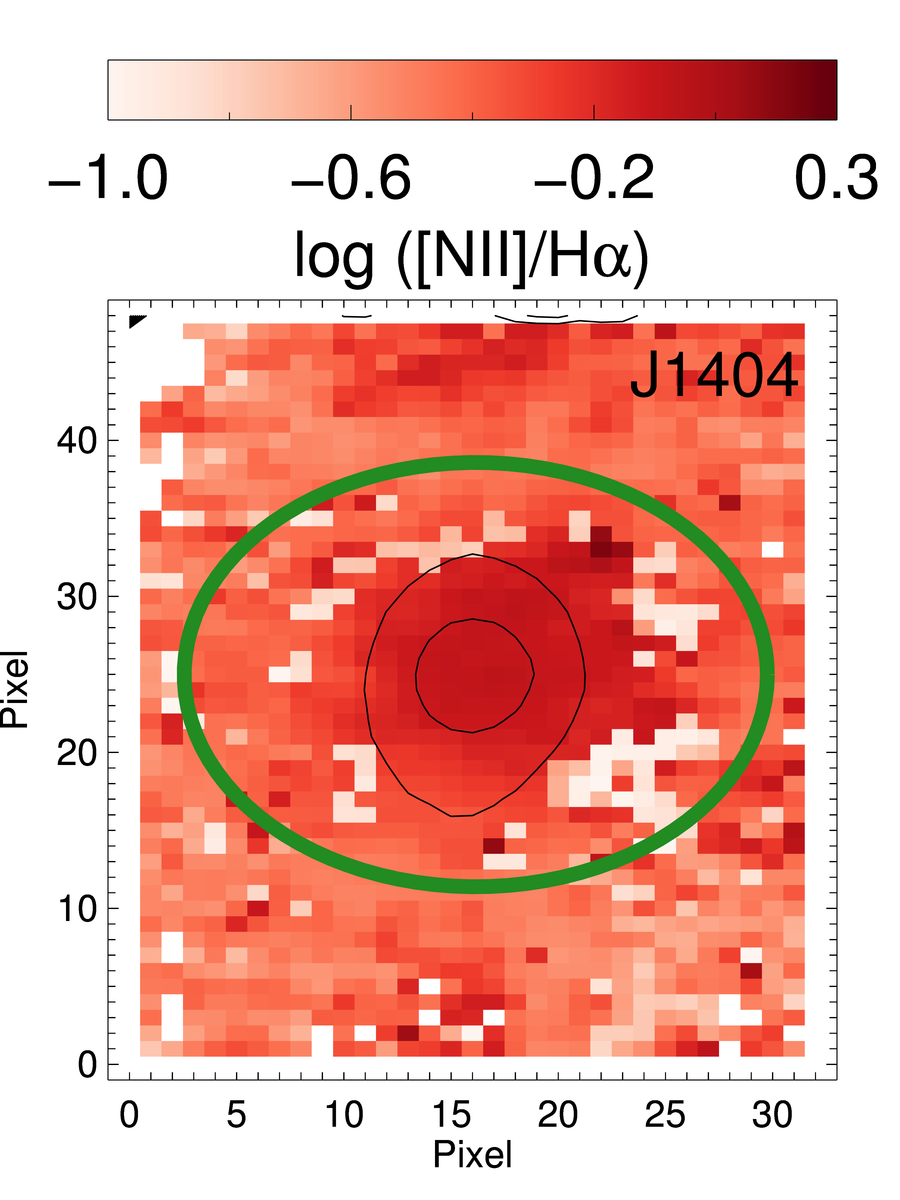}}
\raisebox{-0.5\height}{\includegraphics[width=0.16\textwidth,angle=0,trim={60 65 20 40},clip]{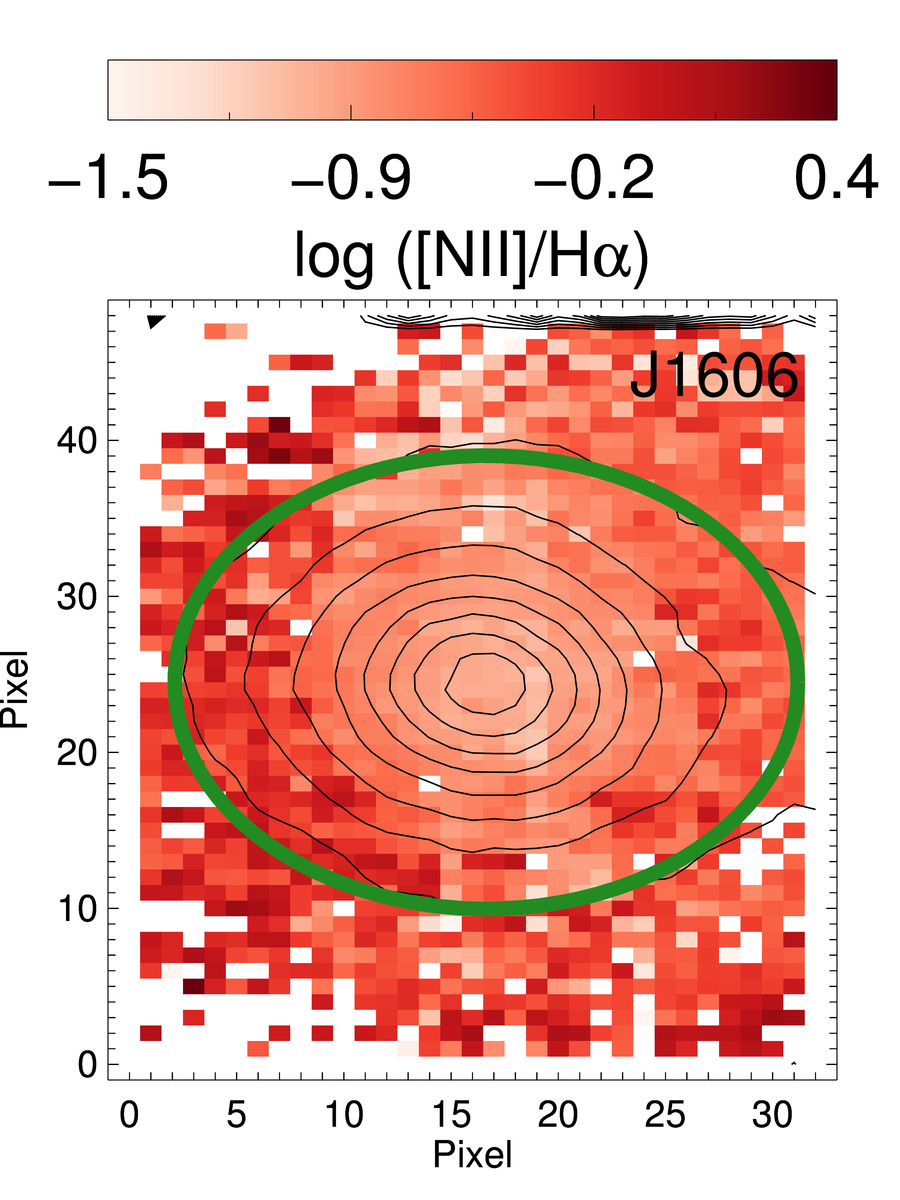}}
\raisebox{-0.5\height}{\includegraphics[width=0.16\textwidth,angle=0,trim={60 65 20 40},clip]{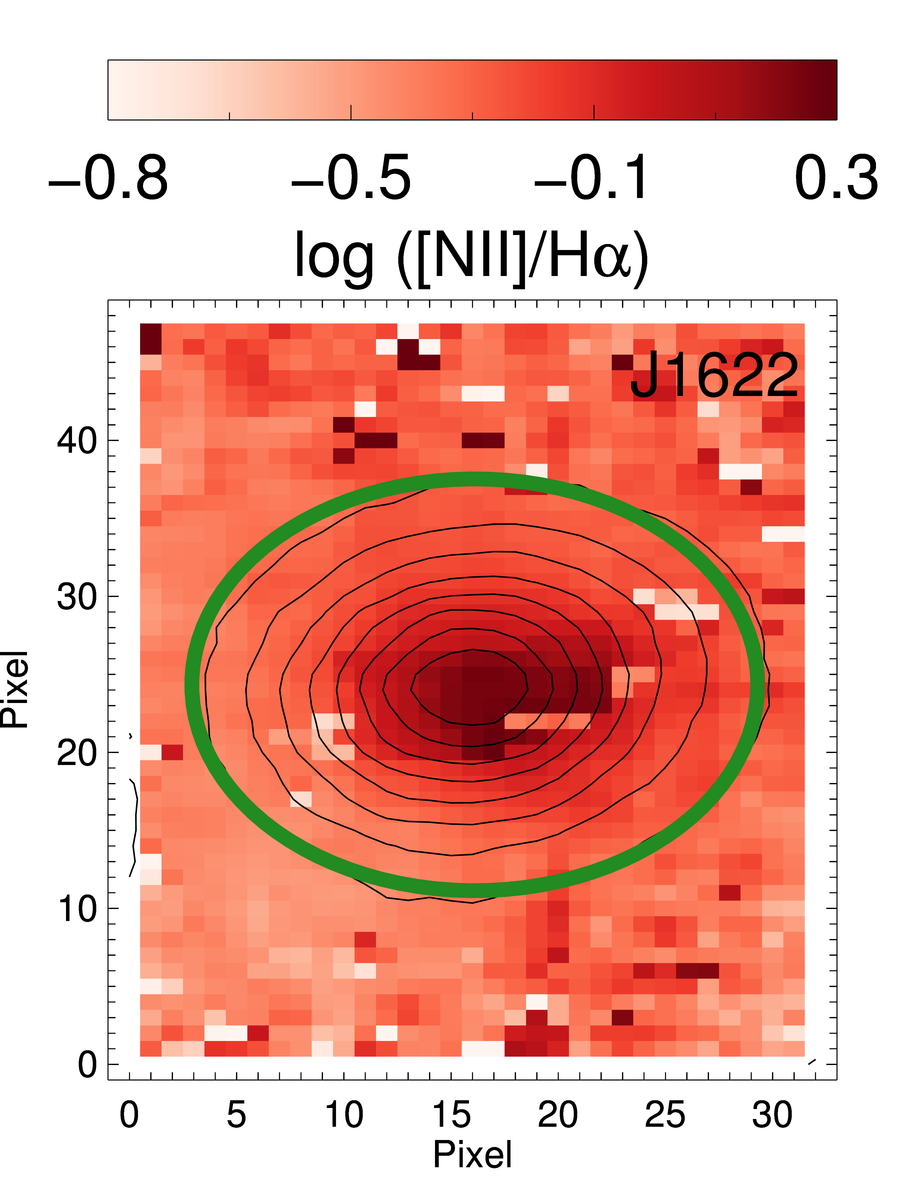}}
\raisebox{-0.5\height}{\includegraphics[width=0.16\textwidth,angle=0,trim={60 65 20 40},clip]{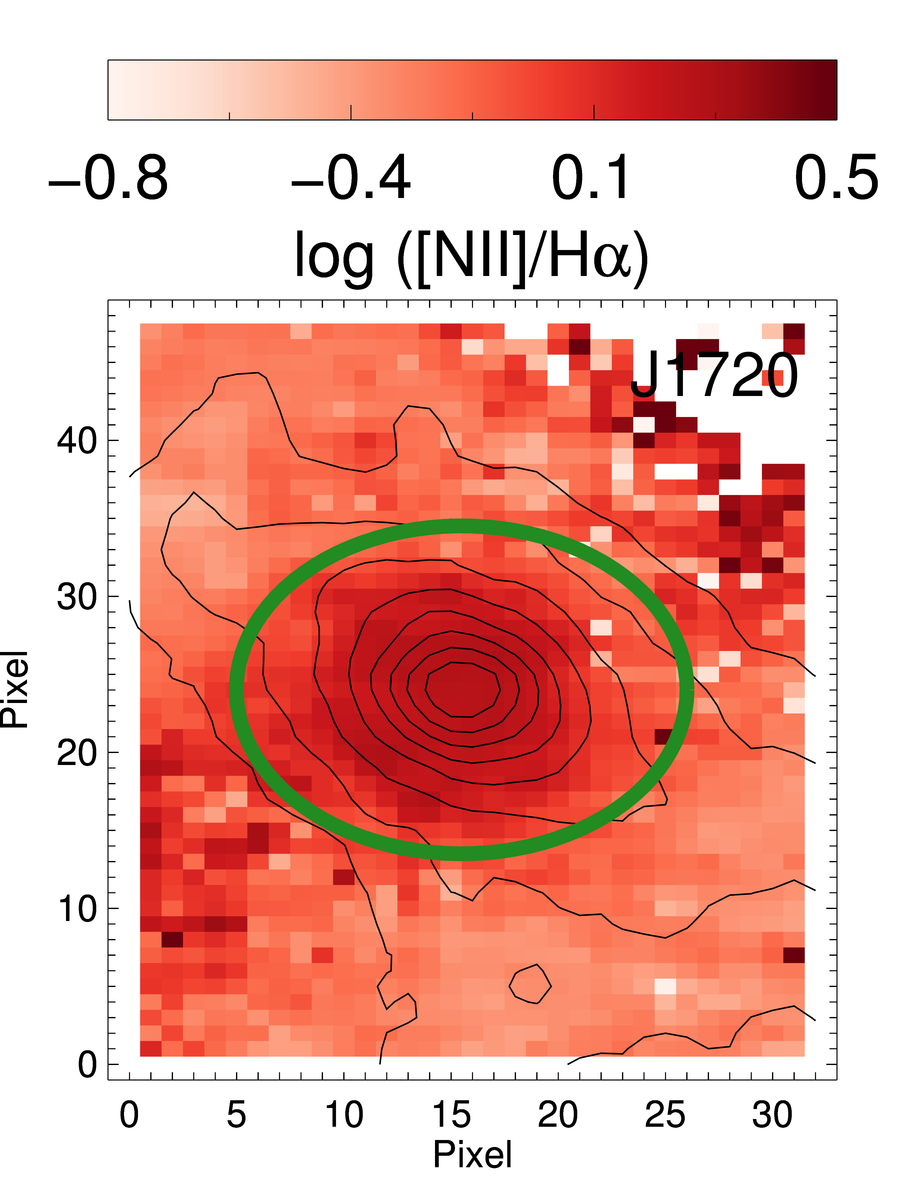}}\\
\caption{\OIII/H$\beta$ (top) and \NII/H$\alpha$ (bottom) flux ratios for the total emission lines. Contours show the continuum flux at 10\% intervals from the peak. The adopted S/N limit is 1 for H$\beta$ and 3 for \OIII\, H$\alpha$, and \NII. White spaxels indicate weak or non-detected emission lines. Green circles show the kinematically defined size of the outflow as calculated in Paper I. {The xy coordinates are defined in units of pixels.}\textbf{ North is up and East is left for J0918, J1606, J1622, and J1720. North is left and East is up for J1135 and J1404.}}
\label{fig:all_fluxratio_map}
\end{center}
\end{figure*}

\begin{figure*}[tbp]
\begin{center}
\raisebox{-0.5\height}{\includegraphics[width=0.8\textwidth,angle=0,trim={0 60 24 5},clip]{BPT_class_map_head.jpg}}\\
\vspace{5pt}
\includegraphics[width=0.4\textwidth,angle=0,trim={20 5 20 10},clip]{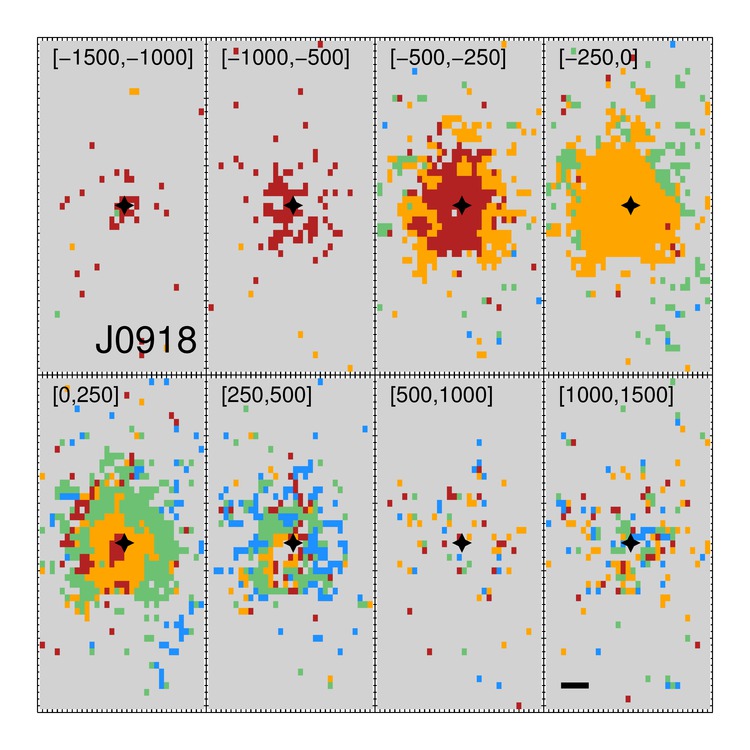}
\includegraphics[width=0.4\textwidth,angle=0,trim={20 5 20 10},clip]{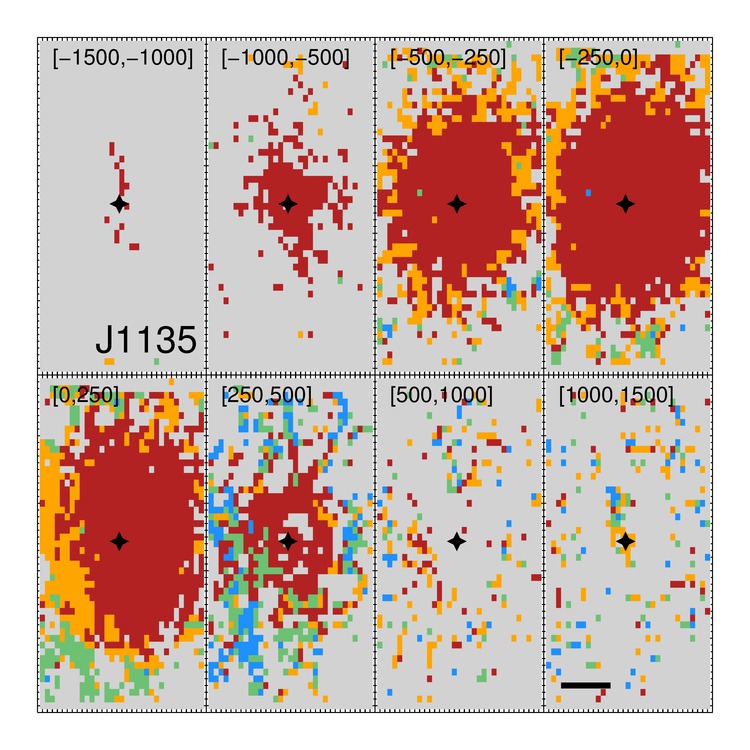}\\
\includegraphics[width=0.4\textwidth,angle=0,trim={20 5 20 10},clip]{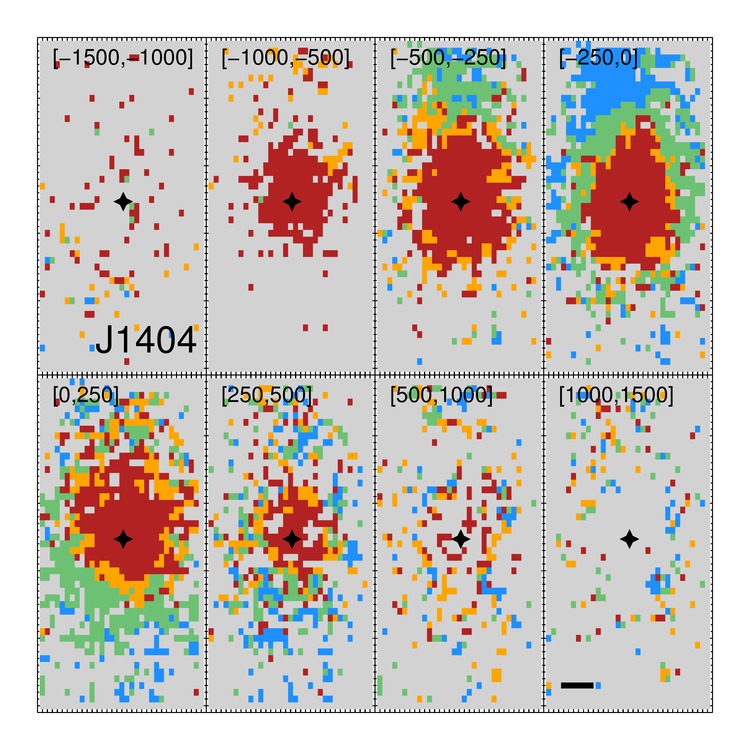}
\includegraphics[width=0.4\textwidth,angle=0,trim={20 5 20 10},clip]{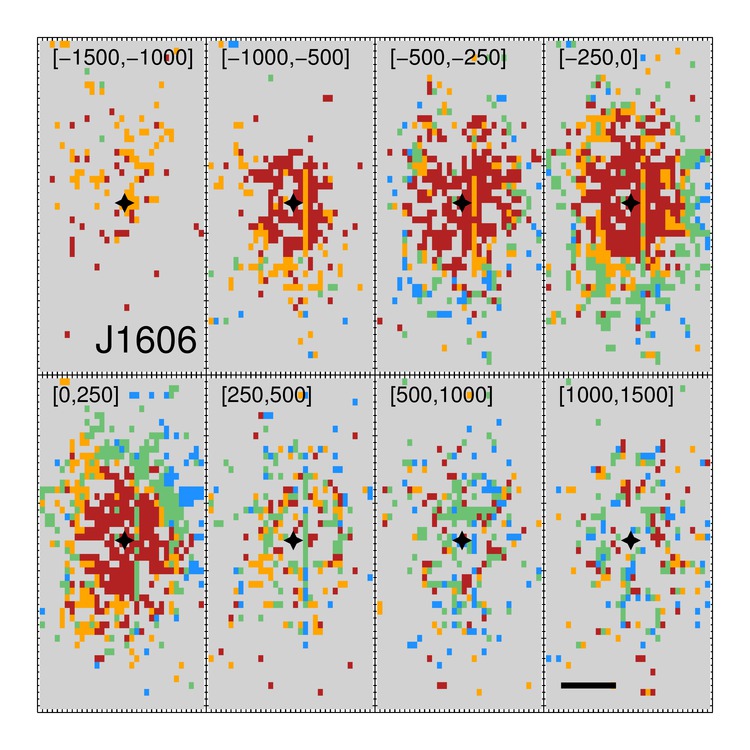}\\
\includegraphics[width=0.4\textwidth,angle=0,trim={20 5 20 10},clip]{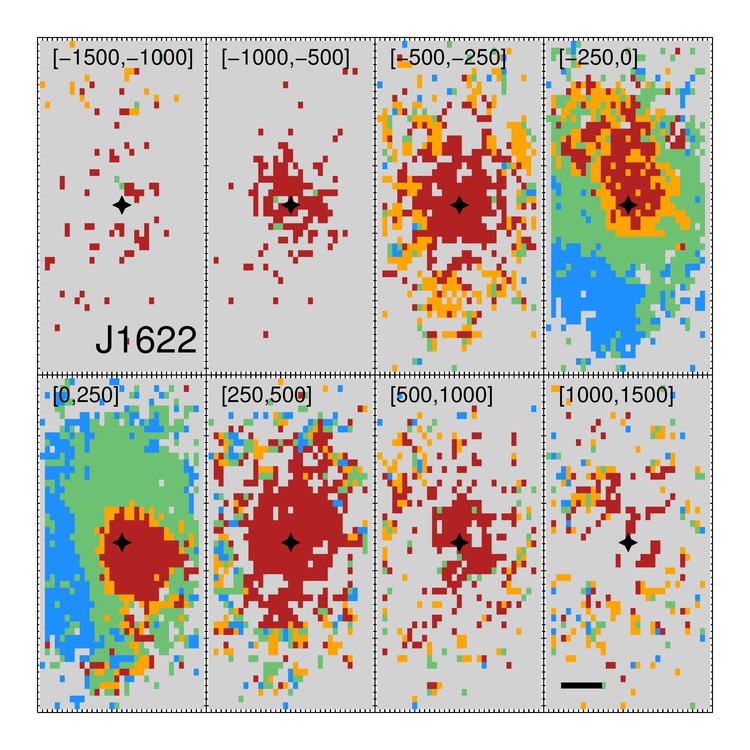}
\includegraphics[width=0.4\textwidth,angle=0,trim={20 5 20 10},clip]{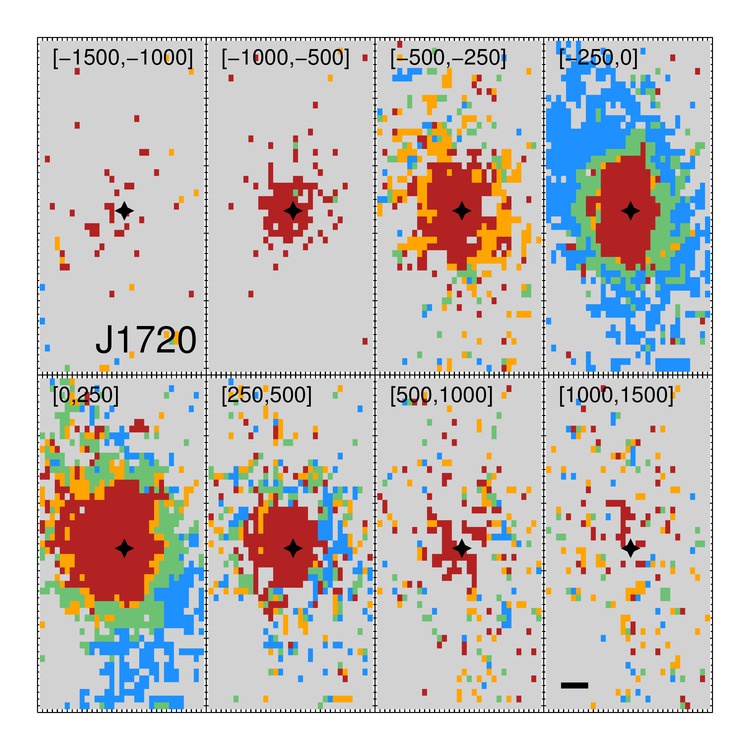}
\caption{BPT classification maps in individual velocity channels, ranging from -1500 to 1500 km s$^{-1}$ and with channel widths of 500 and 250 km s$^{-1}$. S/N limits are 1 for H$\beta$ and 3 for \OIII, H$\alpha$, and \NII. Gray spaxels indicate weak or non-detected emission lines. {Black bars in the lower right panel of each source mark a physical distance of 1 kpc at that redshift.} \textbf{The orientation of the maps is the same as in Figures 1 and 4.}}
\label{fig:bpt_map_chan}
\end{center}
\end{figure*}

The \OIII/H$\beta$ flux ratio is sensitive to the hardness of the ionization field (e.g., \citealt{Kewley2006}). The clear negative gradient with radial distance seen for the \OIII/H$\beta$ ratio (Fig. \ref{fig:bpt_rad}, top) reflects the radially decreasing dominance of AGN photoionization. Nonetheless, within the central kpc, all sources except one (J0918) show flux ratios typical of Seyfert-like emission. Beyond $\sim3$ kpc, the \OIII/H$\beta$ flux ratio indicates ionization from either HII regions or shocks. We do not observe a plateau of the \OIII/H$\beta$ flux ratio as a function of radius, e.g., reported in \citet{Liu2013b}. This may imply that we are not resolving the inner structure of these outflows given their lower \OIII\ luminosities and, consequently, smaller outflows sizes (\citealt{Liu2013b} report outflows size $\sim10-20$ kpc, compared to 1.3-2.1 kpc for our sources).

We consider the difference between the flux ratios of the broad and narrow emission components (Fig. \ref{fig:bpt_rad}, bottom), finding a consistent positive difference between the two that however decreases radially. This implies a higher ionization parameter for the gas entrained in the outflow and therefore that the broad component of the emission line is physically connected to the photoionizing emission of the AGN.

In the case of the \NII/H$\alpha$ flux ratio profile (Fig. \ref{fig:bpt_rad2}), there is a much wider range for any given radial distance compared to the \OIII/H$\beta$ flux ratio, reflected by the large error bars. Unlike \OIII/H$\beta$, the \NII/H$\alpha$ ratio does not show a consistent behavior, with four sources showing declining radial profiles (J0918, J1404, J1622, and J1720), one source showing a fairly flat radial profile (J1135), and one source with a clearly rising trend (J1606). The latter can be due to the challenging decomposition of the H$\alpha$ complex due to the very broad H$\alpha$ component in the central part of the AGN ($<1$ kpc). At radii $\gtrsim2$ kpc, all sources show fairly constant \NII/H$\alpha$ flux ratios, which combined with the negative radial trend of the \OIII/H$\beta$ flux ratio may be indicative of the termination region of the outflow	and the increasing importance of photoionization from massive stars. 

The difference of the \NII/H$\alpha$ flux ratio between the broad and the narrow components of the emission lines (Fig. \ref{fig:bpt_rad2}, bottom) shows differences that take consistently positive values for all radii where broad components are detected. There is  a general decreasing trend except for J1135, although the differential radial profiles for \NII/H$\alpha$ show even larger uncertainties implying significant scatter among co-centric spaxels. This is a result of the distinct kinematic structures seen in the spatially resolved H$\alpha$ velocity and velocity dispersion maps presented in Paper I. 

The spatial distribution of the \OIII/H$\beta$ flux ratio is relatively concentrated (Fig. \ref{fig:all_fluxratio_map}). We do not observe any clear spatial asymmetries, consistent with the photometric and kinematic maps of \OIII\ (Paper I). This in turn indicates that the observed \OIII\ emission is a combination of PSF-smeared emission originating at scales below the spatial resolution of our observations (e.g., \citealt{Husemann2015}) and a larger-scale \OIII\ emission component of lower ionization parameter, potentially not associated with the outflow. As a result any asymmetric morphological or kinematic features associated with the launching of the outflows (e.g., a bicone) are not observed.

In contrast, the \NII/H$\alpha$ flux ratio maps show much more extended and less symmetric distributions than \OIII/H$\beta$. This implies that a considerable fraction of the H$\alpha$ and \NII\ emission arise at larger spatial scales than \OIII . For 4 sources (J0918, J1404, J1622, J1720) \NII/H$\alpha$ ratios are high at the center of the galaxy (log values $>0$). We detect a ring of elevated \NII/H$\alpha$ flux ratio for J1606 but only modest values for the central part\footnote{We expect that contamination to \NII\ from the very broad --Type 1-- H$\alpha$ component is not significant, given the distance of the ring from the center (more than 2 times the seeing FWHM, 0\farcs6 for J1606).}. We detect an elongated feature of increased \NII/H$\alpha$ ratio for J1622, which coincides with a similar feature seen in the kinematics maps of H$\alpha$ (Paper I) and the Seyfert-like `protrusion' noted in the BPT maps of the same source in Fig. \ref{fig:all_bpt_map}. In the Appendix we explore some of these peculiar features. 

\subsubsection{Velocity channel BPT maps}
We explore the connection between the extreme kinematic (broad) component and its driving mechanism by plotting spatially and kinematically resolved BPT maps (Fig. \ref{fig:bpt_map_chan}). We first define velocity channels ranging from -1500 to 1500 km s$^{-1}$ around the systemic velocity of each source, with channel widths ranging between 250 and 500 km s$^{-1}$. Within each velocity channel, we calculate the flux of each of the four emission lines used in the BPT classification. We can then calculate the \OIII/H$\beta$ and \NII/H$\alpha$ flux ratios for each velocity channel.

For increasingly negative velocities AGN photoionization gradually dominates the BPT map, with the highest negative velocity channels showing predominantly Seyfert-like emission that is however concentrated. For example, the AGN component in J1720 is concentrated within $<1$ kpc diameter in the -1000 to -500 km s$^{-1}$ velocity channel. In contrast, for increasingly positive velocities we see both a strong Seyfert-like component but also contribution from star-forming and composite regions. Photoionization from HII and composite regions is the strongest in the two channels bracketing the systemic velocity. Unlike the other five sources, J0918 shows a strong LINER- and composite-like emission around its systemic velocity, with a strong Seyfert-like component only emerging in the negative velocity channels ($<-250$ km s$^{-1}$). Finally, we observe clear rotation patterns for J1404, J1606, J1622, and J1720 in the two central velocity channels, mostly for the non-AGN classified emission (composite and star formation). Note that sources with large \OIII\ velocity dispersions but modest velocities (e.g., J1622) show similar spatial distributions of BPT classification in different velocity channels. In contrast, sources with large negative velocities (e.g., J1135) clearly show a strong AGN component in negative velocity channels but no clear AGN detection in the positive velocity channels.

\begin{figure}[tbp]
\begin{center}
\includegraphics[width=0.23\textwidth,angle=0,trim={20 5 20 10},clip]{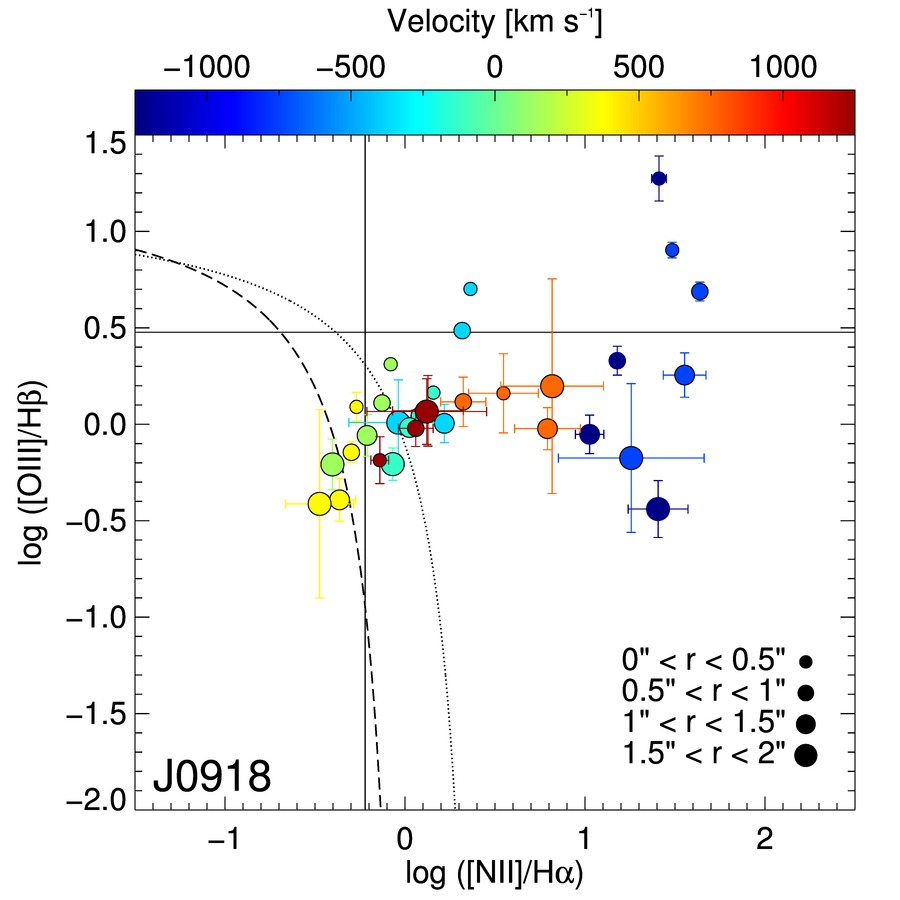}
\includegraphics[width=0.23\textwidth,angle=0,trim={20 5 20 10},clip]{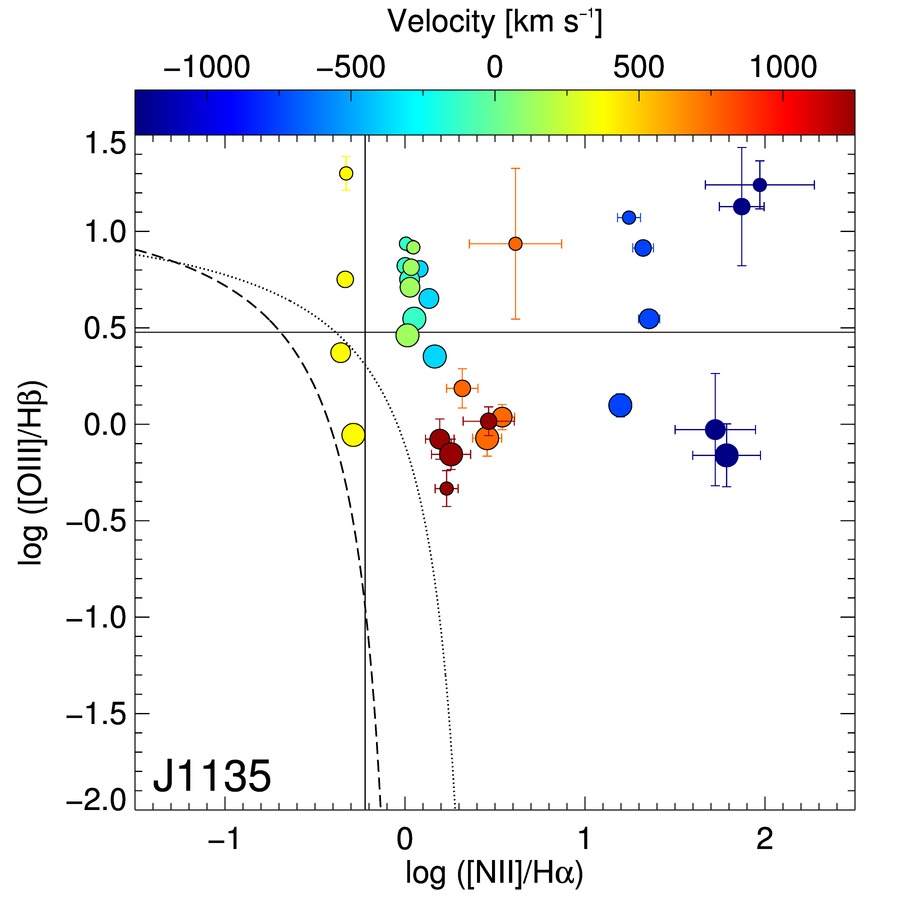}\\
\includegraphics[width=0.23\textwidth,angle=0,trim={20 5 20 10},clip]{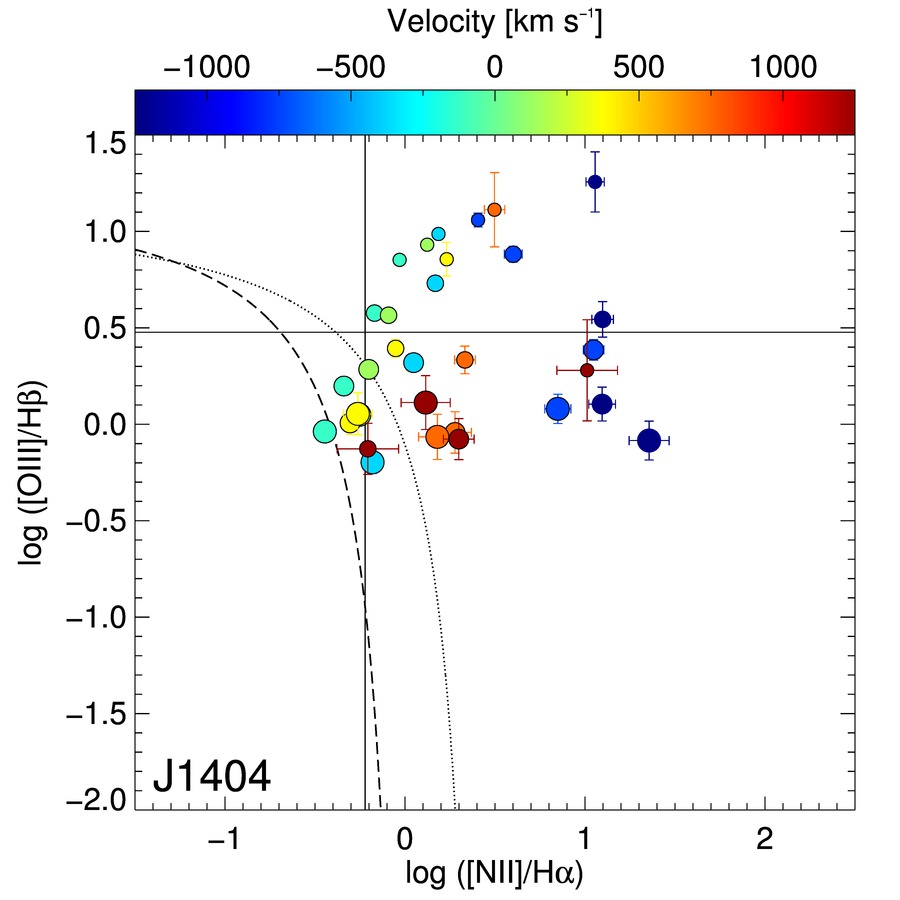}
\includegraphics[width=0.23\textwidth,angle=0,trim={20 5 20 10},clip]{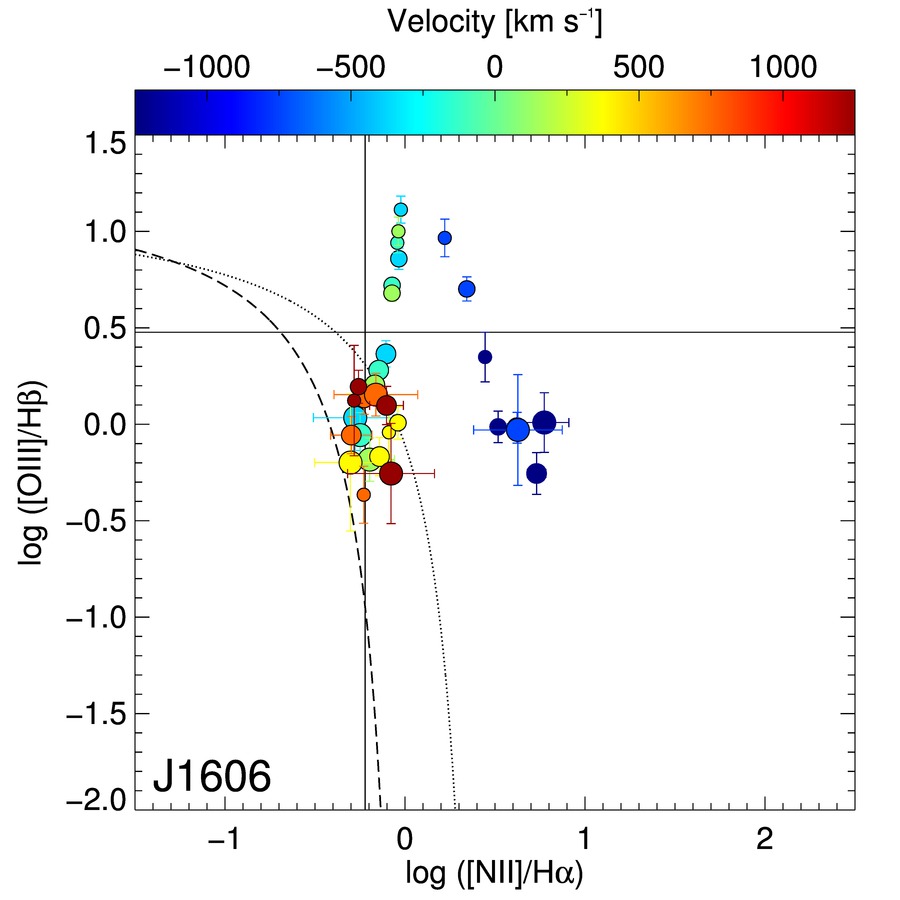}\\
\includegraphics[width=0.23\textwidth,angle=0,trim={20 5 20 10},clip]{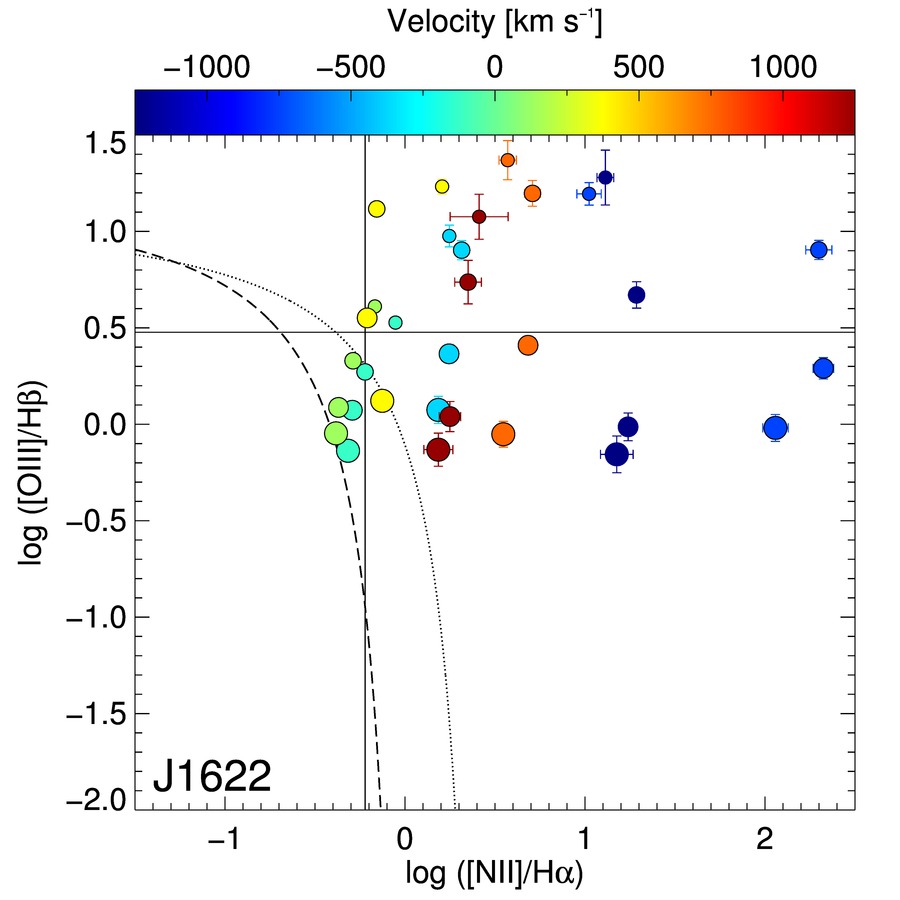}
\includegraphics[width=0.23\textwidth,angle=0,trim={20 5 20 10},clip]{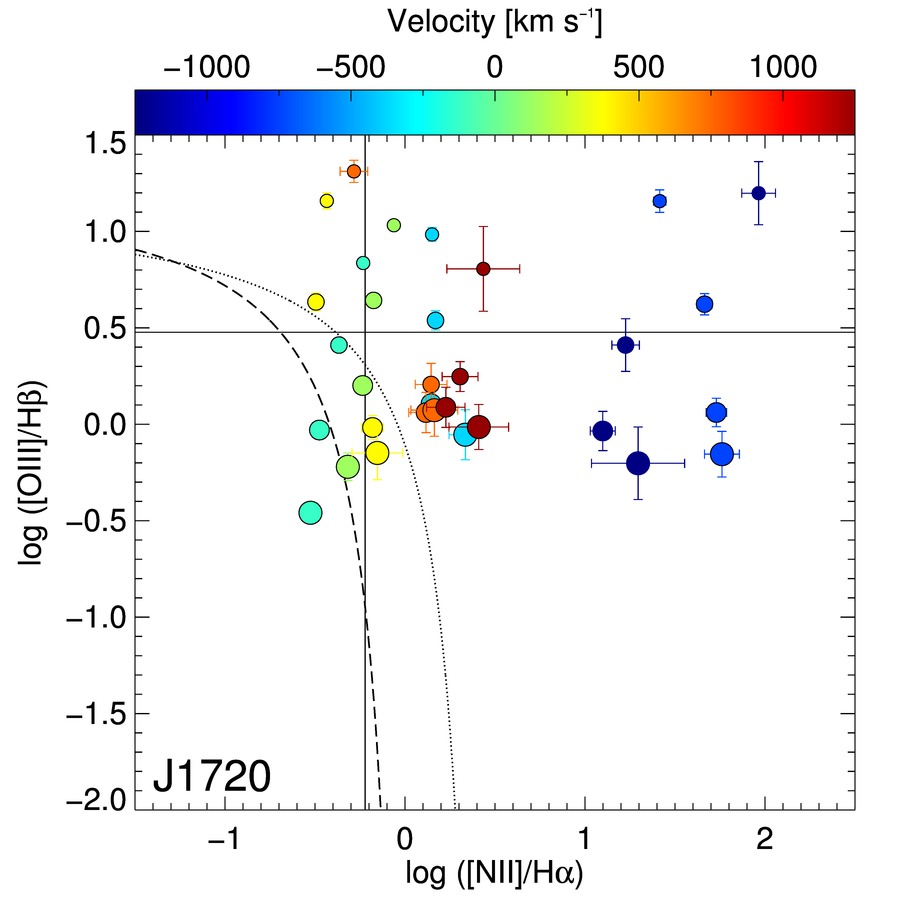}
\caption{BPT diagrams showing mean flux ratios within concentric annuli of 0\farcs5 width, increasing radial distance (different size symbols), and for different velocity channels (color coding). The demarkation lines come from \citet{Kewley2001} (dotted curved line) and \citet{Kauffmann2003} (dashed line). {Vertical and horizontal solid lines come from \citet{Ho1997b} and denote typical flux ratio limits used to identify LINER-like and Seyfert-like emission.} Uncertainties plotted reflect the statistical scatter within each annulus.}
\label{fig:all_bpt_chan}
\end{center}
\end{figure}

Next, we project the BPT velocity channel maps of Fig. \ref{fig:bpt_map_chan} onto the actual BPT diagram in Fig. \ref{fig:all_bpt_chan}. We spatially divide the IFU into annuli of 0\farcs5 width each and calculate the mean flux ratio values for each annulus (symbol size) for a given velocity channel (color scale). Considering emission from the most negative to the most positive velocities leads to a diagonal movement on the BPT diagram, generally from the upper right toward the lower left (from higher to lower ionization parameter). Conversely, emission from increasingly larger radii results in a more vertical movement on the BPT diagram, generally from the upper to the lower part of the BPT. 

More specifically, emission in velocity channels near zero velocity (cyan to yellow) is predominantly classified as either composite or HII-like region. On the other hand, the emission at most negative velocities (dark blue symbols) is predominantly classified as Seyfert-like, with the outer annuli shifting towards the LINER region. This result is due to the negative radial gradient of the \OIII/H$\beta$ flux ratio (Fig. \ref{fig:bpt_rad}) and the fairly flat \NII/H$\alpha$ radial profiles (Fig. \ref{fig:bpt_rad2}). 

\subsection{Velocity-velocity dispersion diagram}

\begin{figure*}[tbp]
\begin{center}
\includegraphics[width=0.3\textwidth,angle=0,trim={0 30 25 25},clip]{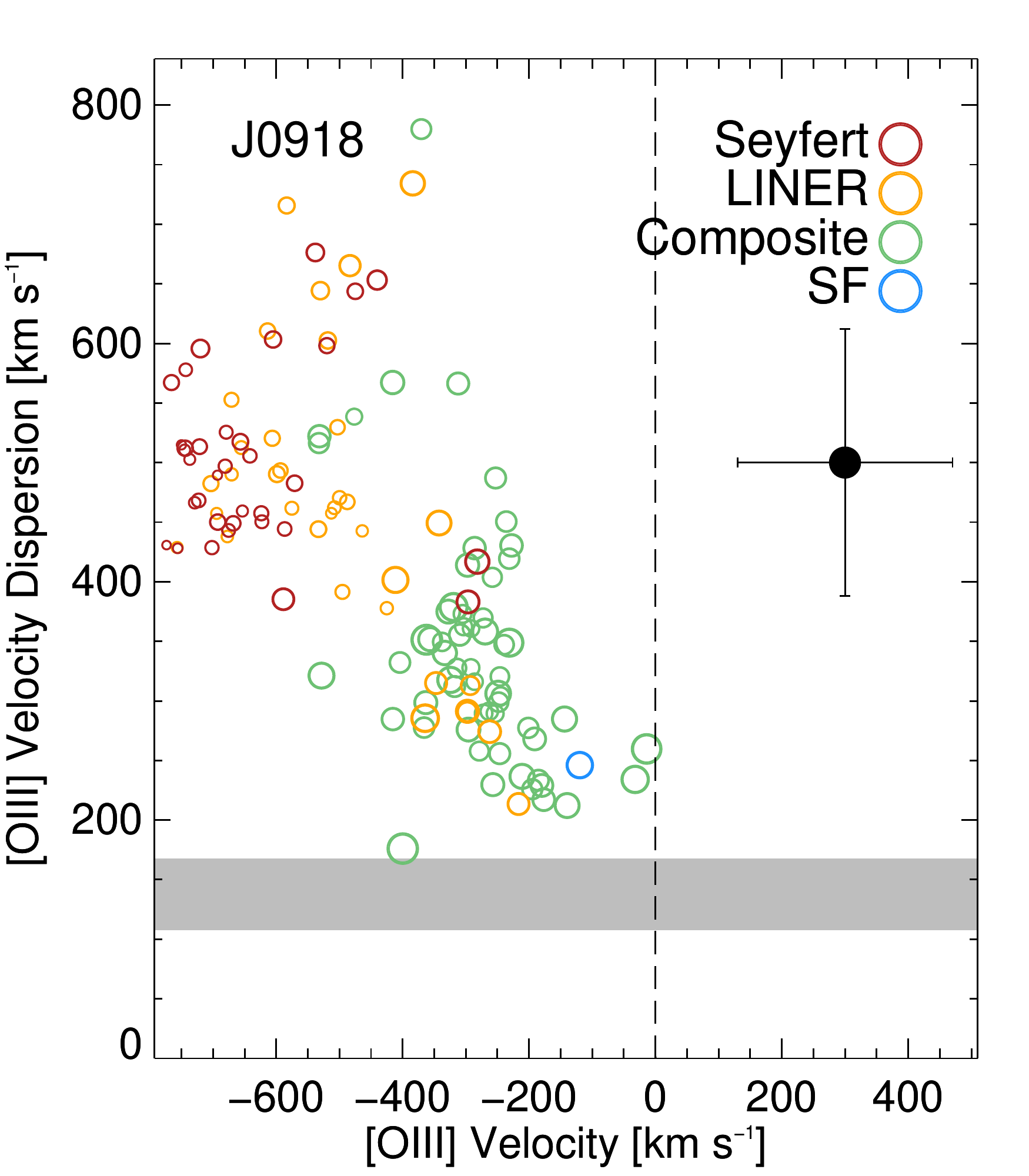}
\includegraphics[width=0.3\textwidth,angle=0,trim={0 30 25 25},clip]{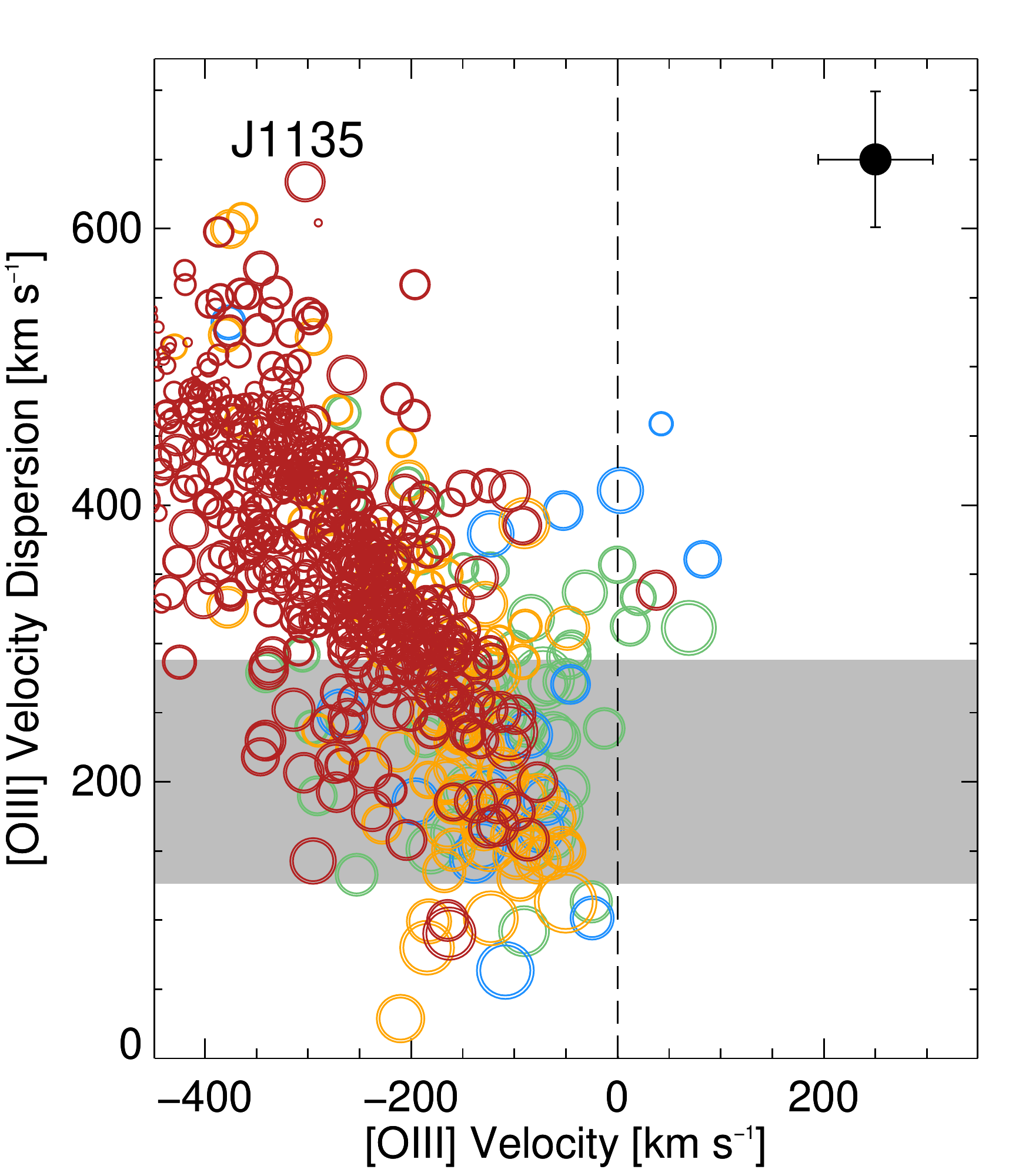}
\includegraphics[width=0.3\textwidth,angle=0,trim={0 30 25 25},clip]{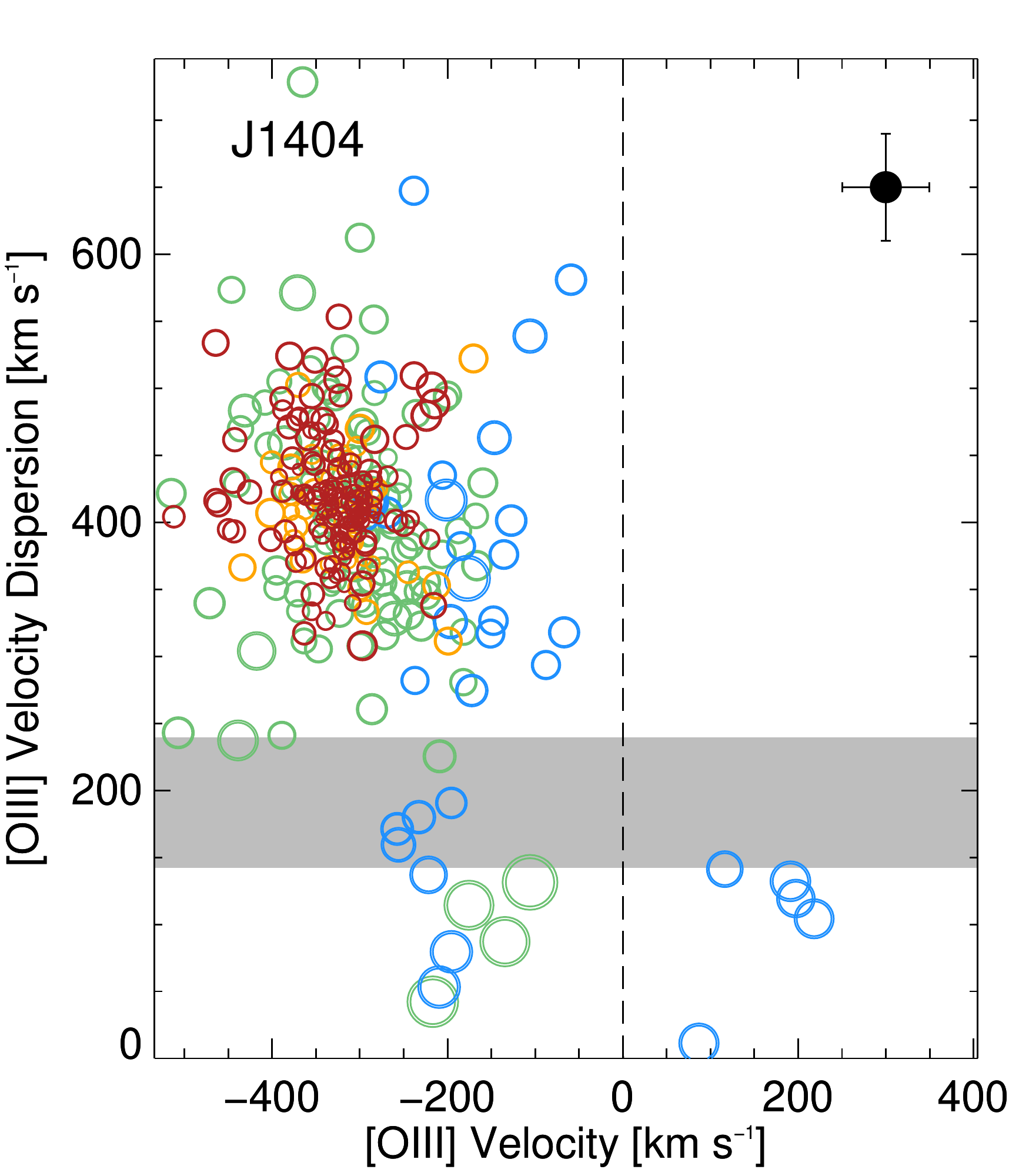}\\
\includegraphics[width=0.3\textwidth,angle=0,trim={0 0 25 25},clip]{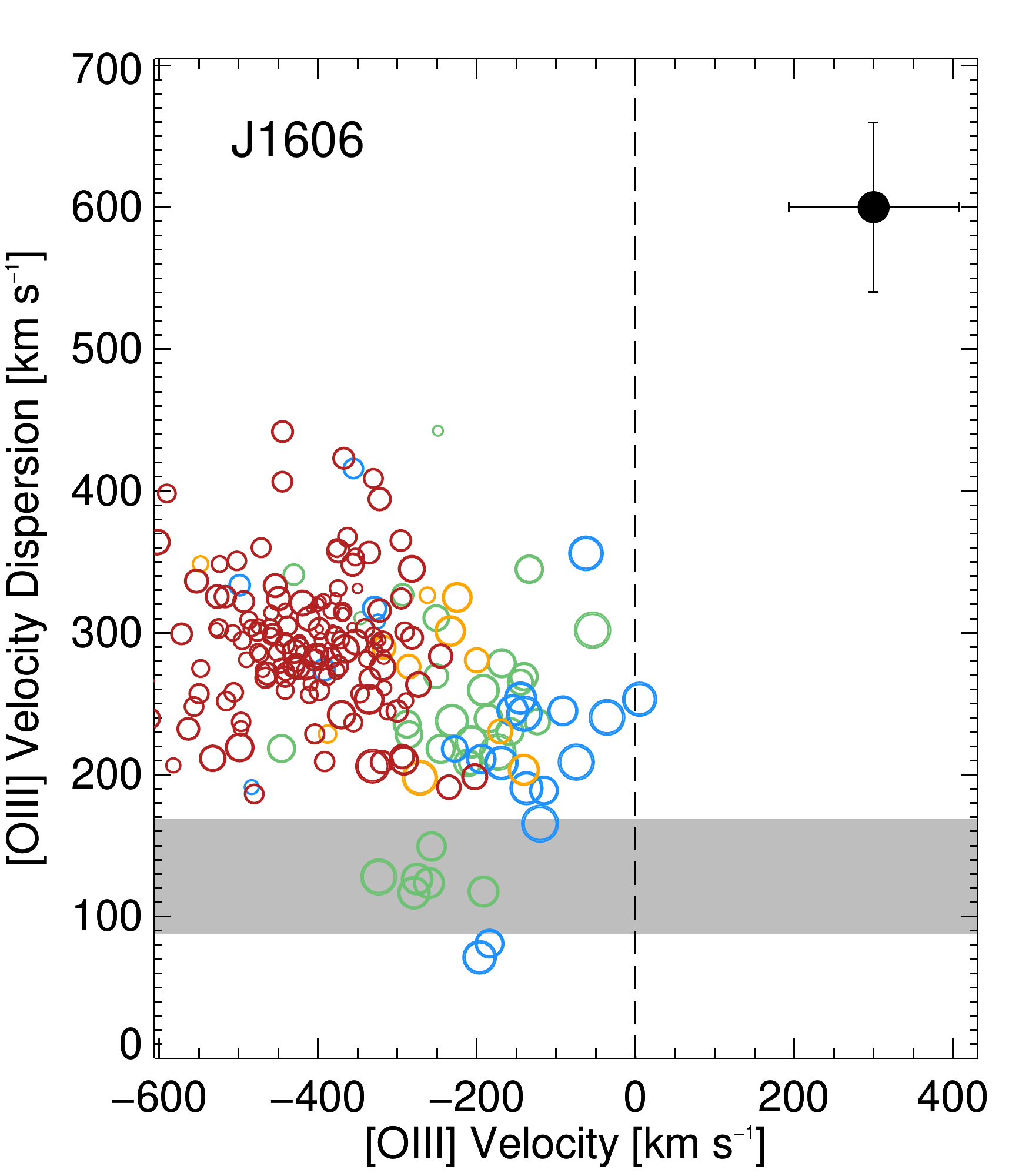}
\includegraphics[width=0.3\textwidth,angle=0,trim={0 0 25 25},clip]{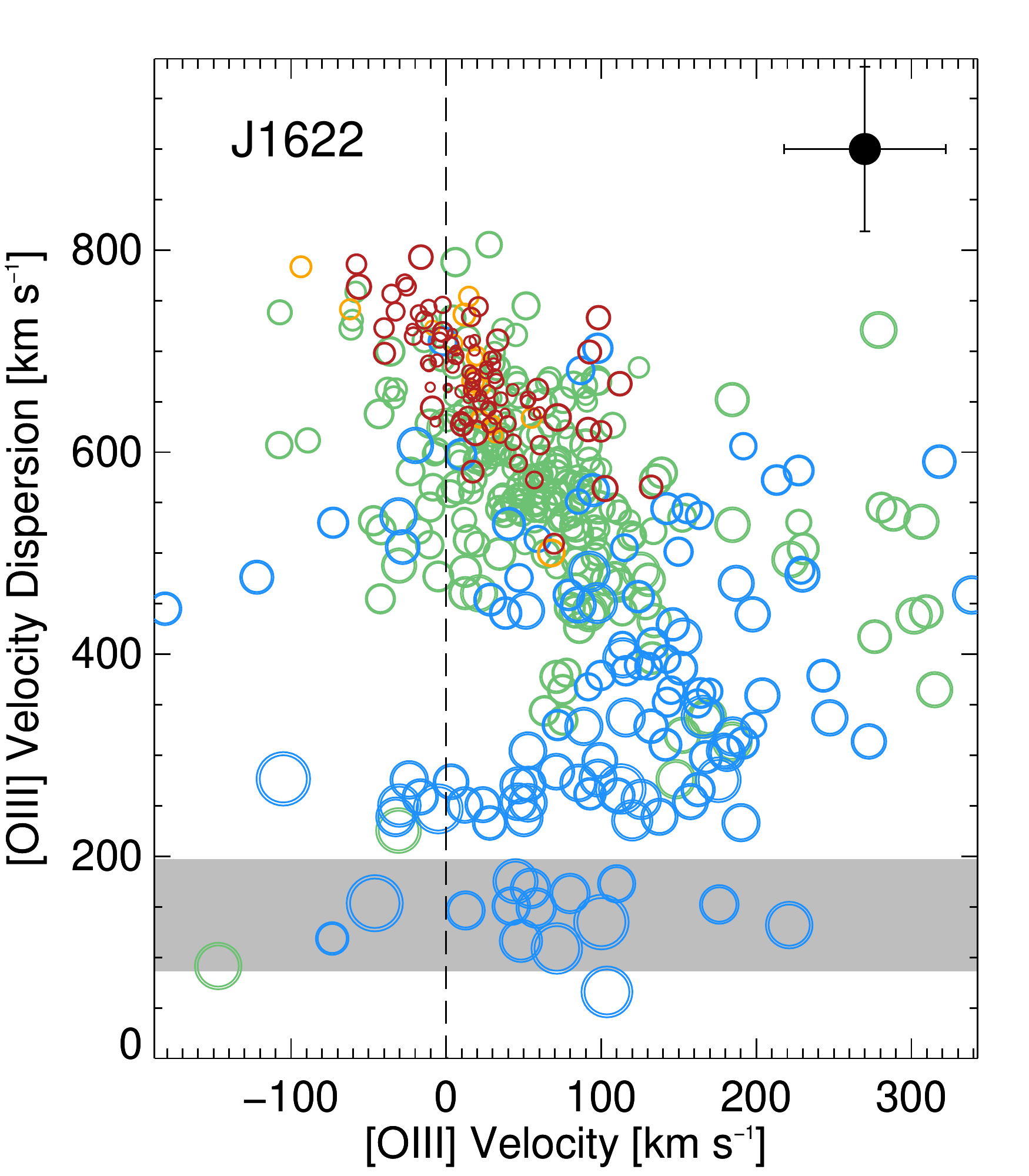}
\includegraphics[width=0.3\textwidth,angle=0,trim={0 0 25 25},clip]{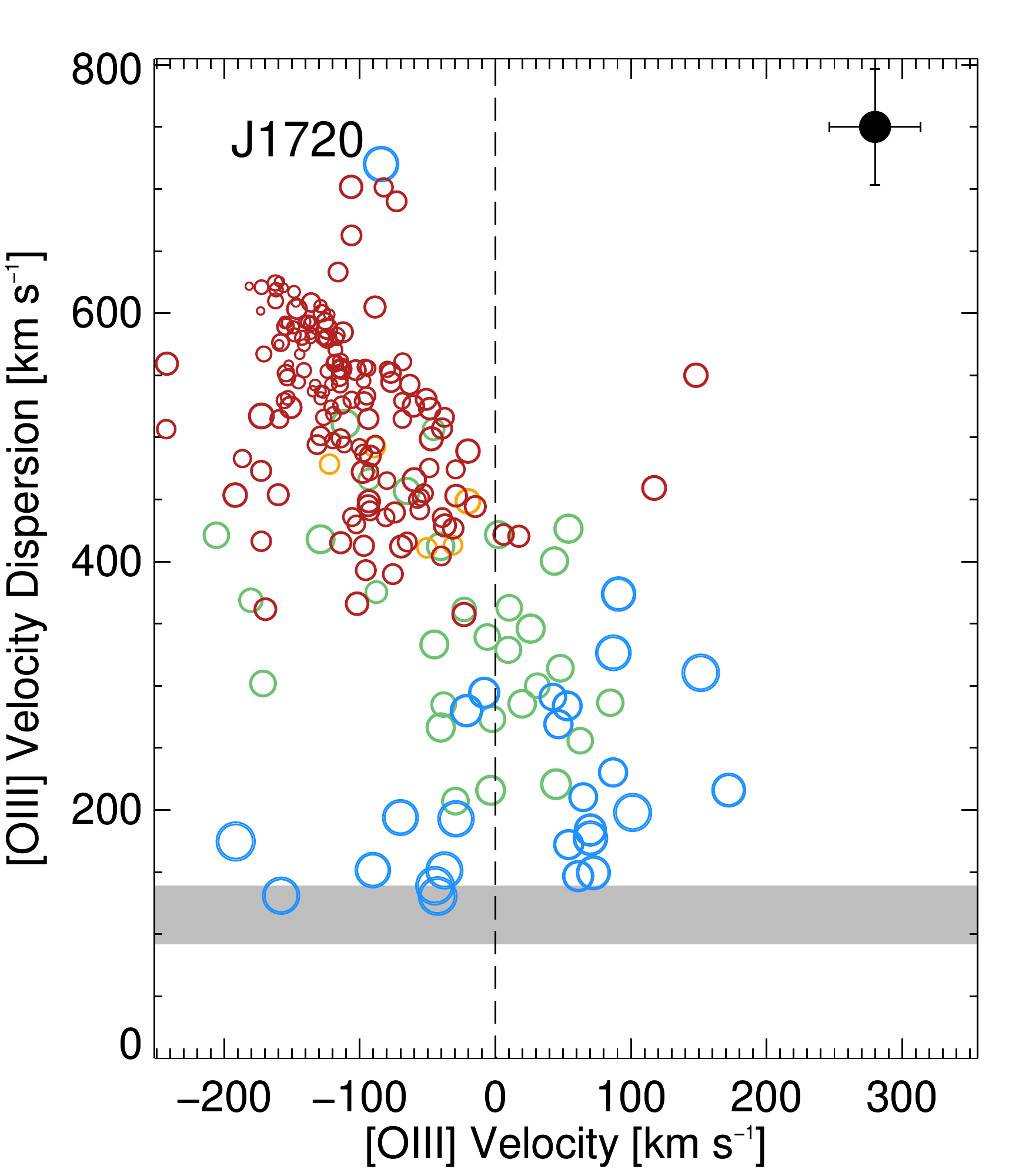}
\caption{Velocity dispersion versus velocity for the broad component of the \OIII\ emission line, for differently classified spaxels based on their narrow emission components (colors as in Fig. \ref{fig:all_bpt_map}). The size of the symbols denotes the radial distance from the continuum peak (i.e., larger symbols indicate spaxels at larger radii). The gray-shaded area shows the stellar velocity dispersion and 3$\sigma$ error from SDSS. The vertical dashed line shows a zero velocity for reference. Spaxels with emission lines below an S/N limit of 1 for H$\beta$ and 3 for \OIII, H$\alpha$, and \NII\ are not plotted.}
\label{fig:all_vvd_oiii}
\end{center}
\end{figure*}

We show the Velocity-Velocity Dispersion (VVD) diagram for the broad component of the \OIII\ emission line in Fig. \ref{fig:all_vvd_oiii}. The radial distance of each spaxel to the galaxy core is noted through the size of each symbol. We use the BPT classification of the narrow components (which reflect the systemic kinematics and photoionization field) to separate spaxels in Seyfert, LINER, composite, and star formation classes (respectively red, orange, green, and blue in Fig. \ref{fig:all_vvd_oiii}). Note that the following conclusions do not change if we instead color-code spaxels using the BPT classification based on the total line profiles. 

As we discussed in Paper I, we observe a significant range of velocity dispersions, which reach down to (and occasionally below) the stellar velocity dispersion, $\sigma_{*}$, measured over the 3\arcsec\ SDSS fiber (gray shaded areas in Fig. \ref{fig:all_vvd_oiii}). Spaxels with emission classified as Seyfert-like lie almost exclusively (for 5 out of 6 sources) in the upper left corner of the VVD diagram, showing high velocity dispersions (2-5 times $\sigma_{*}$) and large negative velocities (up to $-800$ km s$^{-1}$). Yet, we observe that within the spaxels classified as Seyfert-like, those at larger radial distances (larger symbols) show smaller velocity dispersions and velocities, reflecting the decreasing radial profiles of the \OIII\ velocity and velocity dispersion presented in Paper I (particularly for J0918, J1135, and J1720). Spaxels classified as LINER-like show diverse properties that however show continuity with respect to the spaxels classified as Seyfert-like. A dominant fraction of the spaxels classified as LINER-like is in the same region of the VVD diagram as Seyfert-like spaxels (upper left corner) with the rest lying at lower velocity dispersions and velocities consistent with the stellar velocities (e.g., J1135, J1622). This may indicate that although these spaxels are classified as LINER-like in terms of photoionization, their kinematic properties appear to separate them into two sub-groups. While for some sources LINER-like spaxels show a kinematic stratification with respect to their radial distance (e.g., J1135), this is not always the case (J0918, J1622).

Spaxels classified as composite and star-forming region bridge the Seyfert-like part of the VVD diagram with the part of the diagram associated with stellar-like kinematics. J1606 and J1720 clearly exemplify this, with composite-like spaxels closest to the nucleus (and presumably with the highest AGN contribution, e.g., \citealt{Davies2014}) at velocity dispersions $\sim2\sigma_{*}$ and negative velocities, while the ones at the largest distances lie close to $\sigma_{*}$ and a zero velocity. Spaxels dominated by stellar photoionization are mostly around zero velocity but can reach up to a couple hundreds km s$^{-1}$ velocities and dispersions up to $\sim2\sigma_{*}$. We may therefore be observing ionized gas affected by the stellar rotation and the gravitational potential of the galaxy. Exception to the above is J1622, where most spaxels show redshifted emission ($0<v_{\mathrm{\OIII}}<300$ km s$^{-1}$) but a wide range of velocity dispersions. 

\begin{deluxetable}{c c c c c c c c c}
\tabletypesize{\footnotesize}
\tablecolumns{9}
\tablewidth{0pt}
\tablecaption{Broad \OIII\ VVD flux-weighted mean values for different BPT classes. \label{tab:vvd_median}}
\tablehead{\multirow{3}{*}{ID}	& \multicolumn{2}{c}{Seyfert}	&	\multicolumn{2}{c}{LINER}	&	\multicolumn{2}{c}{Comp}	& \multicolumn{2}{c}{SF}	\\
 & v & $\sigma$ & v & $\sigma$ & v & $\sigma$ & v & $\sigma$ \\
 & \multicolumn{8}{c}{[km s$^{-1}$]} \\
\colhead{(1)} & \colhead{(2)} & \colhead{(3)} & \colhead{(4)} & \colhead{(5)} & \colhead{(6)} & \colhead{(7)} & \colhead{(8)} & \colhead{(9)}}
\startdata
J0918     & $-$736     &  529     & $-$523     &  451     & $-$281     &  338     &    \nodata     &   \nodata \\ 
J1135     & $-$341     &  446     & $-$208     &  332     & $-$122     &  309     &   23     &  430 \\ 
J1404     & $-$321     &  408     & $-$313     &  401     & $-$317     &  415     & $-$172     &  388 \\ 
J1606     & $-$389     &  299     & $-$276     &  296     & $-$252     &  314     & $-$353     &  298 \\ 
J1622     &   $-$4     &  696     &   14     &  671     &   42     &  599     &  129     &  405 \\ 
J1720     & $-$137     &  570     &  $-$60     &  437     &  $-$28     &  368     &   31     &  281 \\ 
\enddata
\tablecomments{Flux-weighted mean values of velocity and velocity dispersion of broad \OIII\ for the differently classified spaxels, as shown in Fig. \ref{fig:all_vvd_oiii}.}
\end{deluxetable}

\begin{deluxetable}{c c c c c c c c c}
\tabletypesize{\footnotesize}
\tablecolumns{9}
\tablewidth{0pt}
\tablecaption{Broad H$\alpha$ VVD flux-weighted mean values for different BPT classes. \label{tab:vvd_median_Ha}}
\tablehead{\multirow{2}{*}{ID}	& \multicolumn{2}{c}{Seyfert}	&	\multicolumn{2}{c}{LINER}	&	\multicolumn{2}{c}{Comp}	& \multicolumn{2}{c}{SF}	\\
  & v & $\sigma$ & v & $\sigma$ & v & $\sigma$ & v & $\sigma$ \\
 & \multicolumn{8}{c}{[km s$^{-1}$]} \\
\colhead{(1)} & \colhead{(2)} & \colhead{(3)} & \colhead{(4)} & \colhead{(5)} & \colhead{(6)} & \colhead{(7)} & \colhead{(8)} & \colhead{(9)}}
\startdata
J0918     & $-$101     &  239     &  $-$79     &  227     &  $-$80     &  197     &  $-$30     &  147 \\ 
J1135     &  $-$86     &  201     &  $-$40     &  196     &   $-$8     &  164     & $-$276     &  335 \\ 
J1404     & $-$316     &  434     & $-$187     &  550     &    6     &  574     &   $-$8     &  293 \\ 
J1606     & $-$170     &  133     &  $-$87     &  105     & $-$180     &  110     & $-$199     &  128 \\ 
J1622     & $-$115     &  409     &   92     &  426     &   35     &  282     &   45     &  224 \\ 
J1720     &    4     &  288     &    3     &  223     &   10     &  221     &  $-$22     &  193 \\ 
\enddata
\tablecomments{Flux-weighted mean values of velocity and velocity dispersion of broad H$\alpha$ for the differently classified spaxels based on their narrow emission components.}
\end{deluxetable}

The VVD diagrams of the broad H$\alpha$ component show qualitatively similar patterns, as was established in Paper I, but reaching less extreme values of velocity and velocity dispersion and we therefore choose not to present them. VVD diagrams based on the narrow \OIII\ and H$\alpha$ components are mostly dominated by rotation, with Seyfert-like spaxels showing the highest velocity dispersions (central spaxels) and velocities close to the systemic velocity. Conversely, star-forming spaxels show the lowest velocity dispersions (outer spaxels) but a wide range of both positive and negative velocities, reflecting rotation. Composite and LINER-like spaxels lie in-between. In Tables \ref{tab:vvd_median} and \ref{tab:vvd_median_Ha} we provide the flux-weighted mean velocities and velocity dispersions of the \OIII\ and H$\alpha$ broad components for the sub-samples of spaxels in the different BPT classes.

\section{Physical properties of the AGN outflows}
\label{sec:physical}

\subsection{Defining the outflow}
\label{sec:cases}
{In Paper I, we established the presence of outflows in all our targets, based on the decomposed kinematics maps of} \OIII\ {(see Figs. 4-6 in Paper I). Table \ref{tab:sample} gives the flux-weighted mean velocities and velocity dispersions for} \OIII\ {and H$\alpha$ broad components, with the former showing velocity dispersions that reach up to 800 km s$^{-1}$. In order to understand the impact of these outflows on their host galaxies, we first revisit the definition of its size.} 

Our dataset allows us to identify the region affected by the AGN outflow in at least two ways. We can use the BPT classification to isolate spaxels that are being photoionized by the AGN and use them to derive the mass and the kinetic energy of the outflow. Caveats of this method include the fact that in the outer part of the IFU we do not detect all 4 emission lines necessary for the BPT classification. Additionally, beyond the photoionization, shocks can also play an important role in sustaining the AGN-driven outflow, especially beyond the central region of the galaxy. Alternatively, we can use the decomposed broad \OIII\ and H$\alpha$ components to constrain the mass and energy of the outflow. This method however also has caveats, most prominent one being that the emission line profile decomposition (especially for H$\alpha$) can suffer from degeneracies and uncertainties. In addition, as we clearly showed in Paper I, even the broad component of \OIII\ converges to stellar-like kinematics at distances $>2.5-3$ times the effective radius, $r_{\mathrm{eff}}$, i.e., the radius within which half of the total \OIII\ or H$\alpha$ flux is contained.
In the following, we are going to employ three different methods to calculate the basic physical properties of the observed outflows: 
\begin{itemize}
\item total mass and kinetic energy over the IFU FoV ($5\farcs0\times3\farcs5$, Case I) {and using the total emission line profile}, 
\item total quantities over all spaxels with AGN-like emission (Case II) {and using the decomposed emission line profile}, and 
\item total quantities within the kinematically defined outflow size (Case III), {again using the decomposed emission line profile}. 
\end{itemize}
Case I is the simplest assumption and is equivalent to deriving quantities based on spatially integrated spectra. In this case, upper limits for the mass and energy of the outflow are derived since contamination from motions due to the gravitational potential and emission from gas photoionized by massive stars is not accounted for. For Case II, we consider LINER emission as AGN-like, while for spaxels classified as composite we make the simplifying assumption that half (0.5) of the emission is due to AGN photoionization. Finally, for Case III we adopt the kinematic size of the outflows (the distance at which the outflow kinematics become comparable to the stellar kinematics, as defined in Paper I), within which we sum the relevant quantities. 

{For Cases II and III, we use the decomposed emission line profile to calculate the kinetic energy of the narrow and broad components separately. For} \OIII\ {these two are then added to get the total kinetic energy, while for H$\alpha$ only the kinetic energy based on the broad component is considered. This allows us to both account for the much larger velocity of the broad component in} \OIII {, while also disregarding the contribution of the narrow component (which is dominated by the gravitational potential of the galaxy, Paper I) for the H$\alpha$-based quantities.}

\subsection{Mass and kinetic energy}
\label{sec:energy}
\subsubsection{Definitions}
The mass of the ionized gas can be derived based on the luminosity of either \OIII\ or H$\alpha$ and the electron density, N$_e$. We derive the latter based on the \SII\ $\lambda$6716, 6731\AA\ doublet flux ratio (e.g., \citealt{Osterbrock1989}). Radial profiles of the electron density are shown in Fig. \ref{fig:all_ne} in the Appendix and median values within one r$_{\mathrm{eff}}^{\mathrm{\OIII}}$ are between 200 and 800 cm$^{-3}$. These values are roughly consistent (although at the upper limit) with the usually assumed (e.g., \citealt{Liu2013}, \citealt{Harrison2014}, \citealt{Carniani2015}) or measured values in the literature (e.g., \citealt{Holt2006}, \citealt{Westoby2012}, \citealt{Rodriguez2013}), 100-500 cm$^{-3}$.

Assuming case B recombination (\citealt{Osterbrock1989}), we calculate the ionized gas mass based on the measured H$\alpha$ luminosity at each spaxel:
\begin{equation}
\label{eq:massha}
\frac{M_{H\alpha}}{M_{\sun}}=9.73\times10^{8}\left(\frac{L_{H\alpha}}{10^{43} \mathrm{erg\, s}^{-1}}\right)\left(\frac{N_{e}}{100\, cm^{-3}}\right)^{-1}	
\end{equation}
Alternatively, we can calculate the ionized gas mass based on the measured \OIII\ luminosity:
\begin{equation}
\label{eq:massoiii}
\frac{M_{\OIII}}{M_{\sun}}=0.4\times10^{8}\left(\frac{L_{\mathrm{\OIII}}}{10^{43} \mathrm{erg\,s}^{-1}}\right)\left(\frac{N_{e}}{100\, cm^{-3}}\right)^{-1},
\end{equation}
where we have followed the calculations of \citet{Carniani2015}\footnote{\citet{Carniani2015} use the PyNeb tool (\citealt{Luridiana2015}) to calculate the \OIII\ emissivity, assuming a narrow-line region of temperature T$=10^4$ K and electron density N$_{e}=500$ cm$^{-3}$.} and assumed an \OIII\ emissivity j$_{\mathrm{\OIII}}=3.4\cdot10^{-21}$ erg s$^{-1}$ cm$^{-3}$, solar metallicity, a condensation factor C=$<n_{e}>^{2}/<n_{e}^{2}>=1$ (\citealt{Cano2012}), and neglecting the mass contributed by elements heavier than helium (the mass fraction of heavier elements is negligibly small, e.g., \citealt{Asplund2005}). The inferred ionized gas masses depend on the assumed metallicity of the gas. Low metallicity AGN (below solar) are extremely rare in the local universe (e.g., \citealt{Groves2006}), with most AGN having ionized gas metallicities between 1 and 4 times solar (e.g., \citealt{Storchi1998,Nagao2002,Shin2013}). For the most metal-rich AGNs, masses may be lower by up to 25\%.

The kinetic energy can then be estimated simply as:
\begin{equation}
E_{\mathrm{kin}}=\frac{1}{2}M_{\mathrm{out}}v_{\mathrm{out}}^{2}.
\end{equation}
{The velocity term in the above equation is the bulk velocity of the outflow and is not trivial to determine due to projection effects, the unknown geometry of the outflow, dust extinction, and the role of turbulent motions within the outflow. These effects impact the estimation of the physical properties of outflows both based on integrated spectra (e.g., SDSS) but also for IFU data due to the finite spatial resolution of our instruments and the projected nature of our observations. We can circumvent these uncertainties by rewriting the above equation as:}
\begin{equation}
E_{\mathrm{kin}}=\frac{1}{2}M_{\mathrm{out}}(v_{\mathrm{rad}}^{2}+\sigma^{2}),
\label{eq:kina}
\end{equation}
where $v_{\mathrm{rad}}$ and $\sigma$ are the measured velocity and velocity dispersion for \OIII\ or H$\alpha$. {The first term, $v_{\mbox{rad}}$, is representative of the bulk velocity of the outflow but suffers from projection effects. The second term, $\sigma$, reflects the spread of velocities along the line of sight and can be used to account for the projection effects affecting $v_{\mathrm{rad}}$ and an additional turbulence term that may be significant. $\sigma$ itself suffers however from extinction effects and directly relates to the opening angle of the outflow (\citealt{Bae2016}). A combination of the two has been used in the literature to approximate the true bulk velocity (e.g., \citealt{Storchi2010,Mueller2011}), occasionally with $\sigma$ given larger weight (e.g., \citealt{Harrison2014}) or used solely as a proxy of the outflow bulk velocity (e.g., \citealt{Liu2013}). We choose the simpler approach of Eq. \ref{eq:kina} but note that we also explored alternative prescriptions, which however do not affect our results significantly (up to a factor of a few larger kinetic energies).}

Alternative prescriptions for the bulk velocity and kinetic energy of the outflow have been used in the literature. \citet{Liu2013} find that their W$_{80}$ measurement, the width that contains 80\% of the total flux of an emission line, divided by 1.3 is a good estimation of the bulk velocity of a spherical outflow with extinction due to an obscuring dusty disk. \citet{Rodriguez2013} (following \citealt{Holt2006}) instead use a similar prescription to Eq. \ref{eq:kina} but with a multiplicative factor of 3 for the velocity dispersion component, again assuming a spherical outflow. Note that these two prescriptions are equivalent if the measured velocity, $v_{\mathrm{rad}}$, equals the velocity dispersion of the ionized gas. For the majority of our sources and within the kinematic size of the outflow, the velocity dispersion dominates over the velocity in terms of absolute scale. This implies that alternative prescriptions for the kinetic energy discussed here would result to higher values of at most a factor of $\sim3$.

\begin{deluxetable}{c c c c c c}
\tabletypesize{\footnotesize}
\tablecolumns{6}
\tablewidth{0pt}
\tablecaption{Gas mass and kinetic energy. \label{tab:physics}}
\tablehead{\colhead{ID}	& \colhead{Case}	&	\multicolumn{2}{c}{[OIII]}	&	\multicolumn{2}{c}{H$\alpha$}	\\
 \colhead{ } & \colhead{ } & \colhead{M$_{\mathrm{out}}$} & \colhead{E$_{\mathrm{kin}}$} & \colhead{M$_{\mathrm{out}}$} & \colhead{E$_{\mathrm{kin}}$} \\
\colhead{ } & \colhead{ }	& \colhead{[M$_{\sun}$]}	&	\multicolumn{1}{c}{[erg]}	&	\colhead{[M$_{\sun}$]}	&	\multicolumn{1}{c}{[erg]}  \\
\colhead{(1)} & \colhead{(2)} & \colhead{(3)} & \colhead{(4)} & \colhead{(5)} & \colhead{(6)}}
\startdata
\multirow{3}{*}{J0918} & I &     4.62     &     53.13     &      6.37     &     53.70    \\ 
 & II &     4.58     &     53.12     &      5.95     &     53.68     \\ 
 & III &     4.59     &     53.11     &      5.98     &     53.71     \\ 
\multirow{3}{*}{J1135} & I &     5.37     &     53.49     &      6.55     &     54.19    \\ 
 & II &     5.37     &     53.52     &      6.29     &     54.09      \\ 
 & III &     5.37     &     53.52     &      6.04     &     53.82      \\ 
\multirow{3}{*}{J1404} & I &     5.21     &     53.57     &      6.59     &     54.70    \\ 
 & II &     5.14     &     53.61     &      6.07     &     54.54      \\ 
 & III &     5.17     &     53.62     &      6.06     &     54.62      \\ 
\multirow{3}{*}{J1606} & I &     4.73     &     52.76     &      6.39     &     55.60    \\ 
 & II &     4.49     &     52.54     &      5.02     &     52.87      \\ 
 & III &     4.73     &     52.76     &      5.06     &     53.12      \\ 
\multirow{3}{*}{J1622} & I &     5.17     &     53.71     &      6.85     &     54.27    \\ 
 & II &     5.05     &     53.97     &      5.95     &     54.08      \\ 
 & III &     5.01     &     53.69     &      5.86     &     54.10      \\ 
\multirow{3}{*}{J1720} & I &     5.14     &     53.73     &      6.67     &     53.80    \\ 
 & II &     5.06     &     54.03     &      5.81     &     53.62      \\ 
 & III &     5.06     &     54.00     &      5.79     &     53.56      \\ 
\enddata
\tablecomments{Column (1): source ID, Column (2): the case considered for the calculations (I - Total, II - BPT, III - radius, see text for details), Columns (3-4): the mass and kinetic energy (using Eq. \ref{eq:kina}) based on \OIII, Columns (5-6): the mass and kinetic energy based on H$\alpha$. All quantities given in logarithmic scales. For Case I, kinetic energies are calculated based on the total profile of the emission lines. For Cases II and III, we instead sum the kinetic energies of the narrow and broad components separately.}
\end{deluxetable}

\begin{figure*}[hbpt]
\begin{center}
\includegraphics[width=0.33\textwidth,angle=0,trim={0 60 0 10},clip]{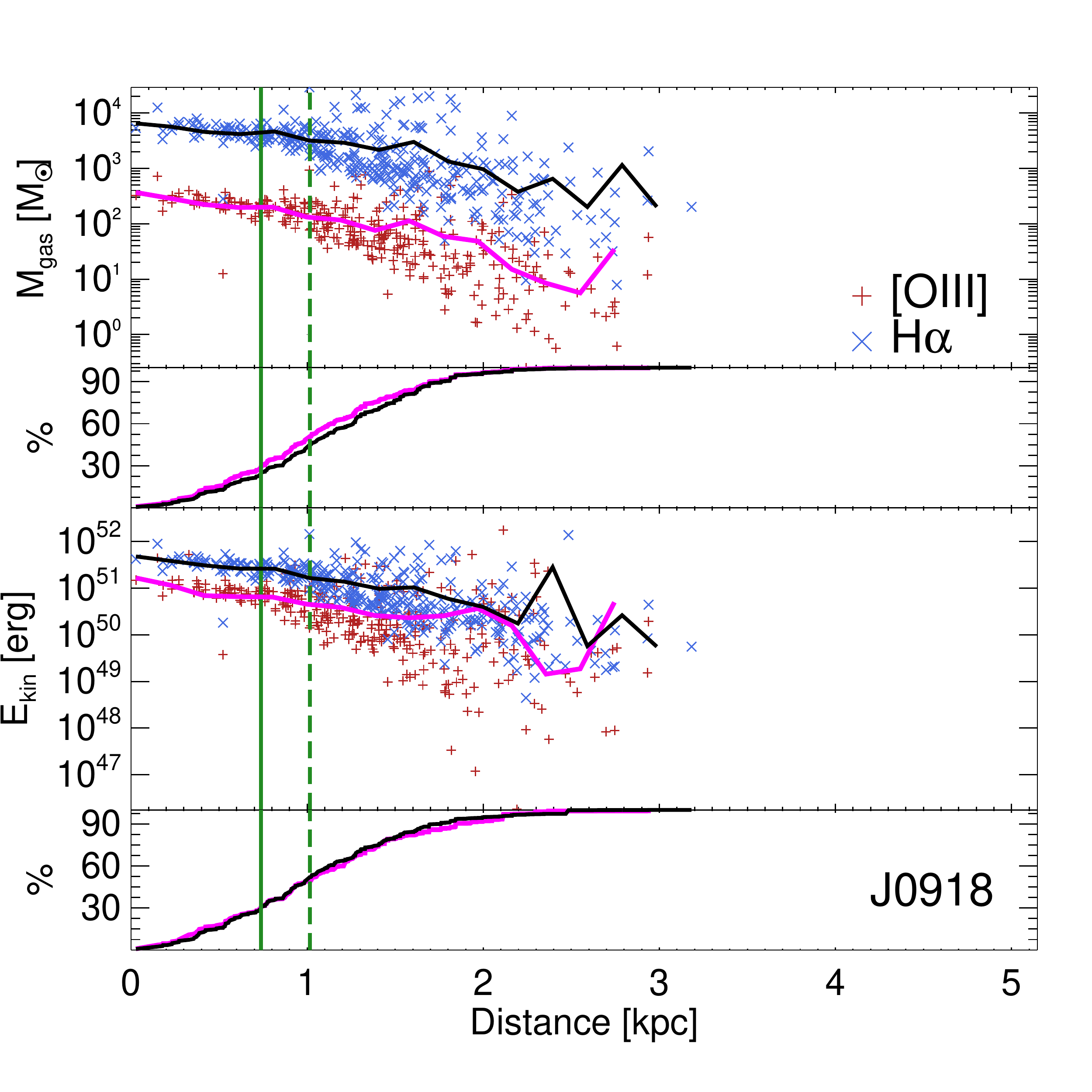}
\includegraphics[width=0.33\textwidth,angle=0,trim={0 60 0 10},clip]{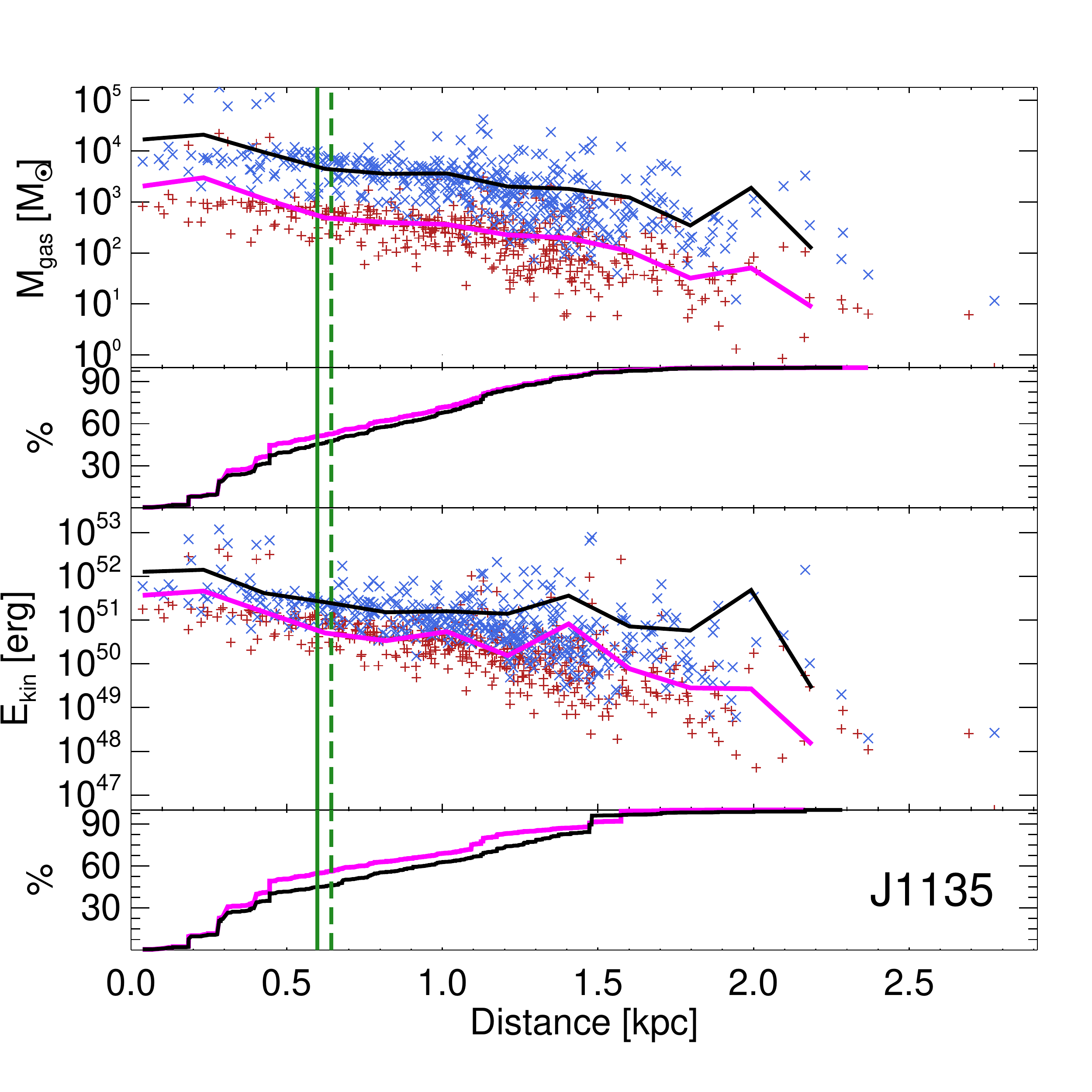}
\includegraphics[width=0.33\textwidth,angle=0,trim={0 60 0 10},clip]{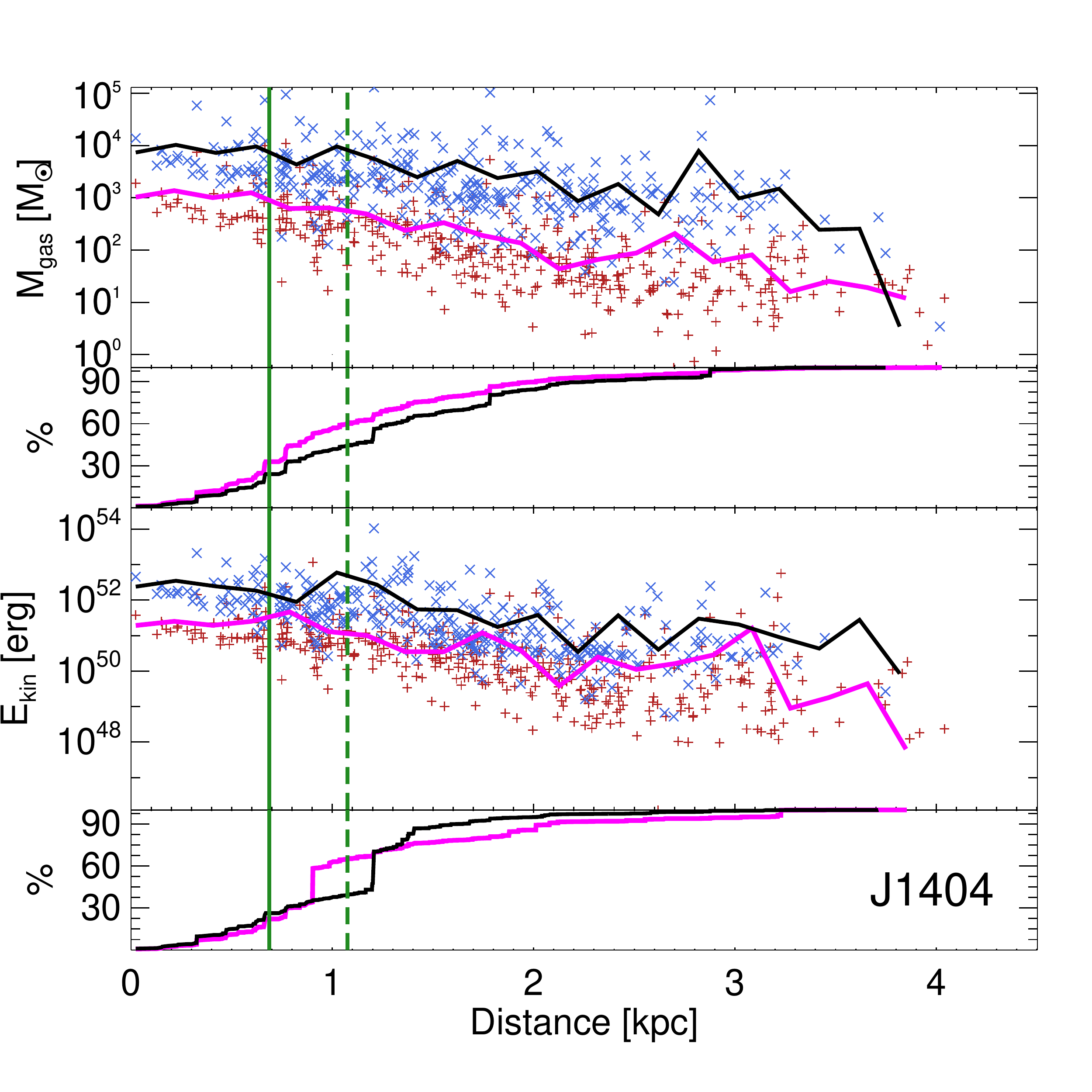}
\includegraphics[width=0.33\textwidth,angle=0,trim={0 30 0 10},clip]{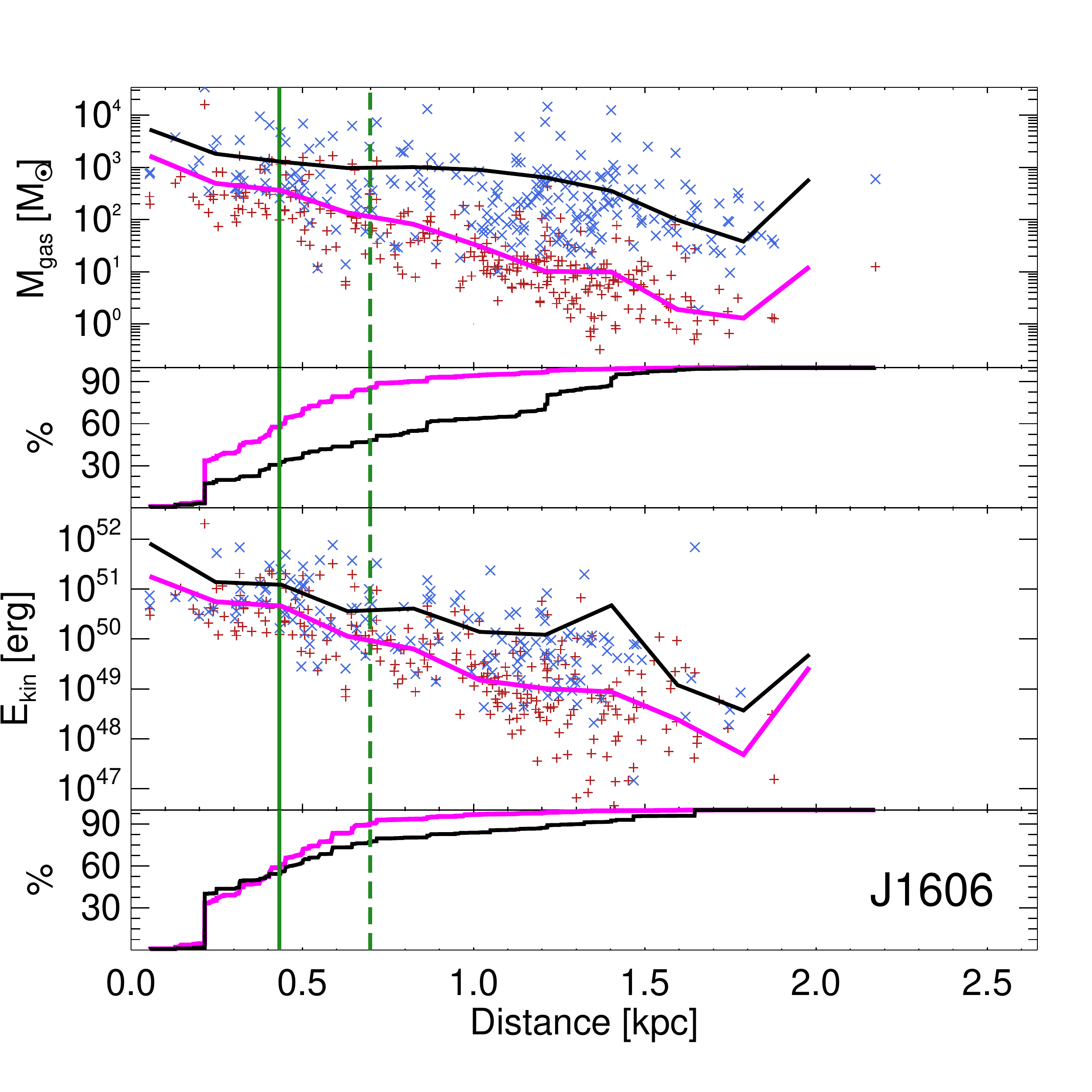}
\includegraphics[width=0.33\textwidth,angle=0,trim={0 30 0 10},clip]{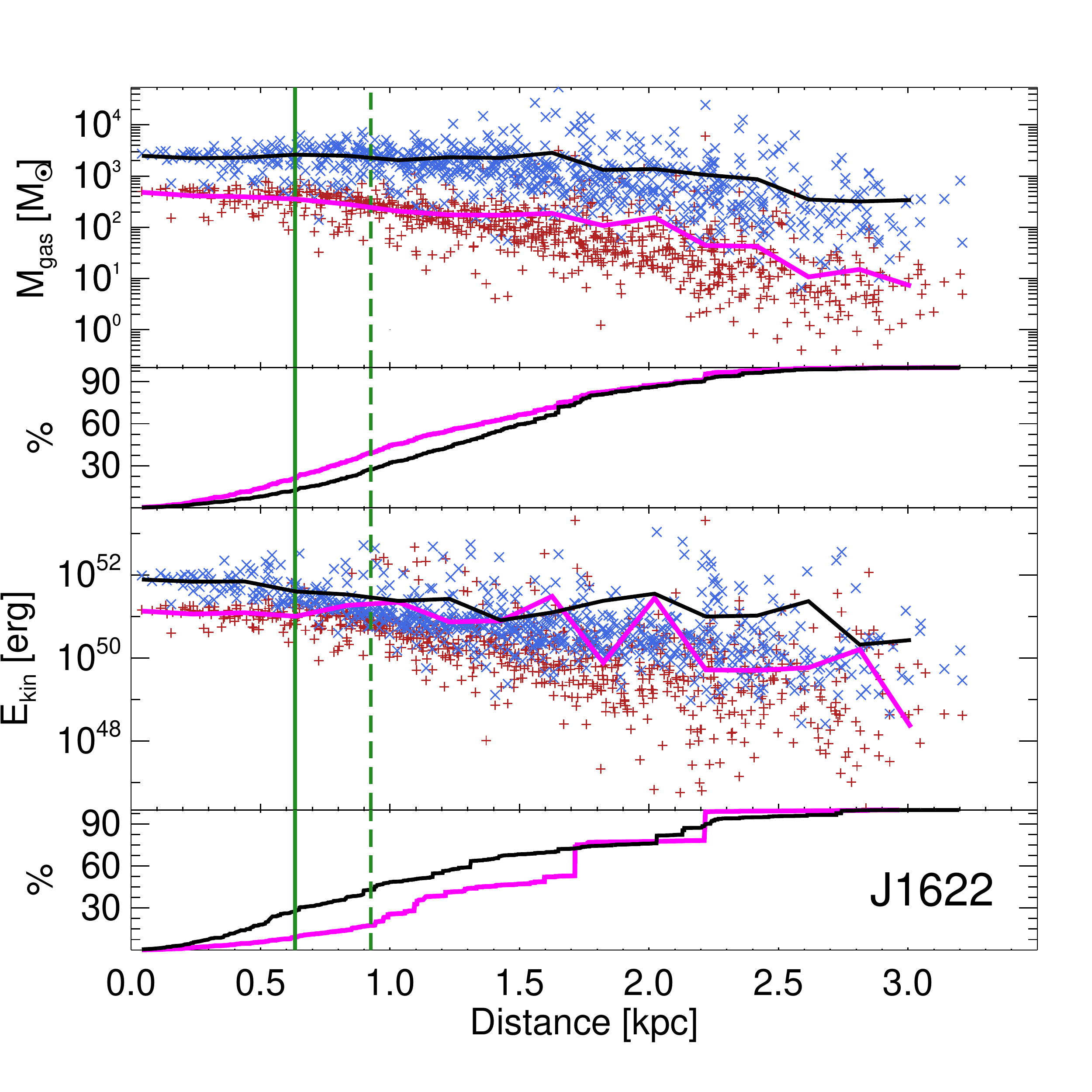}
\includegraphics[width=0.33\textwidth,angle=0,trim={0 30 0 10},clip]{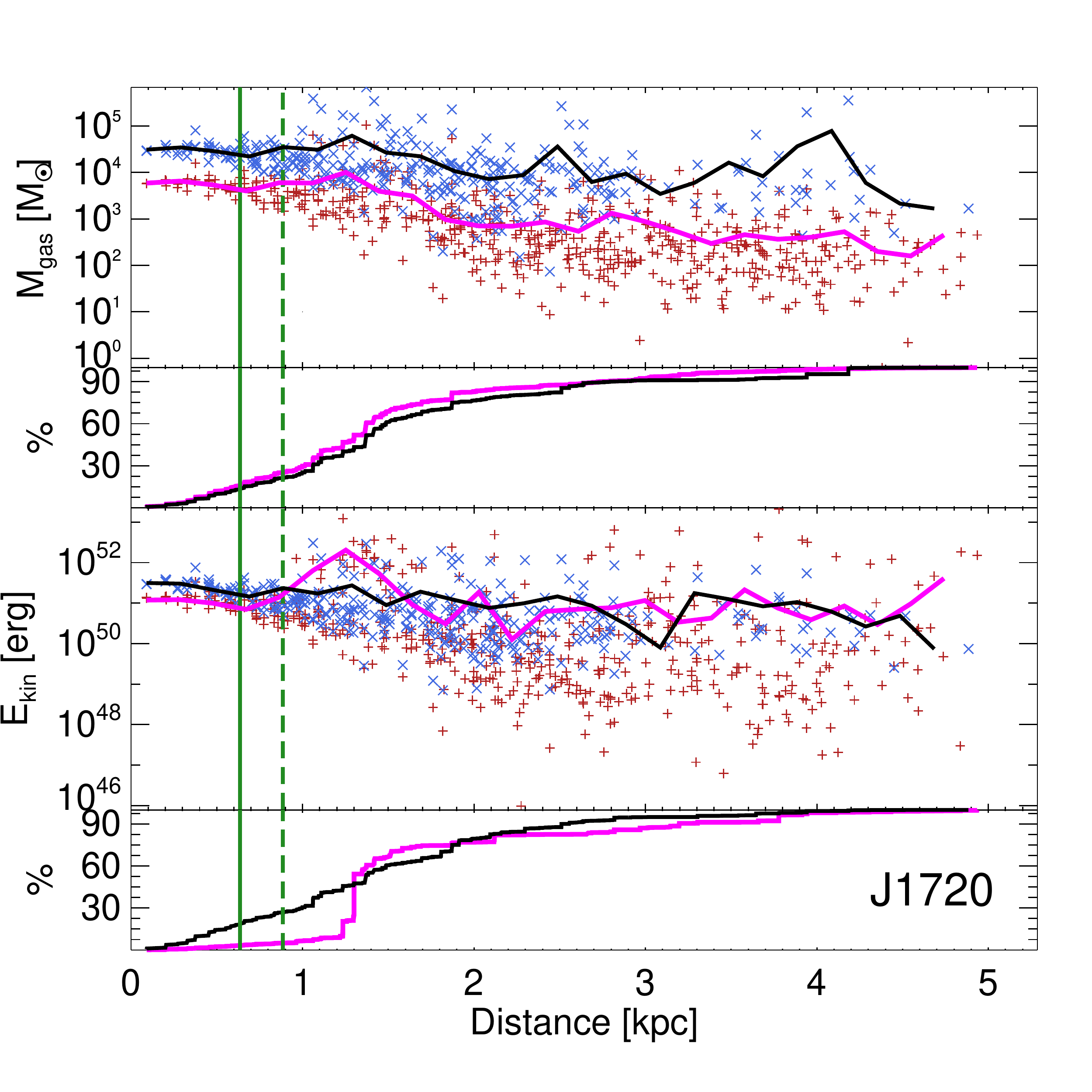}
\caption{Radial profiles of the ionized gas mass (first row) and kinetic energy (third row) based on \OIII\ (red crosses and magenta lines) and the broad H$\alpha$ emission (blue x's and black lines). Solid lines show the average values in bins of distance. We also plot the cumulative mass and kinetic energy as a function of distance in the second and fourth-row panels. Vertical solid and dashed green lines show the r$_{\mathrm{eff}}$ for \OIII\ and H$\alpha$.}
\label{fig:rad_prof_mass}
\end{center}
\end{figure*}

We perform the above calculations for each spaxel, using the respective local properties (velocity, velocity dispersion, electron density, luminosity) of that spaxel. This allows a more robust estimation of the outflow mass and energy by circumventing the need to use averaged or integrated properties. \OIII -based ionized gas masses are between $10^{4.5}$ and $10^{5.5}$ M$_{\sun}$ and kinetic energies between $E_{\mathrm{kin}}=10^{52.5}$ and $10^{54}$ ergs\footnote{Presented values are not corrected for dust absorption.}. Similarly, we calculate H$\alpha$-based ionized gas masses between $10^5$ and $10^{6.9}$ M$_{\sun}$ and kinetic energies between $10^{52.8}$ and $10^{54.6}$ ergs\footnote{Note that Case I for J1606 includes the flux of the very broad -- Type 1 -- component of H$\alpha$ and therefore significantly overestimates the H$\alpha$-based kinetic energy.}. In Table \ref{tab:physics} we provide the derived masses and kinetic energies of the outflows for the three cases.

Using the IFU FoV (Case I) gives the largest values, which can deviate from $\sim$0.1 dex up to $\sim0.4$ dex (factor of $\sim$2.5) from Cases II and III\footnote{{Note that for J1404 and J1720 Case I gives a smaller kinetic energy than Cases II and III. This is due to the summing of the decomposed emission line profiles and indicates that the velocity dispersion of the} \OIII\ {broad component is much larger than the velocity dispersion measured from the total profile.}}. On average, H$\alpha$-based masses show larger differences among the three methods due to its extended spatial distribution. In contrast, \OIII -based ionized gas mass estimates do not vary significantly among the three cases. The largest differences are seen between the \OIII - and H$\alpha$-based estimates, with around $\sim1$ dex higher kinetic energies for H$\alpha$ than \OIII\, and $\sim1.5$ dex higher ionized gas masses. This discrepancy between \OIII - and H$\alpha$-based mass estimates has been noted by, e.g., \citealt{Carniani2015} and was attributed to the different volume from which the \OIII\ and H$\alpha$ emission arises (for a stratified narrow line region, \OIII\ is expected to be more centrally concentrated and fill a smaller volume than H$\alpha$). 

\subsubsection{Mass and Energy Radial profiles}
Next, we show the radial profiles of the mass and kinetic energy of the ionized gas based on \OIII\ and H$\alpha$ (Fig. \ref{fig:rad_prof_mass}). In Paper I we concluded that both broad and narrow \OIII\ components are affected by the outflow while the narrow H$\alpha$ component appears to follow the gravitational potential. Based on this finding, we use the sum of the narrow and broad emission components of \OIII\ and the broad H$\alpha$ emission component for the mass and kinetic energy calculation. There are clear negative gradients with radial distance for both quantities, although the slope of the gradient differs among sources. This is not surprising based on the radial decrease of the velocity, velocity dispersion, and surface brightness. The cumulative mass and kinetic energy as a function of radial distance (Fig. \ref{fig:rad_prof_mass}, bottom panels) reveal that the outflow entrains from $\sim30\%$ up to $\sim90\%$ of its total mass and kinetic energy within one effective radius. For 3 sources (J1622, J0918, and J1135), both mass and kinetic energy cumulative radial profiles follow each other closely. For J1720, the second lowest \OIII\ luminosity source in our sample, the \OIII -based cumulative energy radial profile shows a jump at radii $\sim0.6$ kpc, which can be attributed to an apparent increase of the velocity dispersion (see Fig. 8 in Paper I). Contrary, the cumulative radial distribution of the \OIII -based kinetic energy of J1135 is found above the one derived based on H$\alpha$, out to $\sim1.5$ kpc. Note that for the majority of the sources, both H$\alpha$ and \OIII -based kinetic energy radial profiles show similar values, despite the differences in their mass radial profiles. This is because despite the larger gas mass implied by the H$\alpha$ luminosity, the kinematics of \OIII\ are far more extreme.  

Previous studies have also calculated the kinetic energy of AGN-driven outflows assuming an energy-conserving bubble that expands in a uniform medium (e.g., \citealt{Heckman1990,Nesvadba2006,Harrison2014}). This gives orders of magnitude higher energy outflow rates than the ones calculated above, as it is proportional to the size of the outflow squared and the velocity cubed and does not take the ionized gas mass into account. This was originally proposed for starburst-driven super-winds that expand perpendicular to the stellar disk. It is however unclear whether the energy-conservation assumption, as well as the usually assumed ambient electron densities of 0.5 cm$^{-3}$ are valid for AGN-driven outflows that should be more concentrated, can very well expand at any angle with respect to the stellar disk (e.g., \citealt{Lena2015}), and may in fact be momentum-driven (e.g., \citealt{Zubovas2014b}). For these reasons, we do not consider the energy-conserving bubble scenario. For the following, we use the mass and kinetic energy estimates based on \OIII\ and H$\alpha$ as lower and upper limits for the mass and energy content of the outflows.

\subsubsection{The relative importance of the narrow and broad components}
\begin{figure}[bpt]
\begin{center}
\includegraphics[width=0.45\textwidth,angle=0,trim={10 0 15 10},clip]{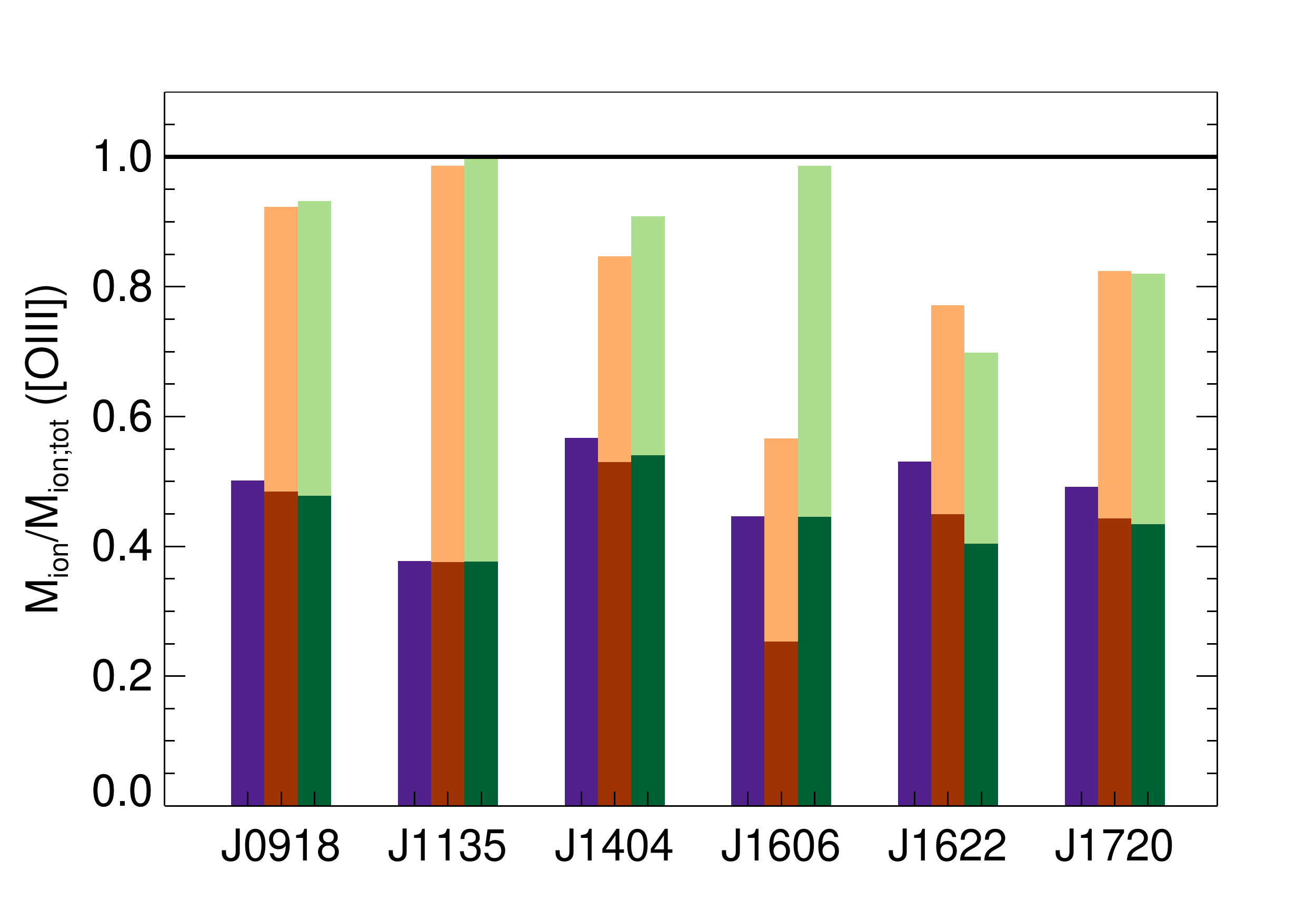}
\includegraphics[width=0.45\textwidth,angle=0,trim={10 0 15 10},clip]{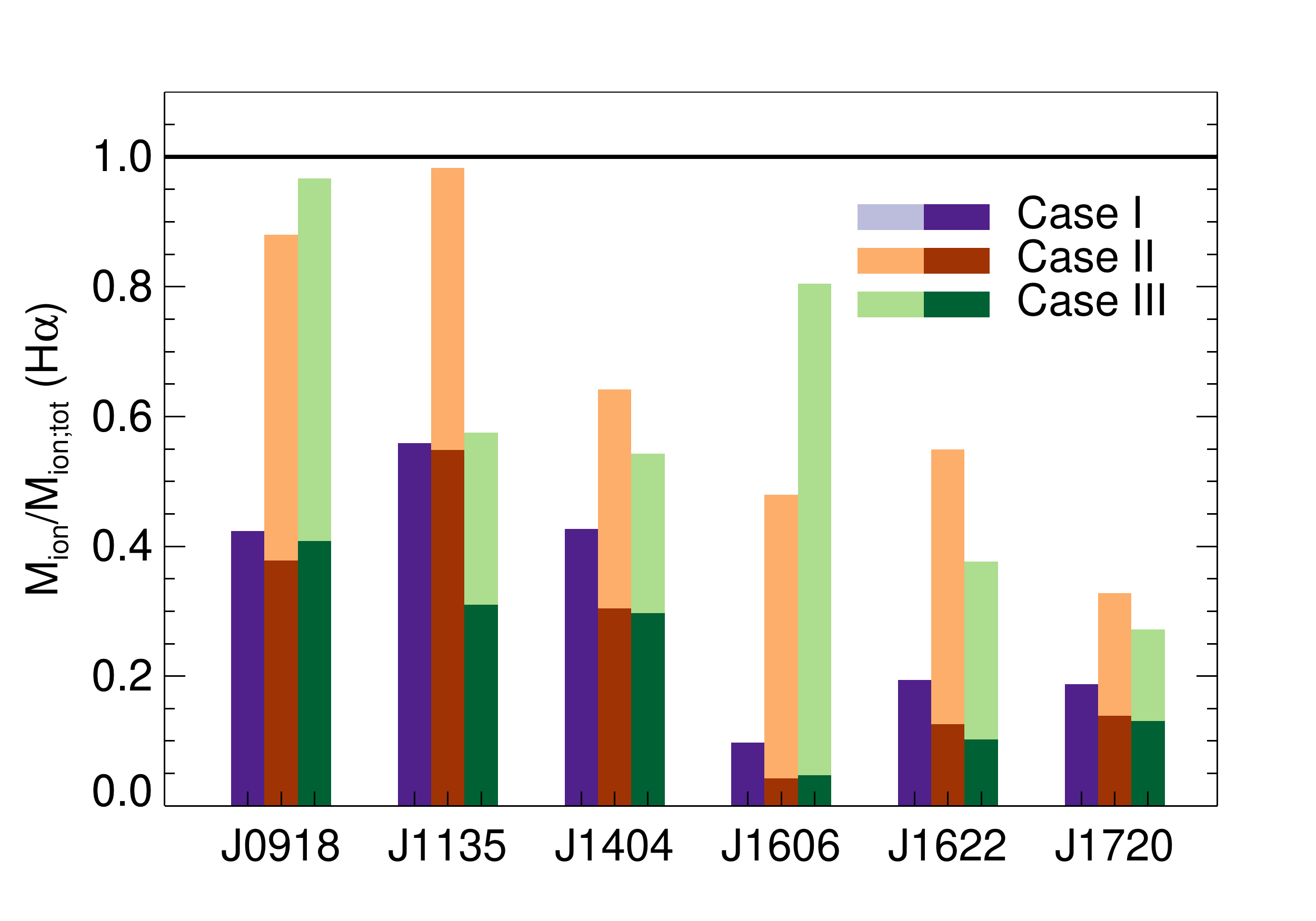}
\caption{Fractional masses with respect to the total ionized gas mass for \OIII\ (top) and H$\alpha$ (bottom) within the IFU FoV (solid line), for the three different Cases (see text), and for the total line profile (light colors) and the broad component only (dark colors).}
\label{fig:frac_mass}
\end{center}
\end{figure}

Next, we investigate the relative contribution of the narrow and broad components of \OIII\ and H$\alpha$ to the mass and kinetic energy calculated with respect to the total emission line profiles and depending on the assumed size of the outflow, i.e., Case I-III (Fig. \ref{fig:frac_mass}).  Regarding the total \OIII -based ionized gas mass (light colors in Fig. \ref{fig:frac_mass}, top), Case II (AGN-classified spaxels only, orange) includes between $\sim60\%$ and $\sim100\%$ of Case I (IFU FoV, equal to 1 for this comparison). Case III (spaxels within the kinematic size of the outflow, green) on the other hand recovers most of the \OIII\ emission ($\sim70-100\%$). Considering the ionized gas mass based on the broad \OIII\ component (dark colors), it accounts for $\sim40-60\%$ of the total \OIII -based mass, with a mean of $\sim50\%$. This fraction does not depend strongly on the Case considered. This reflects the very concentrated spatial distribution of the broad component of \OIII\, which is dominated by the unresolved AGN core.


Compared to \OIII , there are more pronounced differences for H$\alpha$-based masses (Fig. \ref{fig:frac_mass}, bottom) depending on the definition of the outflow size (i.e., Case I-III). Case II gives a range between $\sim30-100\%$ of Case I, with a mean of $\sim65\%$. Case III recovers similar mass, ranging between $\sim30\%$ and $100\%$ of Case I, with a mean of $\sim60\%$. The broad component of the H$\alpha$ line accounts for $\sim5-55\%$ of the total ionized mass with a mean of $\sim30\%$\footnote{For J1606 the mass based on the total profile of H$\alpha$ includes the very broad, Type 1 like, component, which leads to a much smaller fractional mass contributed by the broad component.}, smaller than the fraction represented by the broad \OIII\ component. This comparison reinforces our previous point that using the total H$\alpha$ profile to estimate the outflow properties not only significantly overestimates the outflow size (Paper I) but also the mass of the outflowing gas and the kinetic energy carried by the outflow.

The fractional kinetic energy with respect to that based on the total profile of \OIII\ does not show clear trends with various size estimates (Case I-III), similar to the relative mass measurements. If we use the broad \OIII\ component instead of the total \OIII\ profile, the estimated kinetic energy decreases by 20\% for 4 out of 6 objects, while it increases by a factor of 1.5-2 for the other 2 sources. This indicates that the broad component, owing to its extreme kinematics, dominates the energy budget of the system, even though it contains roughly equal mass with the narrow component. The increase is due to the lower velocities derived based on the total emission profile of \OIII\ compared to those based on just the broad component. 

In contrast, fractional kinetic energy based on H$\alpha$ shows some dependency on whether the BPT information (Case II) or kinematic size of the outflows (Case III) are used. The former gives fractional kinetic energies between $\sim60\%$ and $\sim90\%$ of Case I, while the latter ranges from $\sim40\%$ to $\sim110\%$. The broad H$\alpha$ component clearly dominates the total energy budget, ranging between $\sim90\%$ and $\sim110\%$ of the kinetic energy based on the total H$\alpha$ profile. 

Consequently, for the following we use the total emission profile (both broad and narrow emission components) for \OIII\ but only the broad component for H$\alpha$. This decision is further supported by the fact that for most of our objects, both the narrow and broad components of \OIII\ show similarly blueshifted emission within the central region of the galaxy (see Table 2 of Paper I). We note here that by using the total \OIII\ profile we take a flux-weighted mean of the velocity and velocity dispersion of the broad and narrow components. Alternatively, we could also assume the sum of the two components kinetic energies that would lead to higher \OIII -based kinetic energies of factor $\sim2$. 

\subsection{Integrated properties of the outflow}
\subsubsection{Outflow lifetime}
{The bulk velocity of the outflow ($\sqrt{v_{\mathrm{rad}}^{2}+\sigma^{2}}$, as defined in Eq. \ref{eq:kina}) is calculated based on the flux-weighted velocity and velocity dispersion within the kinematic size of the outflow. The flux-weighted velocity dispersion is corrected for the contamination from the gravitational potential of the galaxy by subtracting in quadrature the stellar velocity dispersion from SDSS. The bulk velocities are then combined with} the kinematic sizes presented in Paper I ($\sim1.3-2.1$ kpc) to derive the lifetimes of the outflows. Flux-weighted mean bulk velocities, $\bar{v}_{\mathrm{out}}$, range between 500 and 750 km s$^{-1}$ for \OIII\ and between 80 and 550 km s$^{-1}$ for H$\alpha$. The expansion timescale is then t$_{\mathrm{out}}=$R$_{\mathrm{out}}/\bar{v}_{\mathrm{out}}$, which is in the range $2-10$ Myr, with consistent \OIII\ and H$\alpha$-based timescales within a factor of a few (Table \ref{tab:rates}). Note that there is an uncertainty associated with these timescales due to the fact that we used a mean bulk velocity and that the outflow velocity is a function of radius.

\begin{deluxetable*}{c c c c c c c c c c c c c}
\tabletypesize{\footnotesize}
\tablecolumns{13}
\tablewidth{0pt}
\tablecaption{Sizes, velocities, masses, mass rates and energy rates of the outflows. \label{tab:rates}}
\tablehead{\multirow{3}{*}{ID}	& \multicolumn{6}{c}{[OIII]}	&	\multicolumn{6}{c}{H$\alpha$}	\\
 \colhead{ } & \colhead{R$_{\mathrm{kin}}$} & \colhead{$v_{\mathrm{bulk}}$} & \colhead{t$_{\mathrm{out}}$} & \colhead{M$_{\mathrm{out}}$} & \colhead{$\dot M_{\mathrm{out}}$} & \colhead{$\dot E_{\mathrm{kin}}$}  & \colhead{R$_{\mathrm{kin}}$} & \colhead{$v_{\mathrm{bulk}}$} & \colhead{t$_{\mathrm{out}}$} & \colhead{M$_{\mathrm{out}}$}  & \colhead{$\dot M_{\mathrm{out}}$} & \colhead{$\dot E_{\mathrm{kin}}$} \\
\colhead{ } & \colhead{[kpc]} & \colhead{[km s$^{-1}$]} & \colhead{[yr]} & \colhead{[M$_{\sun}$]} & \colhead{[M$_{\sun}$ yr$^{-1}$]}	&	\colhead{[erg s$^{-1}$]} & \colhead{[kpc]} & \colhead{[km s$^{-1}$]} & \colhead{[yr]}	&	\colhead{[M$_{\sun}$]}	&	\colhead{[M$_{\sun}$ yr$^{-1}$]}	&	\colhead{[erg s$^{-1}$]}  \\
\colhead{(1)} & \colhead{(2)} & \colhead{(3)} & \colhead{(4)} & \colhead{(5)} & \colhead{(6)} & \colhead{(7)} & \colhead{(8)} & \colhead{(9)} & \colhead{(10)} & \colhead{(11)} & \colhead{(12)} & \colhead{(13)}}
\startdata
J0918 &     1.84     &     739     &     6.4     &      4.6     &     0.02     &     39.2     &     2.03     &     205     &     7.0     &     5.98     &     0.10     &     39.2     \\
J1135 &     1.80     &     528     &     6.5     &      5.4     &     0.07     &     39.5     &     0.77     &      82     &     7.0     &     6.04     &     0.12     &     39.4     \\
J1404 &     2.06     &     491     &     6.6     &      5.2     &     0.04     &     39.5     &     1.61     &     555     &     6.5     &     6.06     &     0.40     &     40.7     \\
J1606 &     1.30     &     472     &     6.4     &      4.7     &     0.02     &     38.8     &     0.70     &     200     &     6.5     &     5.06     &     0.03     &     39.1     \\
J1622 &     1.58     &     696     &     6.3     &      5.0     &     0.05     &     39.8     &     1.39     &     382     &     6.6     &     5.86     &     0.21     &     40.1     \\
J1720 &     1.91     &     635     &     6.5     &      5.1     &     0.04     &     40.0     &     1.77     &     263     &     6.8     &     5.79     &     0.09     &     39.2     \\
\enddata
\tablecomments{Column (1): source ID, Columns (2-7): the kinematic size (from Paper I), flux-weighted bulk velocity, timescale, the mass, mass rate, and the energy rate of the outflow based on the \OIII\ emission, and Columns (8-13): the timescale, the mass, and the energy rate of the outflow based on the H$\alpha$ emission.}
\end{deluxetable*}

\begin{deluxetable*}{l c c c c c c c c c c c}
\tabletypesize{\footnotesize}
\tablecolumns{12}
\tablewidth{0pt}
\tablecaption{Properties of the literature studies used in Section \ref{sec:physical}. \label{tab:lit}}
\tablehead{\colhead{Reference}	&	\colhead{z}	&	\colhead{Type}	& \colhead{logL$_{\mathrm{AGN}}$}	&	\colhead{N}	&	\colhead{v$_{\mbox{[O\,\textsc{iii}]}}$}	&	\colhead{v$_{\mathrm{H\alpha}}$}	&	\colhead{$\sigma_{\mbox{[O\,\textsc{iii}]}}$}	&	\colhead{$\sigma_{H\alpha}$}	&	\colhead{R$_{\mathrm{out}}$} & \colhead{$\log\dot M_{\mathrm{out}}$} & \colhead{$\log\dot E_{\mathrm{out}}$}\\ 
\colhead{ }	&	\colhead{ } 	&	\colhead{ }	&	\colhead{[erg s$^{-1}$]}	&	\colhead{ }	&	\multicolumn{4}{c}{[km s$^{-1}$]}	&	\colhead{[kpc]} & \colhead{M$_{\sun}$ yr$^{-1}$} & \colhead{erg s$^{-1}$} \\
\colhead{(1)} & \colhead{(2)} & \colhead{(3)} & \colhead{(4)} & \colhead{(5)} & \colhead{(6)} & \colhead{(7)} & \colhead{(8)} & \colhead{(9)} & \colhead{(10)} & \colhead{(11)} & \colhead{(12)}}
\startdata
\citet{Schnorr2016}		&$<0.01$	& Type 2  & 43.6  &  1       &\nodata & 330 & \nodata & 185 & 0.2\tna & -2.36	& 43.6\\
\citet{McElroy2015}		& $<0.11$	& Type 2	& 46.1 & 17	& \multicolumn{2}{c}{220} 	& \multicolumn{2}{c}{252}	& 8.0	 & 3.0 & 43.6\\
\citet{Harrison2014}		& $<0.2$	& Type 2	& 45.3 & 16	& 268	& \nodata	& 456	& \nodata	& 2.9	 & 0.7 & 43.2\\
\citet{Schnorr2014}		& $<0.01$ & Type 2  & 43.5 & 1        &\nodata  & 525       & \nodata  & 300       & 0.3\tna & -0.05 & \nodata\\
\citet{Liu2013}			& 0.5 	& Type 2	& 46.3 & 11	& 166	& \nodata	& 426	& \nodata	& 14.1 & 4.0 &	45.2\\
\citet{Brusa2015}\tnb		& 1.5		& Type 2	& 46.2 & 6		& \multicolumn{2}{c}{1660}	& \multicolumn{2}{c}{555}	& 5.0 & \nodata & 43.1	\\
\citet{Greene2011}\tnb	& $<0.5$	& Type 2	& 45.6 & 15	& \nodata	& \nodata	& 213	& \nodata	& 4.1	 & 1.6 & 42.0 \\
\cutinhead{Near-IR\tnd}
\citet{Riffel2015}		& $<0.01$ & Type 2  & 42.3 & 1	& \multicolumn{2}{c}{220} & \multicolumn{2}{c}{200} & 0.3\tna & -0.4\tnc & \nodata\\
\citet{Schonell2014}		& $<0.01$ & Type 2  & 42.5 & 1	& \multicolumn{2}{c}{280} & \multicolumn{2}{c}{100} & 0.15\tna & 1.0 & 41.5\\
\citet{Riffel2013}		& $<0.01$ & Type 2  & 44.3 & 1	& \multicolumn{2}{c}{150} & \multicolumn{2}{c}{165} & 0.3\tna & 0.54 & 40.5\\
\citet{Riffel2011b}		& $<0.01$ & Type 2  & 42.2 & 1	& \multicolumn{2}{c}{130} & \multicolumn{2}{c}{250} & 0.2\tna & 0.78 & 41.4\\
\citet{Riffel2011a}		& $<0.01$ & Type 2  & 43.2 & 1	& \multicolumn{2}{c}{500} & \multicolumn{2}{c}{150} & 0.3\tna & -1.22 & \nodata\\
\citet{Mueller2011}		& $<0.01$ & Type 2  & 44.1 & 6	& \multicolumn{4}{c}{170\tne}   & 0.2\tna & 0.87 & 40.9 \\
\citet{Storchi2010}		& $<0.01$ & Type 2  & 44.9 & 1	& \multicolumn{2}{c}{600} & \multicolumn{2}{c}{200} & 0.4\tnf & 0.08 & 41.4\\
\enddata
\tablecomments{Column (1): reference, Column (2): (median) redshift, Column (3): AGN type, Column (4): (median) AGN bolometric luminosity, Column (5): number of objects in the sample, Column (6, 7): (median) maximum velocities of \mbox{[O\,\textsc{iii}]} and H$\alpha$, Column (8, 9): (median) maximum velocity dispersions of \mbox{[O\,\textsc{iii}]} and H$\alpha$, Column (10): (median) outflow size, Column (11): reported (median) mass outflow rates, and Column (12): reported (median) energy outflow rates. AGN bolometric luminosities, gas kinematics, and outflow properties are measured differently among various studies. We have assumed FWHM=$2\sqrt{2\ln{2}}\times\sigma$ and W$_{80}$=1.088$\times$FWHM to convert literature line width values to velocity dispersions.}
\tablenotetext{a}{Linear scales at which the outflow reaches its maximum velocity.}
\tablenotetext{b}{Based on spatially resolved long-slit observations.}
\tablenotetext{c}{Reported as a lower limit for the mass outflow rate.}
\tablenotetext{d}{A mix of different lines is used to derive the outflow kinematics. These include [S III], [He I], [Fe II], Br$\gamma$, and Pa$\beta$.}
\tablenotetext{e}{Outflow velocity based on the modelling of the IFU data with a biconical outflow.}
\tablenotetext{f}{Outflow size is based on long-slit observations with the Hubble Space Telescope STIS from \citet{Das2005}.}
\end{deluxetable*}

\subsubsection{Mass outflow rates}

{Previous studies have claimed that, for given AGN luminosity and measured outflow properties, different prescriptions for the calculation of the mass outflow rates (i.e., different outflow geometries) provide consistent results within a factor of a few (e.g., \citealt{Maiolino2012,Harrison2014}). However, as we showed previously, the decomposition of the different emission components and choice of emission lines with different ionization levels can lead to significant differences.} Here, we make the simplest assumption of a steady outflow for which $\dot M_{\mathrm{out}}=M_{out}/t_{\mathrm{out}}$.  For our sample, mass outflow rates are between 0.02 and 0.07 M$_{\sun}$ yr$^{-1}$ based on \OIII -calculated quantities and between 0.03 and 0.4 M$_{\sun}$ yr$^{-1}$ based on H$\alpha$ (Table \ref{tab:rates}). 

\begin{figure}[bpt]
\begin{center}
\includegraphics[width=0.4\textwidth,angle=0,trim={5 30 15 10},clip]{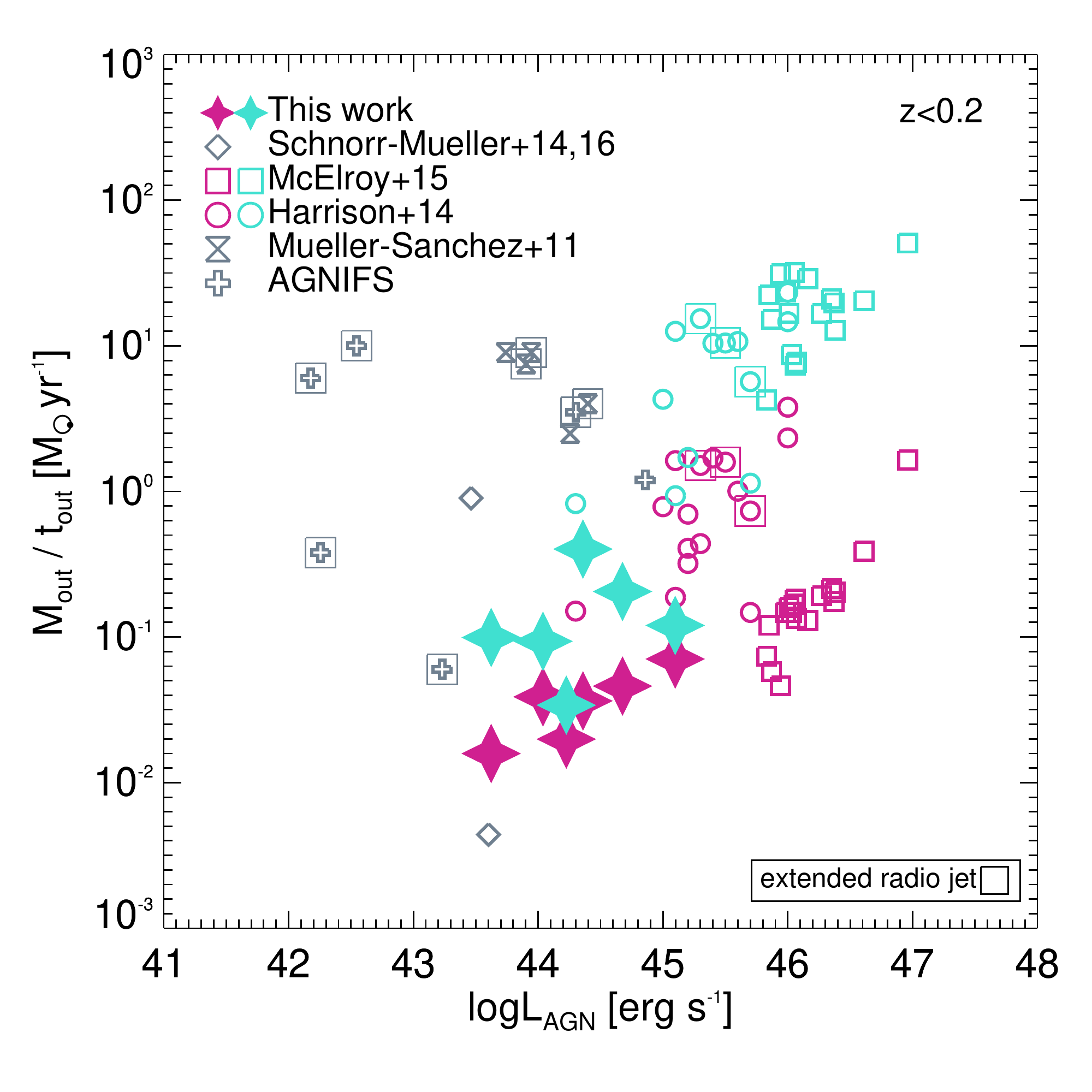}
\includegraphics[width=0.4\textwidth,angle=0,trim={5 30 15 10},clip]{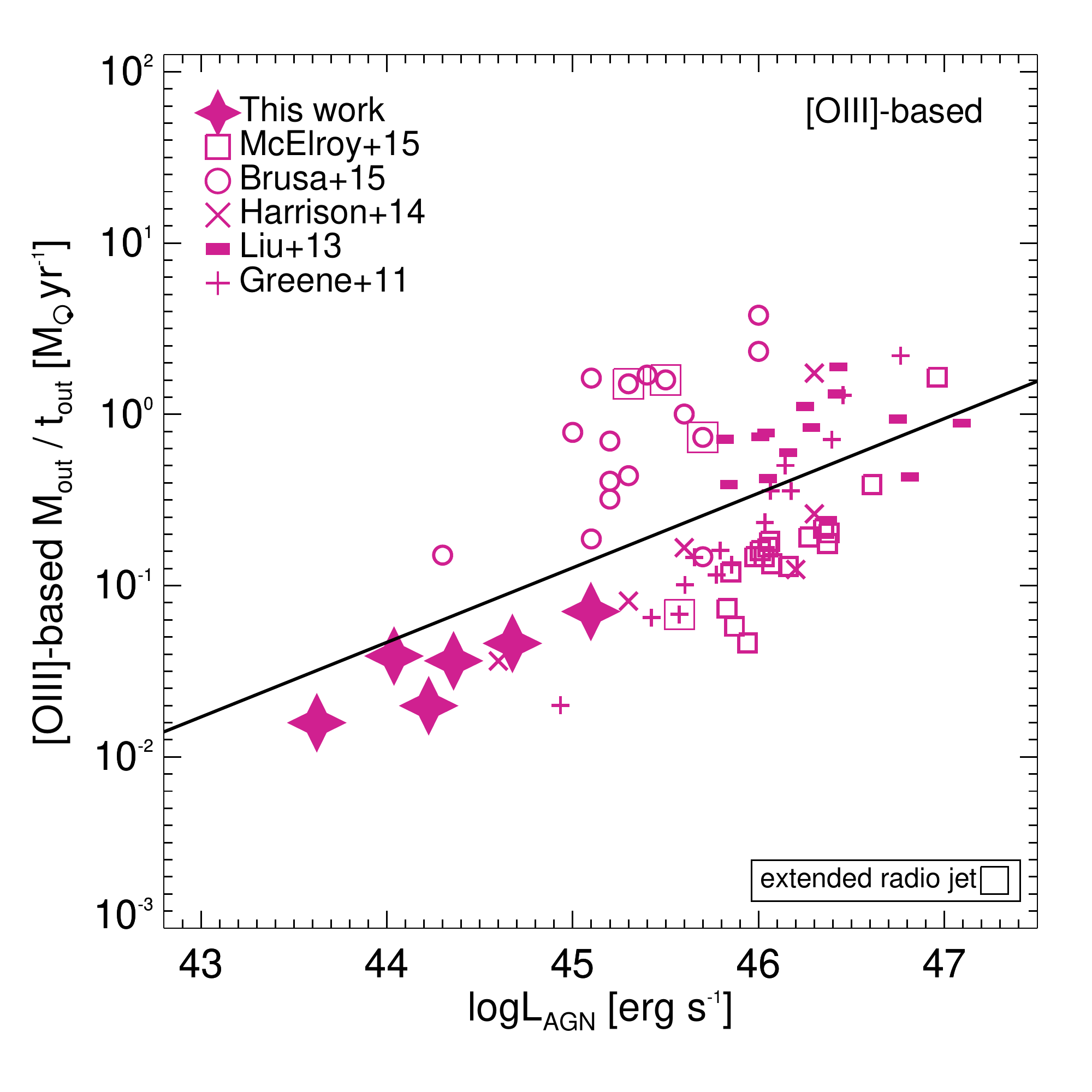}
\caption{{Top: Mass outflow rate versus AGN bolometric luminosity for our sample (stars) and the literature samples of local Type 2 AGN, based on} \OIII\ {(violet), H$\alpha$/H$\beta$ (turquoise), and other low ionization emission lines (gray). Literature samples are drawn from \citet{Schnorr2016,Schnorr2014,McElroy2015,Brusa2015,Harrison2014,Mueller2011} and the AGNIFS program (i.e., \citealt{Storchi2010,Riffel2011a,Riffel2011b,Riffel2013,Schonell2014,Riffel2015}). Bottom: Same as top panel but focusing only on} \OIII {-based mass outflow rate estimates, also adding high(er) redshift and luminosity Type 2 samples from \citet{Brusa2015,Liu2013,Greene2011}. The solid black line shows a linear regression fit to the total combined sample. Big squares indicate sources with extended radio jets aligned with the ionized gas outflows.} }
\label{fig:lit_Mdot}
\end{center}
\end{figure}

We also compile AGNs with IFU or spatially-resolved long-slit spectroscopy from previous literature studies (Table \ref{tab:lit}) to increase the dynamic range of the AGN bolometric luminosity, outflow mass, and energy. {Literature mass outflow rates range several orders of magnitude as Table \ref{tab:lit} shows, from as low as $\sim10^{-3}$ M$_{\sun}$ yr$^{-1}$ (\citealt{Schnorr2016}) to as high as $\sim10^{4}$ M$_{\sun}$ yr$^{-1}$ (\citealt{Liu2013}). There are several factors that can lead to this wide spread of mass outflow rates, even for AGN of same or similar luminosities:}
\begin{itemize}
\item {The species of ionized gas used to measure the ionized gas mass. We showed significant differences between} \OIII\ {and H$\alpha$ of a factor of 10 or more. Similarly, we expect differences with ionized gas emitting in the NIR (e.g., [Fe II]) that potentially relate to the ionization level of each species and the stratification of the AGN narrow-line region.}
\item {The assumed electron density. Previous studies have used electron densities that range from as low as 1.2 cm$^{-3}$ (e.g., \citealt{Liu2013}) to as high as 500 cm$^{-3}$ (e.g., \citealt{Harrison2014}, here). This can lead to differences of up to a factor of 400. Here we find median electron densities $\sim 500$ cm$^{-3}$ within the central kiloparsec of our sources and this is the value we adopt for our calculations.}
\item {The adopted outflow geometry. While there is a general consensus that ionized gas outflows are un-collimated (unlike radio jets), the opening angle of the outflow and the filling factor of the gas within can lead to differences in the estimated mass outflow rates of a factor of several.}
\item {The definition of the outflow velocity. In the literature, the FWHM and the W$_{80}$ multiplied by some (model-dependent) factor have been used as proxies for the outflow velocity. Both are larger by a factor of 2 to 3 than the velocity dispersion, used here.}
\item {Strong contribution from other, kinematically separate, components in the galaxy, such as its gravitational potential, a radio jet, or outflows driven by young stars.}
\end{itemize}
{The combined effect of the above can lead to differences in estimated mass outflow rates of more than three orders of magnitude. To avoid this problem, we use the total} \OIII\ {and H$\alpha$ luminosities and the ionized gas velocity and velocity dispersion\footnote{{We have assumed FWHM=$2\sqrt{2\ln{2}}\times\sigma$ and W$_{80}$=1.088$\times$FWHM to convert literature line width values in FWHM or W$_{80}$ to velocity dispersions.}} from the respective papers to recalculate mass outflow rates based on Eqs. \ref{eq:massha} and \ref{eq:massoiii} (assuming N$_{e}=500$ cm$^{-3}$). We typically use the maximum velocity and velocity dispersion reported in the respective papers. For studies using different emission lines (i.e., \citealt{Schnorr2016,Riffel2015,Schonell2014,Schnorr2014,Riffel2013,Riffel2011b,Riffel2011a,Mueller2011,Storchi2010}), we adopt the values of mass outflow rates reported in the respective papers. We underline here that by adopting the simplest prescription for the calculation of the mass outflow rate and due to the effects discussed above, the values shown in Fig. \ref{fig:lit_Mdot} can differ from those originally reported in the literature.}

{Among Type 2 AGN in the local Universe (Fig. \ref{fig:lit_Mdot}, top), our sample populates the low luminosity tail of the distribution, comparable with the galaxies of \citet{Mueller2011}\footnote{{We use the dust un-corrected AGN bolometric luminosities to compare our sources with other literature samples in order to circumvent possible uncertainties due to the dust extinction correction. We define the AGN bolometric luminosity as 3500 times the} \OIII\ {luminosity, following \citet{Heckman2004}.}}\textsuperscript{,}\footnote{{We have excluded NGC 2992 from the sample of \citet{Mueller2011} as its outflow appears to be powered by both the AGN and a starburst component. The reported outflow scale and mass outflow rate of NGC 2992 are 0.9 kpc and 120 M$_{\sun}$ yr$^{-1}$, respectively, well above what is expected for its AGN bolometric luminosity and more in line with starburst-driven outflows.}}. We find a rough trend for more luminous AGN having higher mass outflow rates, but with significant scatter around the implied relation. Comparison between H$\alpha$/H$\beta$ and} \OIII {-based mass outflow rates reveals differences of one order of magnitude (our sample, \citealt{Harrison2014}) or more (\citealt{McElroy2015}). This is presumably due to the volume effects discussed in Section \ref{sec:energy} and contamination from other kinematic components (including the gravitational potential and star formation). The scatter in mass outflow rates appears higher at lower AGN luminosities, with sources in the AGNIFS sample (e.g., \citealt{Storchi2010}) ranging roughly three orders of magnitude (note that all these studies use the same NIR emission lines, mainly [Fe II], same electron density, and similar outflow geometry assumptions). Sources with extended radio jets that appear aligned with the ionized gas outflows (squared circles in Fig. \ref{fig:lit_Mdot}) should be considered as upper limits, since radio jets can deposit mechanical energy into the ISM (e.g., \citealt{Morganti2013}) and can therefore (partly) power the observed outflows. Taking this into account, most low luminosity AGN can be reconciled with the relation implied by the Balmer line-based measurements from this paper and the works of \citet{McElroy2015} and \citet{Harrison2014}.} 

{Next, we compare the outflow rates of the local samples with those found at higher redshifts and higher AGN luminosities (Fig. \ref{fig:lit_Mdot}, bottom), focusing only on} \OIII {-based values. With the addition of more sources, we find} there is a correlation between the mass outflow rate and the AGN bolometric luminosity (Kendall's correlation $\tau$ of 0.32 at a confidence of p$<0.0001$ and a sub-linear slope of $0.37\pm0.07$). We stress that this correlation is not trivially expected. While the ionized gas mass correlates with the AGN bolometric luminosity, the electron density, the outflow size, and outflow bulk velocity all enter into the calculation of the mass outflow rate. 


\subsubsection{Kinetic energy outflow rates}

\begin{figure}[bpt]
\begin{center}
\includegraphics[width=0.45\textwidth,angle=0,trim={25 30 15 10},clip]{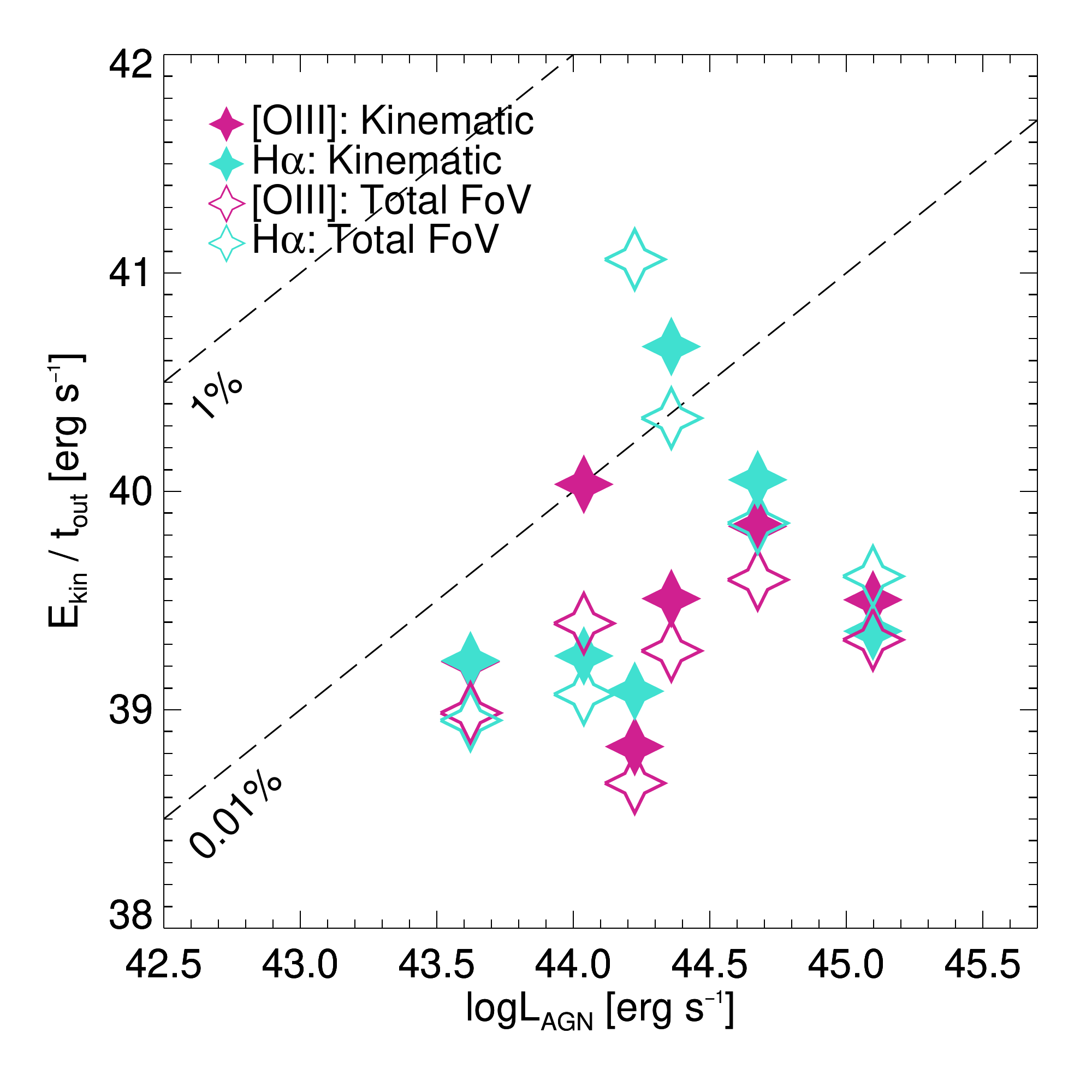}
\includegraphics[width=0.45\textwidth,angle=0,trim={25 30 15 10},clip]{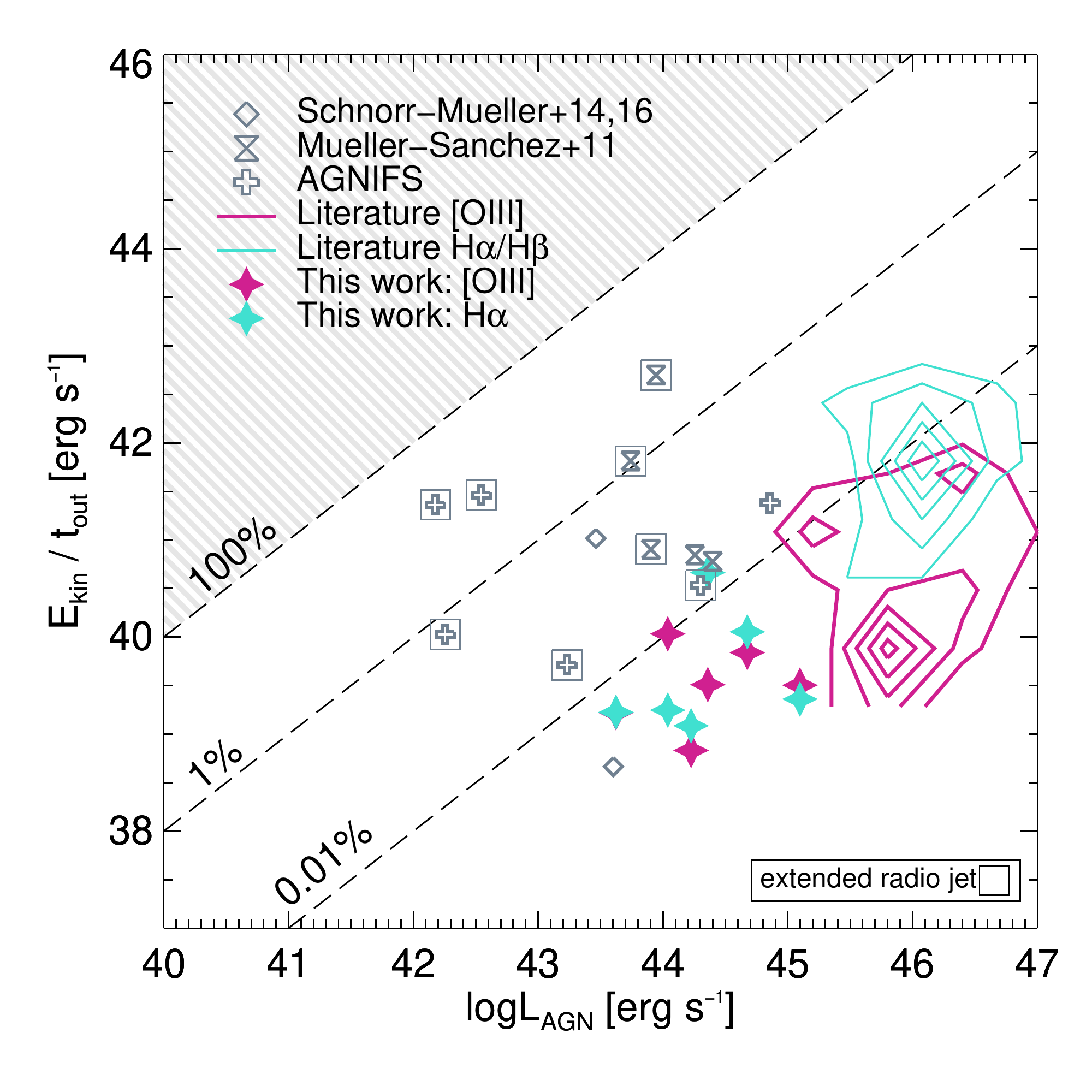}
\caption{Kinetic energy output of the outflow as a function of the AGN bolometric luminosity. Top: Measurements based on \OIII\ (violet) and H$\alpha$ (turquoise), using kinematic sizes (filled stars) and the IFU FoV (open stars), respectively. Bottom: Same as top panel but now also showing {high luminosity/high redshift} literature samples as contours for \OIII - and H$\alpha$/H$\beta$-based measurements. {Low luminosity/low redshift literature samples are shown with the same symbol and color notation as Fig. \ref{fig:lit_Mdot} (top)}. Diagonal lines denote energy conversion efficiency factors. The gray shaded area marks the region where the outflow kinetic energy output would exceed the AGN bolometric luminosity. {Large squares indicate sources with extended radio jets aligned with the ionized gas outflows.}}
\label{fig:lit_kin_energy}
\end{center}
\end{figure}

We calculate the integrated kinetic energy outflow rates (kinetic luminosities) assuming that $\dot E_{\mathrm{kin}}=E_{\mathrm{kin}}/t_{\mathrm{out}}$ and summing over all spaxels within the outflow region. Energy outflow rates range between 10$^{39}$ and 10$^{40}$ erg s$^{-1}$ for both \OIII\ and H$\alpha$ (Table \ref{tab:rates}), reflecting the \OIII\ luminosity dynamic range probed by our sample. We plot the outflow kinetic energy output, $\dot E_{\mathrm{kin}}$, as a function of the AGN bolometric luminosity in Fig. \ref{fig:lit_kin_energy} (top). Our sources lie mostly around the 0.01\% line and below 1\% efficiency, depending on whether we use the IFU FoV (open symbols) or the kinematic size of the outflow (filled symbols), thus containing a very small fraction of the total AGN energy output. 

H$\alpha$-based measurements (turquoise symbols) are generally consistent with \OIII-based measurements (violet symbols). Kinetic energy output measurements based on the IFU FoV are on average smaller than those based on kinematic sizes, a result of larger outflow sizes (Fig. 11 in Paper I) and therefore longer outflow timescales. Note that the flux-weighted bulk velocity is dominated by the core spaxels and therefore contributes little to the differences between the gold and black stars in Fig. \ref{fig:lit_kin_energy}. 

We present the conglomeration of all literature studies of high luminosity Type 2 AGN in the form of contours, as we are mainly interested in the range of outflow energy output rates (Fig. \ref{fig:lit_kin_energy}, bottom), {while local Type 2 AGN of similar luminosity to our sample are plotted individually. We note again that the values plotted in Fig. \ref{fig:lit_kin_energy} are not the reported energy outflow rates in the literature but instead are recalculated estimates based on the} \OIII\ {and Balmer line luminosities and kinematics (velocity and velocity dispersion) reported in individual papers, using Eqs. \ref{eq:massha}, \ref{eq:massoiii}, \ref{eq:kina}, and assuming a uniform electron density of 500 cm$^{-3}$. As in Fig. \ref{fig:lit_Mdot}, for studies without this information we use the reported values in the respective papers.} 

Literature sources have efficiencies lower than 0.01\%, consistent with our results, and may suffer from an overestimation of the size of their outflows, an underestimation of their bulk outflow velocities, and consequently an overestimation of their outflow timescales. Correcting for these effects may bring these objects closer to the 0.01\% line, as is clearly exhibited by the on average 0.2-0.5 dex higher values between golden and black stars in Fig. \ref{fig:lit_kin_energy}. Kinetic energy outputs from the literature based on H$\alpha$/H$\beta$ lie on average above the \OIII -based ones and are consistent with the trend in our sample. 

{As in the case of the mass outflow rates, low luminosity AGN from the literature show very different energy outflow rates, ranging from as high as $\sim10$\% to as low as $\sim0.0001$\%. In fact, some of these sources show very high energy conversion efficiencies consistent with previously reported molecular outflows (e.g., \citealt{Cicone2014}), despite the much higher mass content in the molecular phase of the ISM. As such, we caution that given the potential importance of mechanical energy injection from radio jets (and potential contamination from ongoing star formation), the true radiatively-driven energy injection rate from the AGN may be significantly lower.}

{The measured efficiencies for the combined sample of outflows are on average lower than those reported in previous studies of more luminous Type 2 AGN. The same effects we discussed in the previous section also apply here, with differences in the way kinematics are measured having more weight due to the $v_{out}^2$ in Eq. \ref{eq:kina}. We therefore stress that these are order of magnitude calculations but argue that our careful kinematic and spatial decomposition offers a more accurate estimation of the mass and kinetic energy outflow rates in these objects.} It is therefore reassuring that the energy conversion efficiencies calculated for our sources are roughly consistent with the values reported recently by \citet{Ebrero2016} based on gas kinematics derived from X-ray and UV absorbers in Seyfert Type 1 AGNs ($\sim0.03-0.2\%$ of the AGN bolometric luminosity).

\section{Discussion}
\label{sec:discussion}

\subsection{Emission line decomposition and the true outflow properties}
Decomposition of the emission lines, identification of the different kinematic components, and proper estimation of the outflow size are crucial in the accurate determination of the ionized gas mass entrained in the outflow (as shown in the previous sections). Integrating luminosities over the IFU FoV results in minimal differences of mass for \OIII\ but can lead to significant overestimation of the mass calculated based on H$\alpha$. Similarly, using photometric methods to estimate the size of the outflow, as we showed in Paper I, leads to an overestimation by a factor of 2 for \OIII\ and by a larger factor for H$\alpha$ compared to the kinematic size. We argued that the broad component of \OIII\ within the kinematically defined size of the outflow presents the best proxy for the total ionized gas mass in the outflow.

While failure to consider the kinematics decomposition and proper outflow size leads to an overall overestimation of the measured mass of the ionized gas in the outflow, the bulk velocity of the outflow is instead underestimated. This is a result of taking into account the kinematics due to the gravitational potential of the galaxy. Failure to accurately constrain the region dominated by the AGN-driven outflow is further aggravated by the fact that ongoing star formation in the outer parts of the galaxy can also present outflows (e.g., \citealt{Lehnert1996,Westmoquette2011,Ho2014}). We showed that emission lines can be fairly reliably decomposed into gravitationally- and outflow-driven components. Together with spatial classification of the photoionization field, this allows the spatial and spectral decomposition of the emission lines and hence an accurate estimation of the bulk velocity of the outflow.

A final parameter that is usually estimated for these outflows is their lifetime, based on which mass and kinetic energy outflow rates are derived. This timescale directly depends on the size and bulk velocity. The flux-weighted bulk velocity of the outflow is dominated by the core where the AGN-driven outflow dominates the kinematics. Therefore this leads to a net overestimation of the outflow lifetime by a factor of a few, for outflow sizes derived based on various photometric methods.

The degree of misestimation of the size, mass, kinetic energy, and outflow lifetime in previous studies is not straightforward to constrain. However, it is important to keep in mind that the effects of size and mass overestimation may apparently cancel out the bulk velocity underestimation: a factor of $2\times2$ overestimation of the mass due to the combined effect of summing over the IFU FoV and considering the full emission line profile would be cancelled out by a factor of 2 underestimation of the kinematics, again due to the consideration of the full emission line profile. Despite this fortuitous coincidence, the individual estimations of the mass and bulk outflow velocity would be wrong by at least a factor of 4.

\subsection{Outflow geometry and the VVD diagram}

\begin{figure}[bpt]
\begin{center}
\includegraphics[width=0.4\textwidth,angle=0,trim={0 5 2 0},clip]{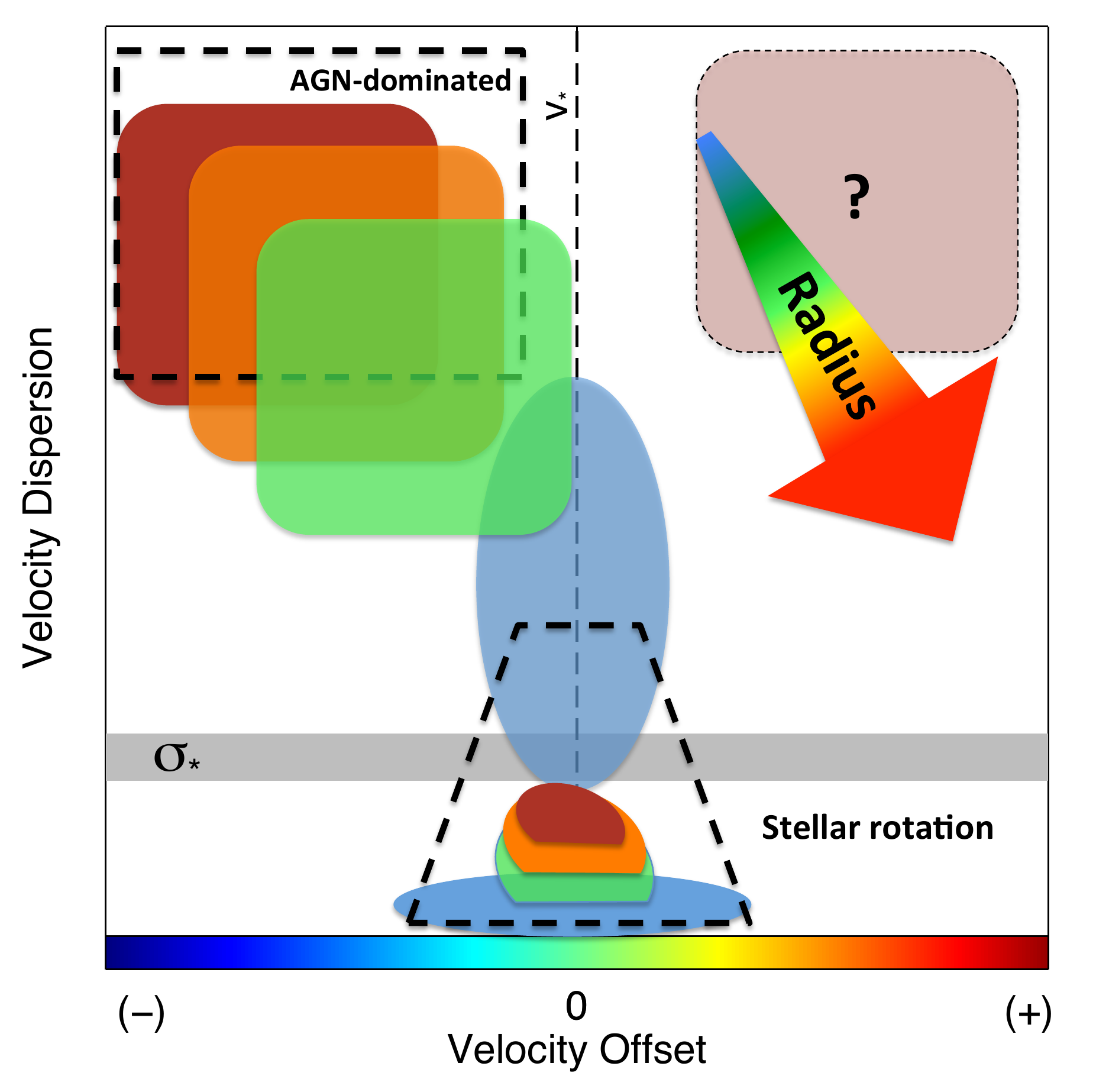}
\caption{A cartoon VVD diagram, summarizing the kinematic components observed in our sample. Colored regions denote different classes in the BPT diagram: Seyfert (red), LINER (cyan), composite (pink), and star-forming (blue). The arrow indicates the observed trend of decreasing velocity and velocity dispersion with distance from the nucleus. Broad and narrow components of the emission lines are clearly separated in comparison with $\sigma_{*}$ (gray-shaded area). Dashed lines define rough regions of the VVD diagram where different processes appear to dominate.}
\label{fig:VVD_cartoon}
\end{center}
\end{figure}

We established that distinct kinematic components (e.g., rotation, outflow) are present in all of the objects studied here and presumably in AGN host galaxies in general. These otherwise apparently cospatial components are clearly separated on the VVD diagram, particularly if we consider the additional information provided by the BPT classification. In Fig. \ref{fig:VVD_cartoon}, we draw a cartoon of the different regions of the VVD diagram and how they reflect the kinematics of the ionized gas. With the stellar kinematics as the reference point, AGN-dominated spaxels lie predominantly in the upper left corner of the plot. Conversely, the gravity-dominated gas lies in the lower part of the VVD. Based on previous results, shocks and stellar outflows do not show preferentially negative velocities but instead are characterized by increased velocity dispersion compared to $\sigma_{*}$  (e.g.,\citealt{Westmoquette2011,Ho2014,Ho2016}). We therefore expect them to dominate in the middle part of the VVD diagram. There is also a clear dependence with radial distance, both in terms of velocity dispersion and velocity. 

The conspicuous absence of sources/spaxels situated in the upper right corner of the VVD diagram constrains the geometry and obscuration of the outflow. We expect some degree of dust extinction for all of our targets, as evidenced by the presence of a rotation-dominated stellar and gaseous (H$\alpha$) disk in all of them. Together with a biconical outflow, this can explain the predominantly blueshifted emission observed (e.g., \citealt{Liu2013,Bae2014,Woo2016,Bae2016}). Given the prevalence of AGN with blueshifted \OIII\ emission (e.g.,\citealt{Woo2016}), this may imply the preference of strong AGN-driven outflows to occur in systems with stellar and gaseous disks in their central regions. 

\subsection{What is driving the outflows?}

The velocities and velocity dispersions, {reported in Paper I and summarized here in Table \ref{tab:sample}}, certainly imply that the AGN energy output is driving these outflows. {Large-scale winds driven by star formation are known to show more modest kinematics (with velocities for non-starburst galaxies comparable to or below the stellar velocities, e.g., \citealt{Veilleux2005,Westmoquette2011,Ho2014,Ho2016}}). Furthermore, at kinetic energy rates of $\sim10^{41}$ erg s$^{-1}$ they would require star formation rates of $\sim7$ M$_{\sun}$ yr$^{-1}$ (\citealt{Veilleux2005}). We calculate H$\alpha$-based star formation rates of $<0.5$ M$_{\sun}$ yr$^{-1}$ using the narrow component of H$\alpha$. Even if we assume that the total H$\alpha$ luminosity is coming from stellar ionization, we still estimate star formation rates of $<1-2$ M$_{\sun}$ yr$^{-1}$. It therefore seems unlikely that the observed outflows are powered by ongoing star formation. Instead, we clearly showed that the region where the outflow kinematics dominate the dynamics of the ionized gas is ionized by the AGN. Moreover, our BPT diagrams based on velocity channels exhibited that the most blueshifted emission appears to be ionized by the AGN. This is further supported by the broad trend for increasing outflow mass and kinetic energy rate with increasing AGN bolometric luminosity.   

There is a host of different models that explain the launching and powering of strong outflows from the AGN energy. It is accepted that the source of the outflow energy is the photons released by the accreting matter close to the event horizon of the supermassive black hole. These initially drive an optically thick inner wind through radiation pressure (e.g., \citealt{King2003,King2010}). However, whether at larger scales the outflows are driven by the injection of energy (in which case the inner wind expands adiabatically while conserving its energy) or momentum (where the total momentum of the inner wind is transferred to the gas after a short phase of radiative cooling) into the ISM is still under debate (e.g., \citealt{Costa2014}). Both scenarios have been discussed extensively in the literature (e.g., \citealt{Koo1992,Silk1998,King2003,King2011,Faucher2012, McQuillin2013,Zubovas2014b,King2015}). It is generally accepted that momentum-driven winds may be relevant close to the black hole ($<$ 1 kpc, e.g., \citealt{Ciotti1997,King2011}), while at larger distances the outflow becomes energy-driven as the cooling time of the outflowing material becomes larger than the flow time. 

As we discussed in Paper I, the outflows observed in our sample have kinematic sizes ranging between 1 and 2 kpc, with the most extreme velocities observed within their central kpc. This would imply that the observed outflows are momentum- rather than energy-driven, even though our data could cover the transition, if any, between the two regimes. Alternatively, \citet{Costa2014} present radial profiles of the outflow velocity derived from analytical models of energy- and momentum-driven outflows. At the scales relevant to our data, energy-driven outflows show a similar acceleration pattern as the one implied by our data (Paper I). It is plausible that we are observing the seeing-convolved momentum-driven regime of these outflows (which should dominate at scales smaller than our seeing size) that is however overlaid on the spatially extended energy-driven regime. The transition between the two appears to be sensitive to the density profile and spatial distribution of the ambient medium (e.g., \citealt{Faucher2012,Zubovas2014b}).

Nonetheless, we cannot exclude the contribution of shocks, either by the AGN itself or by radio jets, to the total energy content of the outflows. All our sources are classified as radio-quiet based on their FIRST 1.4 GHz flux-density measurements. In addition, the majority of the literature sources are also radio-quiet. Therefore, we expect the contribution from radio jets to the observed outflows to be minimal (see also \citealt{Woo2016}). One of our objects, J0918, has a LINER-like nucleus in the spatially resolved BPT map in Fig. \ref{fig:all_bpt_map}, which may be indicative of shock ionization. \citet{McElroy2015} also find evidence for strongly shocked gas, again based on flux ratio measurements. This implies that shocks may play an important role for some AGN-driven outflows, although separating the effect of shock and photoionisation solely based on \NII/H$\alpha$ is challenging.

\subsection{Does the gas escape?}
\label{sec:escape}

Beyond the heating of the ISM by the AGN, effective suppression of star formation in the host galaxy of the AGN during the finite AGN duty cycle ($\sim10^{7-8}$ years, \citealt{Schmidt1966,Marconi2004}; but also see \citealt{Schawinski2015}) necessitates that the gas is ejected from the galaxy. For this  to happen, the outflow should accelerate the entrained gas to velocities larger than the escape velocity of the host galaxy halo. 

%
To explore whether this is possible, we use the simple assumption of a singular isothermal sphere \citep{Rupke2002}, the escape velocity of which is defined at a distance r based on the maximum rotational velocity $v_{c}$, and its truncation radius r$_{\mathrm{max}}$
\begin{equation}
v_{\mathrm{esc}}(r)=\sqrt{2}v_{c}[1+\ln{\frac{r_{\mathrm{max}}}{r}}]^{1/2}.
\end{equation}
The maximum rotational velocity, $v_{\mathrm{c}}$, is directly measured from stellar velocity maps, ranging from 200 to less than 400 km s$^{-1}$ (projected values). The truncation radius, $r_{\mathrm{max}}$, is the radius at which the gas is considered free from the gravitational pull of the galaxy, usually assumed in the literature to be 100 kpc. The distance r is usually defined as the terminating region of the outflow, which in previous studies was assumed to be 10 kpc (e.g., \citealt{Rupke2002,Harrison2012}), therefore having an $r_{\mathrm{max}}/r$ ratio of $\sim10$. The previously assumed outflow size is clearly inappropriate for our sources (we measure kinematic sizes below 2 kpc), and potentially an overestimation also for previous studies. Here, we consider a range of $r_{\mathrm{max}}/r$ ratios from 10, a typical literature value, to 50 (assuming $r_{\mathrm{max}}$ of 100 kpc and r of 2 kpc, an upper limit based on our outflow size measurements). The maximum velocity of the outflow can be approximated as a combination of the measured velocity, $v$, and velocity dispersion, $\sigma$, of the ionized gas. \citet{Rupke2005} define the maximum outflow velocity as $u_{\mathrm{max}}=(|v|+\frac{1}{2}\mathrm{FWHM})=(|v|+1.17\sigma)$. For most of our sources the ionized gas entrained in the outflow should have just enough kinetic energy to escape the gravitational potential of their host galaxies. It is interesting to note that had we considered the full emission profile of \mbox{[O\,\textsc{iii}]} and H$\alpha$, we would have concluded that at least half of the observed outflows cannot expel the gas from the gravitational potential of their host.

Here we have not taken into account projection effects or the stellar velocity dispersion in our estimate of the maximum stellar rotation velocity. Moreover, given that Gemini/GMOS FoV only probes the central part of the galaxy (2-4 kpc), the measured maximum rotation velocities may be underestimated, leading to larger escape velocities. Projection further affects the measured maximum velocity of the gas. For Type 2 AGNs, we can consider a maximum inclination angle of the outflow of $\sim$45$^\circ$ to our line of sight. This would then imply a maximum deprojection factor of $\sim1/\sin{45^\circ}=1.4$. Given these caveats, we cannot offer a definitive answer as to whether the outflowing gas can escape its host galaxy. IFU observations with very wide FoV instruments like MUSE (\citealt{Henault2003}) and CITELE (\citealt{Grandmont2012}) may be able to address these questions more robustly.

\subsection{The potential of outflows as feedback agents}

We showed that the size of these outflows is smaller than a couple of kpc with most of the outflow's energy and mass contained within the very center of the galaxy ($<1$ kpc). Furthermore, while the observed outflows may be able to accelerate the gas to velocities above the escape velocity of their hosts, whether the gas escapes the halo is not clear. \citet{Zubovas2012} find that for AGN-driven outflows to sweep the gas out of their host galaxies they need kinetic energy outputs $\sim5\%$ of the bolometric AGN luminosity (under the assumption of the AGN accreting near its Eddington limit). As we showed in Fig. \ref{fig:lit_kin_energy}, {we calculate} kinetic energy outputs several orders of magnitude below this  theoretical expectation. All the above combined make the impact of these outflows on their surroundings highly debatable.

Perhaps more telling, is that we detect evidence for circumnuclear star formation in all 6 of the objects presented here. Despite these being some of the best candidates in the local universe, it appears that AGN feedback has not managed to quench ongoing star formation even close to the SMBH. Based on their SDSS spectra, these sources actually lie on or above the `Main Sequence' of star formation (e.g., \citealt{Rodighiero2010b,Elbaz2011}), implying that they are globally forming stars at rates consistent with galaxies of similar stellar mass at their present epoch. More intriguingly, we established kinematic links between the termination of the AGN-driven outflows and the circumnuclear star formation (see Appendix). This may imply a different type of feedback, that instead of quenching star formation, it helps to enhance it or even trigger it {(e.g., \citealt{Silk2013})}. This is not a new concept and there are observational evidence for positive feedback at the boundaries of AGN-driven outflows in some AGNs {(e.g., \citealt{Bicknell2000,Croft2006,Cresci2015})}. 

Alternatively, star formation may be a natural consequence of an AGN-driven outflow that experiences instabilities, leading to fragmentation of the ISM shocked by the outflow. The result is a two-phase gas that contains molecular clouds encased in shells of ionized gas (e.g., \citealt{Nayakshin2012,Zubovas2014a}). This is expected to occur within a few kpc from the central heating source, where efficient cooling via brehmsstrahlung and metal line emission is possible. Our detected circumnuclear star formation is at scales between 2-4 kpc and could therefore be the observational signature of these dense molecular clumps, which have in some cases been directly observed at similar scales (e.g., \citealt{Cicone2012}). One caveat to this explanation is that both models and previous observations have focused on quasar-like AGN that were assumed to accrete at near-Eddington accretion rates. This is not always the case for our sources, four of which have rough estimated Eddington ratios $<0.1$. 

A third possibility is that we are simply detecting the underlying star formation of the stellar disk in these galaxies, which is independent from the presence or properties of the AGN-driven outflow. In this scenario, the central part of the star forming disk is either hidden by the projected emission of the AGN or is indeed suppressed. Spatially resolved IFU observations of AGNs residing in matching host galaxies but without prominent outflows will help us elucidate the origin of the detected star formation by investigating both the properties of the ionized gas and the presence and properties of nuclear and circumnuclear star formation. 

A final obstacle relating to the importance of AGN-driven outflows as universal agents of (negative or positive) feedback is the simple fact that even among luminous AGNs, only a fraction shows powerful outflows. From $\sim10\%$ up to $\sim50\%$ of all Type 2 AGN at L$_{\OIII}^{\mathrm{cor}}>10^{42}$ erg s$^{-1}$ do not show extreme kinematics in their \OIII\ line profile\footnote{This depends on the selection criteria for non-outflow AGN. Here we use the selection as \OIII\ velocity $<100$ km s$^{-1}$ and a velocity dispersion cut ranging from $<1.5\sigma_{*}$ down to $<1\sigma_{*}$.}, based on the sample of \citealt{Woo2016}, with this fraction increasing at lower \OIII\ luminosities (also see Figs. 9-11 in \citealt{Woo2016}). The question then arises what determines the emergence of these extreme outflows, beyond the luminosity of the central AGN? It is plausible that the duty cycle of these outflows is shorter than the AGN and therefore at any given time we are missing a fraction of them. However, the calculated timescales of outflows are of the order of a few to a few tens of Myrs (here and also, e.g., \citealt{Harrison2014}), which is roughly consistent with the AGN duty cycle. Alternatively, the gas fraction of the AGN hosts may be the deciding factor. Comparative studies between outflow and control non-outflow AGNs with ALMA will provide a robust handle of the cold gas mass and kinematics and consequently the gas fraction in these objects.

\section{Summary and Conclusions}
\label{sec:conclusions}

Let us summarize our findings:
\begin{itemize}
\item Emission flux ratios reveal a complex photoionization field. While the central parts of the galaxies are dominated by the AGN, all sources exhibit signs of ongoing star formation in the circumnuclear region. If we consider only the broad components of the emission lines, spaxels are exclusively classified as AGN-like (Fig. \ref{fig:all_bpt_map}).
\item There is a strong negative gradient of the \OIII/H$\beta$ flux ratio, with the broad component of the emission lines showing higher \OIII/H$\beta$ ratios (Fig. \ref{fig:bpt_rad}).
\item Blueshifted emission is associated with Seyfert photoionization (except for J1606), while low velocity gas (close to the systemic velocity) is classified as composite or star forming (Fig. \ref{fig:all_bpt_chan}).
\item The VVD diagram clearly separates AGN-driven kinematics from virial motion due to the gravitation potential. The former lie in the upper-left corner, while rotating gas populates the lower part of the VVD (Fig. \ref{fig:all_vvd_oiii}).
\item The mass and kinetic energy of the outflowing ionized gas ranges between 0.1 and $5 \cdot 10^{6}M_{\sun}$ and 10$^{53}$ to 10$^{55}$ ergs, respectively, depending on whether \OIII\ or H$\alpha$ is used. We estimate differences of up to one order of magnitude depending on whether we integrate over the whole IFU FoV or we use the full emission line profiles (Table \ref{tab:physics}).
\item The \OIII -based kinetic energy of the ionized gas is mostly dominated by the outflow but up to 60\% of the H$\alpha$-based kinetic energy may be unrelated to the outflow.
\item There is a positive correlation between L$_{\mathrm{AGN}}$ and the \OIII {-based} mass outflow rate (Fig. \ref{fig:lit_Mdot}).
\item Based on the kinetic energy output rate of the outflows, we calculate conversion efficiencies $<0.1\%$ for most sources in both our sample and those from the literature (Fig. \ref{fig:lit_kin_energy}){, on average lower than previously reported values.}
\end{itemize}

In this paper we established a clear link between the extreme kinematics observed in the \OIII\ and, partially, H$\alpha$-emitting ionized gas and the photoionizing emission from the AGN. We studied the BPT classification of ionized gas at the largest velocities and found it to be dominated by the AGN emission but at the same time concentrated in the central-most part of the galaxy. Together with the clear positive correlations between the mass and kinetic energy outflow rates and the AGN bolometric luminosity, this strongly indicates that,  for the majority of these sources, outflows are driven by the AGN. The fact that we detect both stellar and gaseous rotationally-supported disks in all 6 sources and evidence for ongoing circumnuclear star formation offers intriguing implications on the type of galaxies or conditions required for strong AGN feedback to develop. Vice versa, it is also plausible that the presence of strong AGN-driven outflows enhances circumnuclear star formation. 

Our analysis of both the sizes (in Paper I) and the outflow energetics and observed circumnuclear star formation (here) implies that, at least for this sample, AGN feedback might not impact the growth of their host galaxy significantly. This is an important result given that the {moderate luminosities} we probe are not in the quasar regime and therefore are more representative of the bulk of AGN-dominated galaxies out to the peak of cosmic AGN activity. In following papers we will explore this dataset further, focusing on the stellar population properties and the connection between these AGN-driven outflows and ongoing star formation in their host galaxies with a larger sample. 

\acknowledgments{We thank the anonymous referee for comments that have improved the clarity and presentation of this paper. This research was supported by the National Research Foundation of Korea (NRF) grant funded by the Korea government (MEST) (No. 2016R1A2B3011457 and No. 2010-0027910). This work was supported by K-GMT Science Program (PID: GN-2015A-Q-204) of Korea Astronomy and Space Science Institute (KASI). Based on observations obtained at the Gemini Observatory processed using the Gemini IRAF package, which is operated by the Association of Universities for Research in Astronomy, Inc., under a cooperative agreement with the NSF on behalf of the Gemini partnership: the National Science Foundation (United States), the National Research Council (Canada), CONICYT (Chile), the Australian Research Council (Australia), Minist\'{e}rio da Ci\^{e}ncia, Tecnologia e Inova\c{c}\~{a}o (Brazil) and Ministerio de Ciencia, Tecnolog\'{i}a e Innovaci\'{o}n Productiva (Argentina). This research has made use of NASA's Astrophysics Data System Bibliographic Services. For this research, we have made extensive use of the TOPCAT software (\citealt{Taylor2005}), which is part of the suite of Virtual Observatory tools. This research has made use of the NASA/IPAC Extragalactic Database (NED), which is operated by the Jet Propulsion Laboratory, California Institute of Technology, under contract with the National Aeronautics and Space Administration.}

\bibliographystyle{aa}
\bibliography{bibtex}

\begin{appendix}
\label{sec:appendix}
\vspace{10pt}

\section{Electron density radial profiles}
\label{sec:ne}
\begin{figure}[bpt]
\begin{center}
\includegraphics[width=0.22\textwidth,angle=0,trim={5 65 15 10},clip]{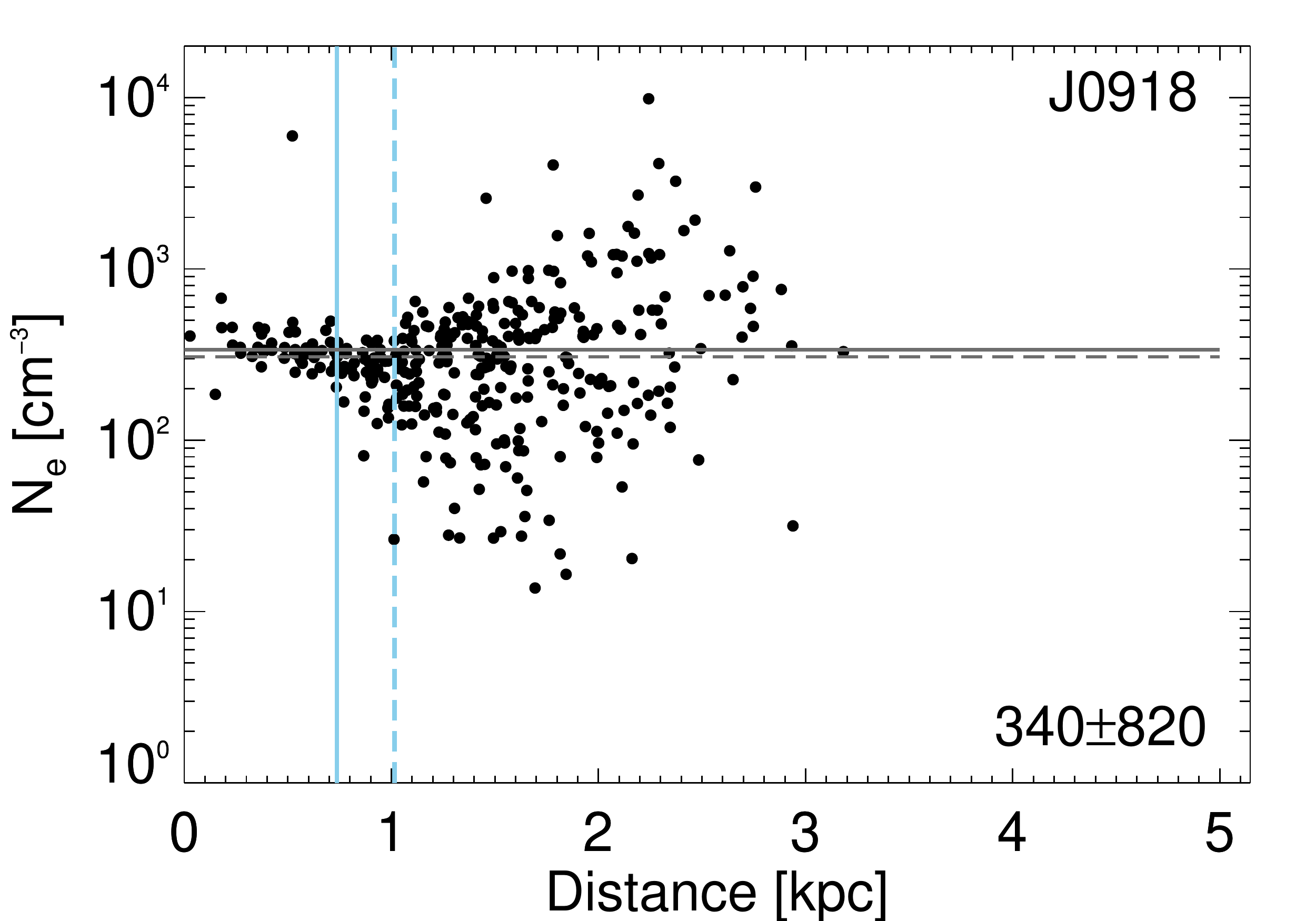}
\includegraphics[width=0.22\textwidth,angle=0,trim={5 65 15 10},clip]{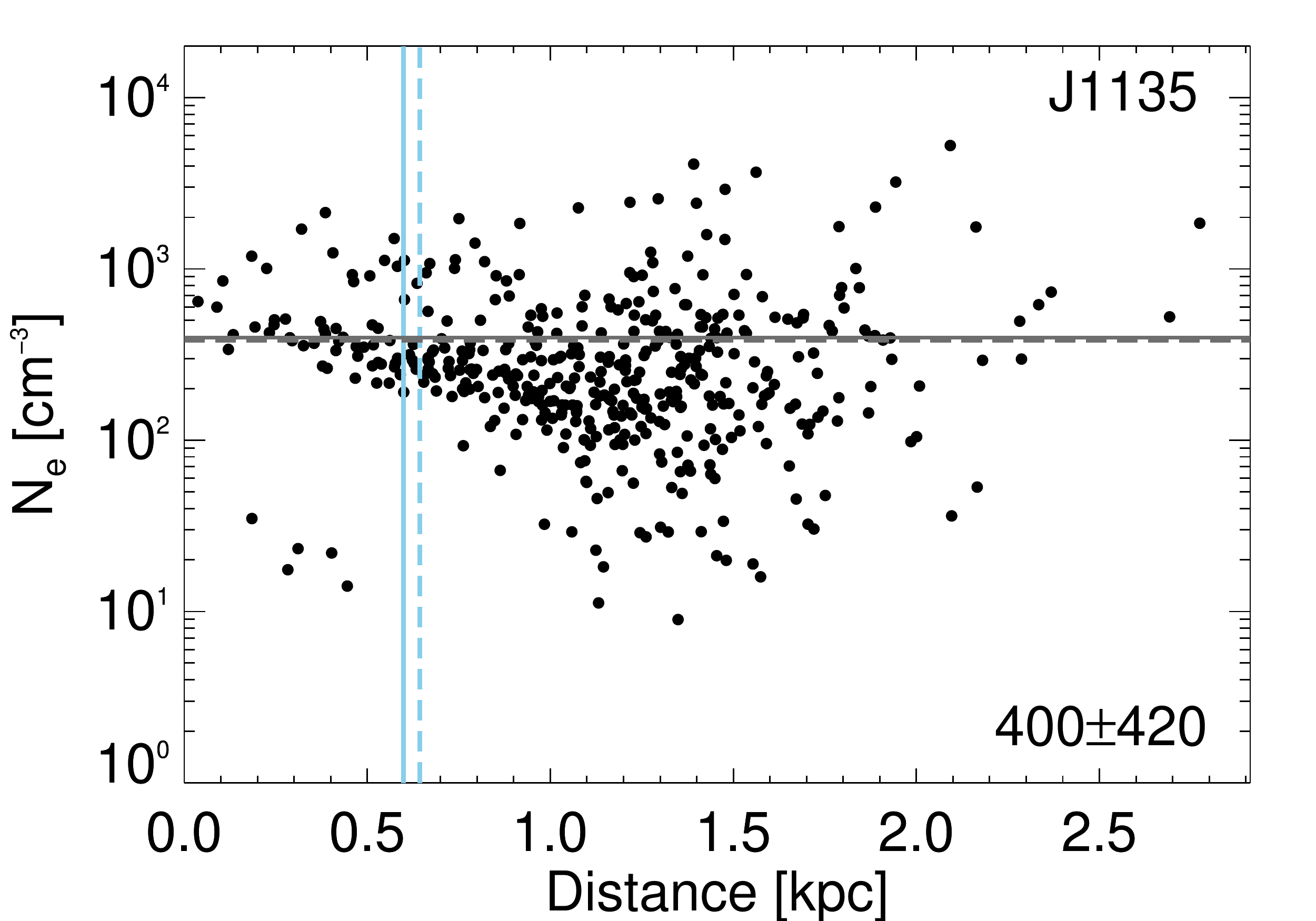}\\
\includegraphics[width=0.22\textwidth,angle=0,trim={5 65 15 10},clip]{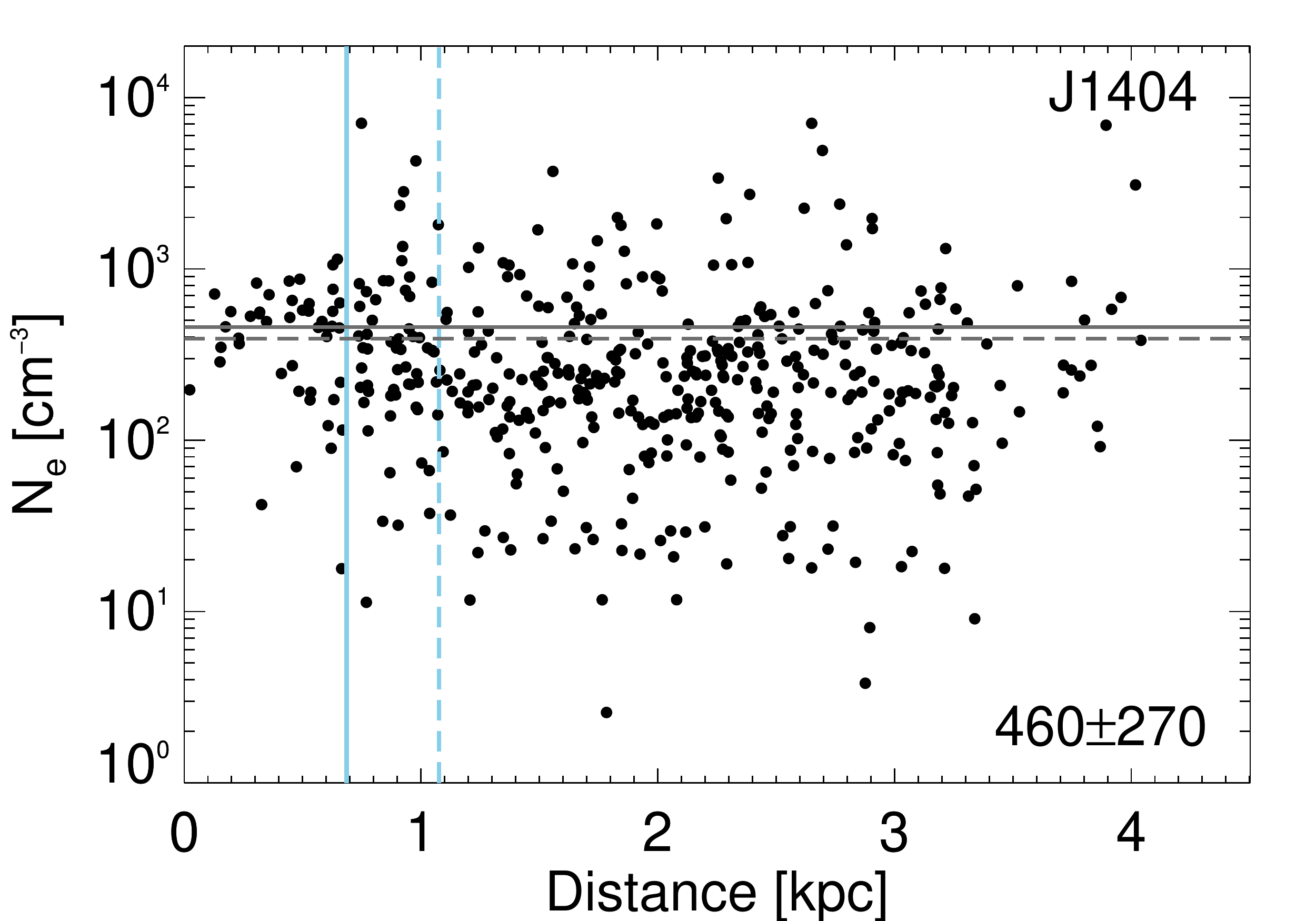}
\includegraphics[width=0.22\textwidth,angle=0,trim={5 0 15 10},clip]{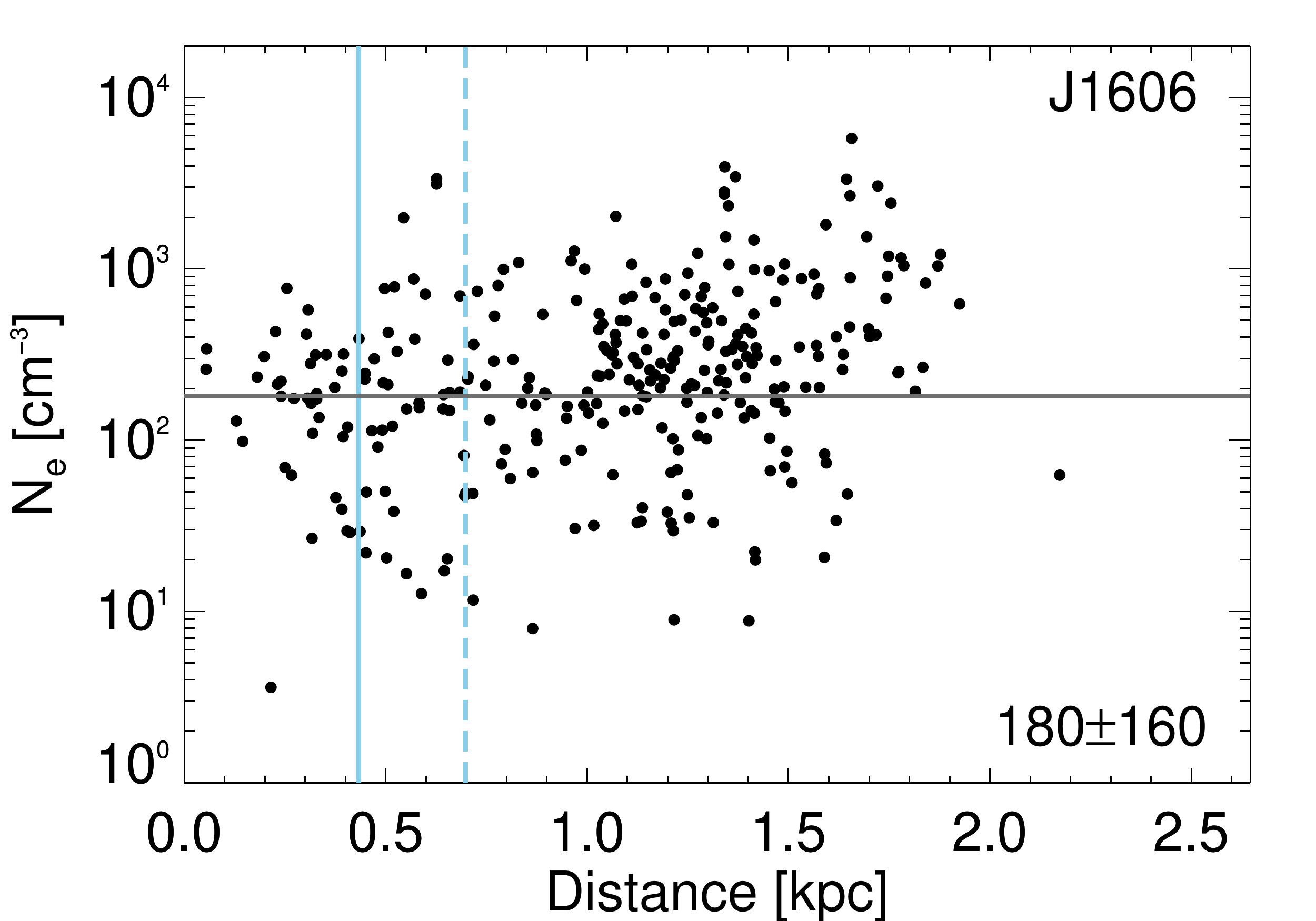}\\
\includegraphics[width=0.22\textwidth,angle=0,trim={5 0 15 10},clip]{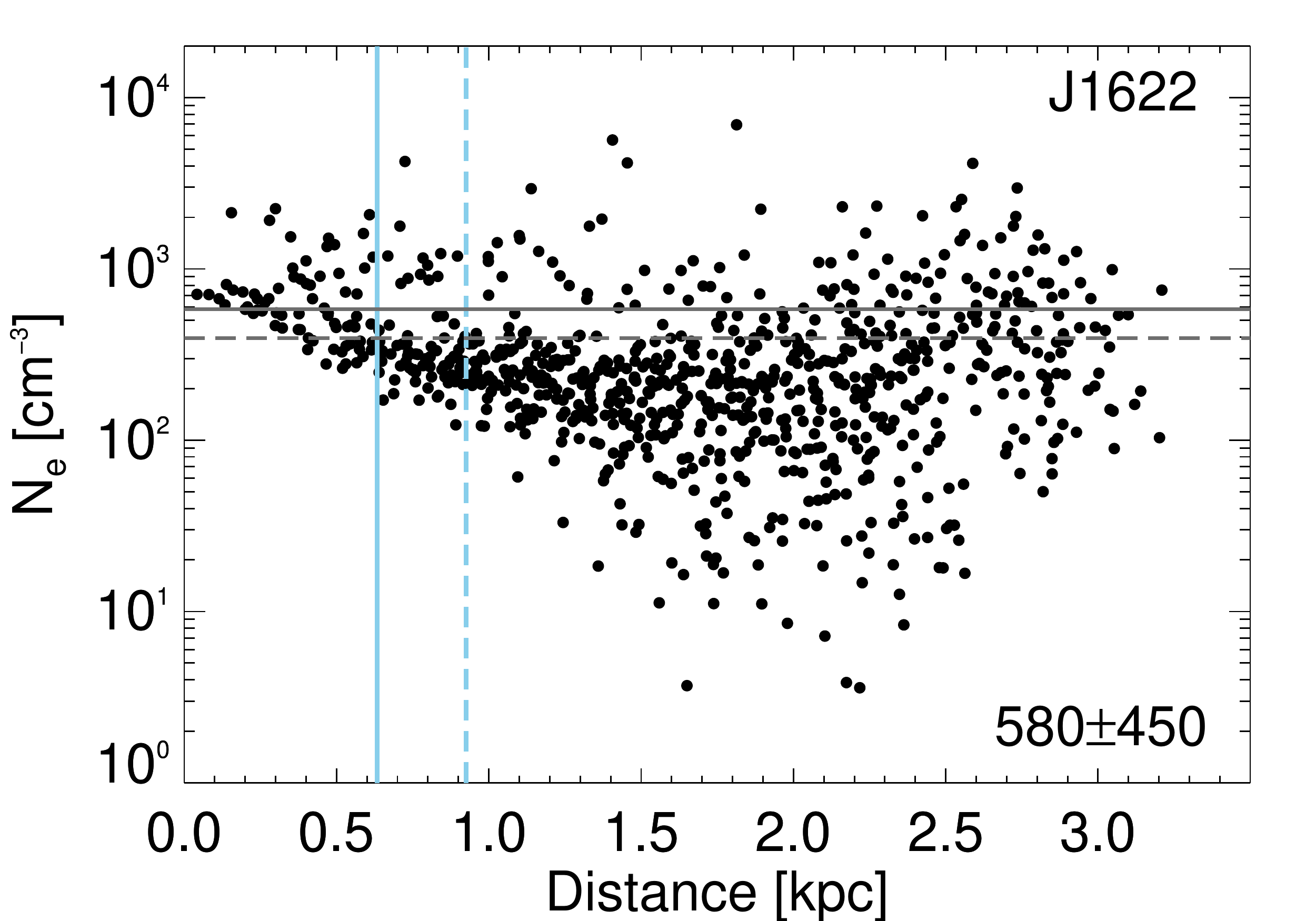}
\includegraphics[width=0.22\textwidth,angle=0,trim={5 0 15 10},clip]{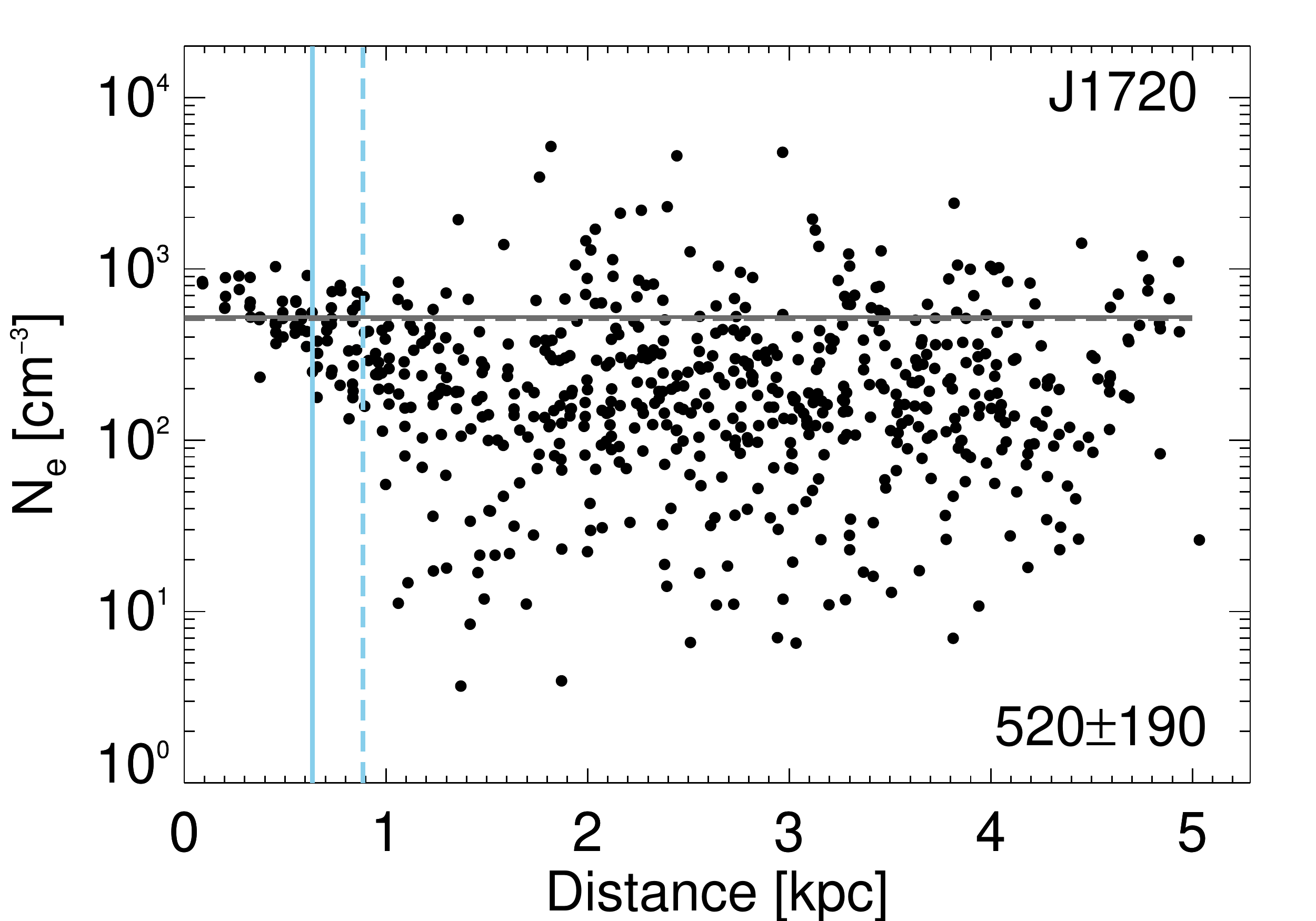}
\caption{Radial profiles of the electron density, N$_{e}$, based on the \SII\ doublet flux ratio. The vertical solid and dashed lines show the effective radius of the broad \OIII\ and broad H$\alpha$ components for each source. The horizontal solid and dashed lines show the median electron densities within one r$_{\mathrm{eff}}^{\mathrm{\OIII}}$ and one r$_{\mathrm{eff}}^{\mathrm{H\alpha}}$. In the lower right corner of each panel we give the median electron density and its standard deviation within one r$_{\mathrm{eff}}^{\mathrm{\OIII}}$. We adopt an S/N limit of $>3$ and flux ratios resulting in densities beyond the critical density range of \SII\ are discarded.}
\label{fig:all_ne}
\end{center}
\end{figure}

In Fig. \ref{fig:all_ne} we provide the radial profiles of the electron density, N$_{e}$, based on the measured flux ratio of the \SII\ doublet. We translate flux ratio values to electron densities based on, e.g., \citet{Osterbrock1989}. Electron densities beyond the critical densities of \SII\ are not considered ($<1$ cm$^{-3}$ and $>10^{4}$ cm$^{-3}$). We only consider spaxels for which both \SII\ doublet lines have S/N$>3$.

Four sources show an initial decreasing trend with radial distance, with typical electron densities at the very center between a few hundred to thousand particles per cm$^{-3}$. Beyond 0.5 kpc we observe increased electron densities and large scatter for a given distance. This is probably a combination of rapidly decreasing S/N for the \SII\ doublet, as well as intrinsically denser ISM outside the galactic core, within the stellar disk. Exception to this is J1404 and J1606, which show a fairly constant electron density but with significant scatter. Median electron densities within one r$_{\mathrm{eff}}^{\mathrm{\OIII}}$ are $\sim500$ cm$^{-3}$ but with very large uncertainties, indicative of the scatter. Given these measurements, the assumption of an expanding shell within an empty medium ($\sim0.5-10$ cm$^{-3}$, e.g., \citealt{Liu2013,Harrison2014}) is not supported and would overestimate the measured kinetic energy contained in the outflows.

\section{Comments on individual sources}
\subsection{J091808+343946}
\label{sec:J0918}

J0918 is mostly classified as a LINER, with a clear AGN core. A strong composite component emerges when considering the "narrow" classification. There appears to exist a half ring of circumnuclear star formation to the north of the nucleus, with AGN dominated spaxels emerging to the south/south-east of the nucleus. Interestingly, the "broad" classification is mostly consistent with LINER emission and shows an absence of clear broad component detection across all four emission lines around the continuum center of the source. Based on the spatially and spectrally resolved BPT diagram (Fig. \ref{fig:all_bpt_chan}), the center of the galaxy appears mostly in the LINER part of the BPT with the only exception the most blueshifted channel, which may be an artifact as \NII\ can be lacking a broad component at the very center therefore pushing the dark blue point to the left of the BPT. 

\subsection{J113549+565708}
\label{sec:J1135}

J1135 is purely dominated by the AGN photoionization. This is perhaps not surprising given that it has the highest \OIII\ luminosity among our sources ($10^{43.1}$ erg s$^{-1}$). Little differences are observed between the total, broad, and narrow component BPT maps. The BPT classification map based on the narrow component of the emission lines shows a clear half circumnuclear ring of LINER/composite emission, which would imply the presence of ongoing star formation. Interestingly, the broad component BPT maps reveals a pocket of LINER-like emission in the South-East of the nucleus. This would roughly coincide with the region of increased velocity and dispersion for the \OIII\ emission (shown in Paper I).

\begin{figure}[hbpt]
\begin{center}
\includegraphics[width=0.2\textwidth,angle=0]{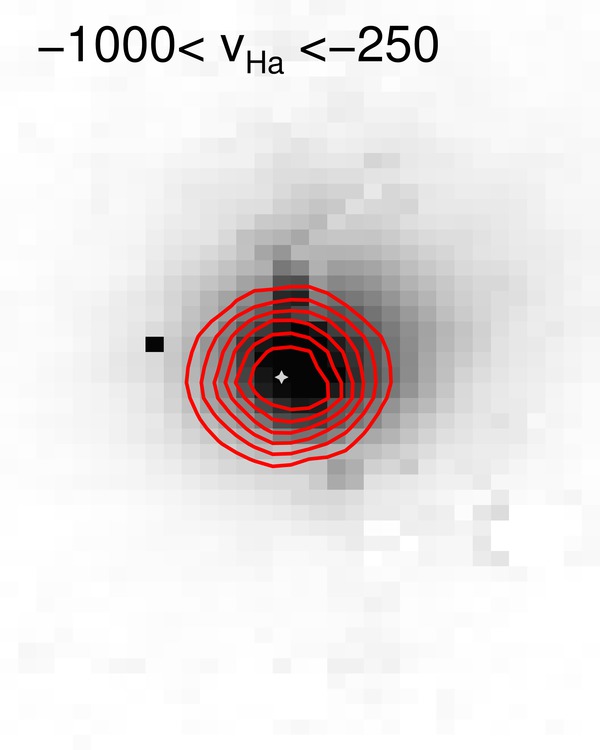}
\includegraphics[width=0.2\textwidth,angle=0]{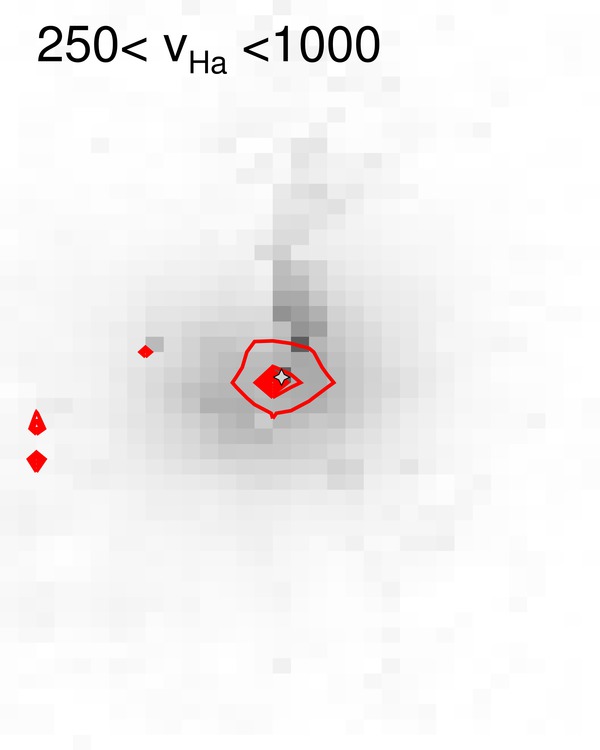}\\
\caption{Total H$\alpha$ flux maps of J1135 in two velocity channels: $-$1000 km s$^{-1}$ to $-$250 km s$^{-1}$ (left) and 250 km s$^{-1}$ to 1000 km s$^{-1}$ (right) overlaid with total \OIII\ flux contours in increments of 10\% of peak flux (red).}
\label{fig:J1135_Hastruct}
\end{center}
\end{figure}

In Fig. \ref{fig:J1135_Hastruct} we plot the total H$\alpha$ flux maps in two velocity channels: $-$1000 km s$^{-1}$ to $-$250 km s$^{-1}$ (top left) and 250 km s$^{-1}$ to 1000 km s$^{-1}$ (top right). For the blueshifted channel we observe a fairly round shape with an obvious protrusion along the South-North direction. Interestingly, a strong feature emerges in the redshifted channel, that follows the same direction as the peculiar structure seen in the H$\alpha$ velocity and dispersion maps. The fact that this feature is predominantly seen in redshifted emission and does not follow the \OIII\ flux distribution may indicate that this outflow is powered by ongoing star formation close to the nucleus. This scenario is corroborated by other circumstantial evidence: J1135 is detected at 1.4 GHz by the FIRST survey (\citealt{Becker1995}) with a k-corrected luminosity (assuming a radio spectral index of 0.8) $L_{\mathrm{1.4GHz}}=10^{38.9}$ erg s$^{-1}$, indicative of ongoing star formation (e.g., \citealt{Mauch2007}); it is heavily obscured in the X-rays and classified as Compton thick (\citealt{LaMassa2014}); and it has strong 3.3 $\mu$m PAH emission (\citealt{Yamada2013}).

\subsection{J140453+532332}
\label{sec:J1404}

The BPT classification map based on the narrow components reveals a strong star formation component, with the AGN at its center and an obvious transition through a composite region to a circumnuclear star formation ring is clearly observed. The circumnuclear ring of star formation starts at the same scale ($\sim1\arcsec$) as the observed ring of elevated velocity dispersion for narrow \OIII\ and broad H$\alpha$ emission (see Paper I). 

\begin{figure}[hbpt]
\begin{center}
\includegraphics[width=0.2\textwidth,angle=0]{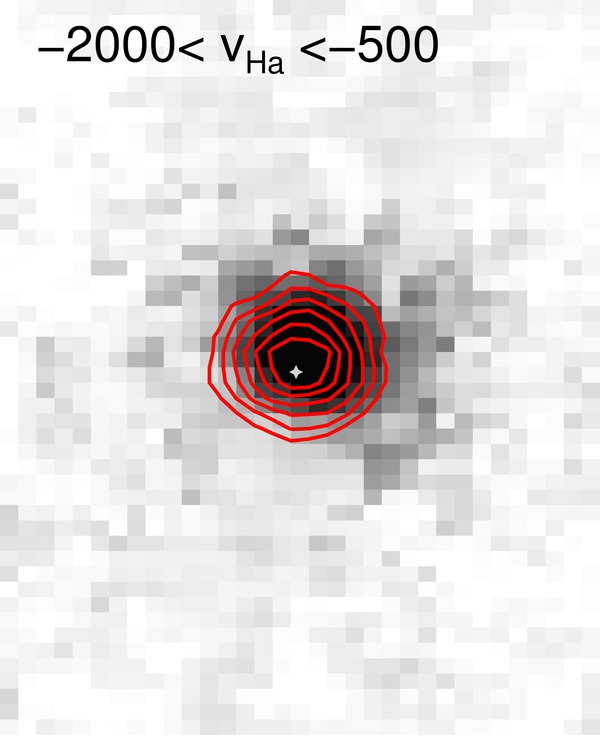}
\includegraphics[width=0.2\textwidth,angle=0]{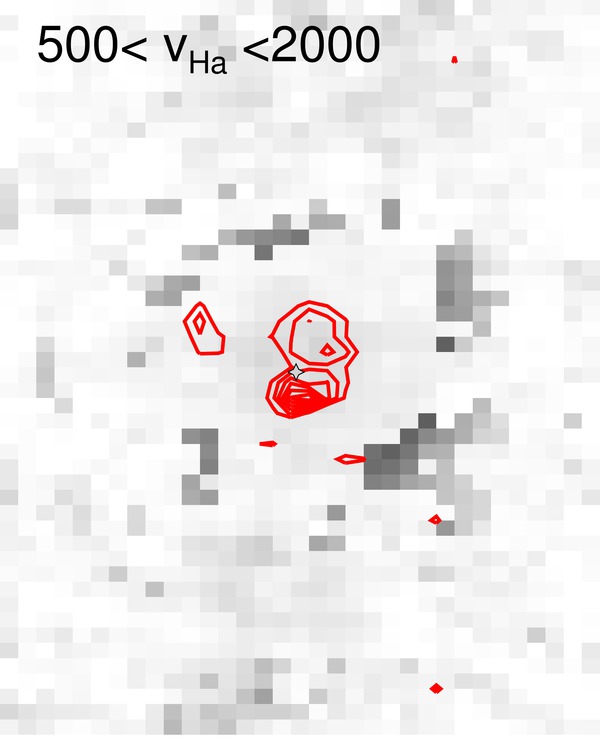}\\
\includegraphics[width=0.2\textwidth,angle=0]{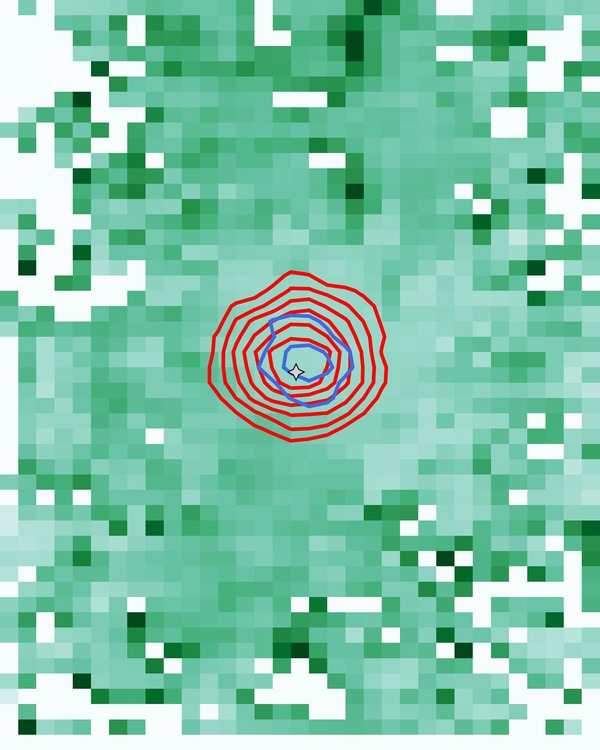}
\includegraphics[width=0.2\textwidth,angle=0]{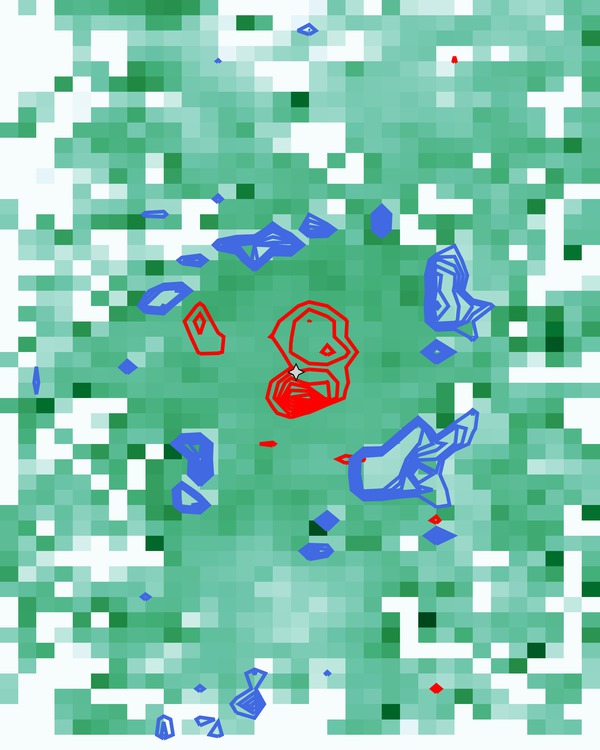}
\caption{Top: Total H$\alpha$ flux maps of J1404 in two velocity channels: $-$2000 km s$^{-1}$ to $-$500 km s$^{-1}$ (left) and 500 km s$^{-1}$ to 2000 km s$^{-1}$ (right), overlaid with \OIII\ flux contours in increments of 10\% of peak flux (red). Bottom: Corresponding \NII/H$\alpha$ flux ratio maps for the two velocity channels, overlaid with H$\alpha$ flux contours in increments of 10\% of the peak flux (blue).}
\label{fig:J1404_ring}
\end{center}
\end{figure}

In Fig. \ref{fig:J1404_ring} we present the H$\alpha$ flux (top) and \NII/H$\alpha$ flux ratio (bottom) plots for two different velocity channels: $-$2000 km s$^{-1}$ to $-$500 km s$^{-1}$ (left) and 500 km s$^{-1}$ to 2000 km s$^{-1}$ (right). The blueshifted channel shows very compact emission for both \OIII\ (red contours) and a weakly detected H$\alpha$ (peak flux 0.14 and 0.07 erg s$^{-1}$ cm$^{-2}$, respectively for \OIII\ and H$\alpha$). The blueshifted velocity channel map of \NII/H$\alpha$ shows a bubble of low [NIII]/H$\alpha$ flux ratio that approximately follows the shape of the blueshifted \OIII\ emission. The redshifted velocity flux map (top right) clearly shows a depression of H$\alpha$ emission at the center, while a ring of strong H$\alpha$ emission emerges at the same scales as the ring in J1404 velocity dispersion map. Moreover, we see a weakly detected \OIII\ component contained within the bubble (peak flux 0.03 and 0.05 erg s$^{-1}$ cm$^{-2}$, respectively for \OIII\ and H$\alpha$). The redshifted velocity channel map of [NIII]/H$\alpha$ shows now a bubble of increased \NII/H$\alpha$ flux ratios, the boundary of which is marked by the ring of H$\alpha$ emission. Beyond the redshifted H$\alpha$ ring the \NII/H$\alpha$ ratio decreases.
The above, together with the VVD and BPT diagram information at hand strongly suggest that we are observing an interaction region, between an outflow powered by the central AGN and traced by the blueshifted \OIII\ emission and a star-forming region and a star formation-driven outflow, which is traced by the redshifted H$\alpha$ emission. The central gas appears shocked by the outgoing AGN outflow, as indicated by the elevated \NII/H$\alpha$ flux ratios within the H$\alpha$ ring.

\subsection{J160652+275539}
\label{sec:J1606}

The classification map (Fig. \ref{fig:all_bpt_map}) based on the BPT diagram is unfortunately sparsely sampled, due to the relatively lower AGN luminosity of the source ($L_{\mathrm{\OIII}}=10^{42.2}$ erg s$^{-1}$, the lowest in our sample) and apparently high extinction. The BPT classification here is particularly difficult due to the very low flux of H$\beta$, with most spaxels not having a significant detection (S/N$<3$). Nevertheless, we observe a qualitatively similar behavior with previous sources. The nucleus is AGN-dominated, while a ring of composite or star formation-like emission encircles it. This ring, at $\sim1.5\arcsec$, roughly coincides with the transition in the kinematics maps of the narrow emission line components from the strongly blueshifted velocity field of the nucleus to the stellar disk-dominated kinematics. The BPT map based on the broad emission component is even more sparsely sampled, with only a few spaxels having a broad component detected in all 4 emission lines. Nonetheless, these spaxels indicate a Seyfert-like nucleus.

\subsection{J162233+395650}
\label{sec:J1622}

The BPT classification based on the narrow component emission is symmetric, with a strong AGN nucleus surrounded by two rings of composite and star formation-dominated regions. The BPT map based on the broad component emission on the other hand is mostly AGN-dominated and appears concentrated around the nucleus, with hints of a LINER-like ring around it. Interestingly, the ring of elevated velocity and dispersion observed in the broad \OIII\ (Paper I) matches, in scale, with the transition region between the AGN and LINER/composite regions in the BPT map. Conversely, the ring seen in the narrow \OIII\ second moment maps matches in scale with the transition region between the composite and star formation-dominated part of the BPT map. The elongated structure of blueshifted broad H$\alpha$ emission matches the extrusion of AGN-classified spaxels within the composite or star formation ring that surrounds the nucleus (Fig. \ref{fig:all_bpt_map}). The same elongated feature shows significantly increased \NII/H$\alpha$ flux ratio, both for the total profile and the broad component, indicative of a shocked ionized gas region. The BPT classification, together with the kinematics of the region imply that we are observing the AGN outflow breaking through into the surrounding, presumably denser, star-forming ISM region.

J1622 is unique in our sample due to its \OIII\ emission being strongly redshifted with respect to the systemic velocity. J1622 is detected in the radio at 1.4 GHz from the FIRST survey, having a radio luminosity L$_{\mathrm{1.4GHz}}=10^{38.6}$ erg s$^{-1}$, corroborating strong star formation. More intriguingly, the galaxy shows strong He$\lambda$4686\AA\ (\citealt{Shirazi2012}) emission, which is a good tracer of stellar outflows from massive stars and particularly Wolf-Rayet stars. This implies that strong contribution from stellar winds may be affecting the measured kinematics. 

\subsection{J172038+294112}
\label{sec:J1720}

The spatial distribution of the BPT classification shows an AGN-dominated nucleus that is encircled by rings of composite and star formation regions. When considering only the narrow emission component, the BPT classification shows a strong star formation region. Some star formation is also detected in the spiral arms, even though the weak H$\beta$ emission line does not allow proper sampling of the BPT classification in the outer parts of the IFU. The BPT map based on the broad emission line components is consistent with the previous sources in that it reveals an AGN-dominated core that appears to be unresolved (compared to the seeing size).




\end{appendix}
\end{document}